\pdfoutput=1
\documentclass[11pt,a4paper]{article}

\usepackage{jheppub}
\usepackage{mystyle}

\title{Anyon Condensation in Virasoro TQFT: Wormhole Factorization}
\author{Shunta Takahashi}
\affiliation{Research Institute for Mathematical Science (RIMS), Kyoto University, Kyoto 606-8502, Japan}
\date{}
\emailAdd{shunta@kurims.kyoto-u.ac.jp}

\abstract{
  Anyon condensation in wormhole geometries is investigated in the Virasoro TQFT (VTQFT) formulation, a proposed
  reformulation of 3d AdS quantum gravity.
  We first review some elementary techniques of VTQFT and summarize a gauging scheme for non-invertible symmetries referred to as anyon condensation.
  We then exhibit that anyon condensation is applicable to VTQFT even though the category of Wilson lines associated with it is not strictly a modular tensor category (MTC) due to the continuously infinite label $p\in\mathbb{R}_+$.
  More specifically, it is shown that the partition function of the wormhole factorizes upon condensing the so-called diagonal condensable anyon $\mathcal{A}=\int_{0}^{\infty}dp\,L_p\boxtimes\overline{L}_p$ in VTQFT.
  The resulting $2$d boundary theory is Liouville CFT by symmetry TFT construction, and to our knowledge, this is among the very few explicit computational examples of gauging \textit{continuous non-invertible} symmetries in the literature.
}


\begin{document}
  \begin{flushright}
    RIMS-1988 \\
  \end{flushright}

  \maketitle
  \flushbottom

  \section{Introduction}
    Seeking well-defined quantum gravity is essential for advancing our understanding of fundamental physics.
    In three-dimensional spacetime, the gravitational dynamics are significantly simplified compared to higher dimensions, providing a tractable model for studying quantum gravitational effects.
    Indeed, Ref.\,\cite{Achucarro:1986uwr} establishes a privileged direct connection between 3d Einstein-Hilbert action with negative cosmological constant and $SL(2,\mathbb{R})\times SL(2,\mathbb{R})$ Chern-Simons action, showing that the level $k$ and the $\mathfrak{sl}(2,\mathbb{R})\times\mathfrak{sl}(2,\mathbb{R})$ gauge fields $A_{\mu}$, $\overline{A}_{\mu}$ are expressed in terms of the Newton constant $G$, the dreibein $e_{\mu}^a$ and the spin connection $\omega_{\mu}^a$ as
    \begin{align}
      k=\frac{l}{4G},\quad A_{\mu}^a=e_{\mu}^a+\frac{1}{l}\omega_{\mu}^{a},\quad\overline{A}_{\mu}^a=e_{\mu}^a-\frac{1}{l}\omega_{\mu}^{a},\qquad(l:\text{AdS radius})
    \end{align}
    which ensures that 3d gravity is topological at the classical level in a sense that it does not depend on the spacetime metric, as Chern-Simons action is manifestly topological.
    The quantization of Chern-Simons theory yields a topological quantum field theory (TQFT) described by the conformal blocks of Wess-Zumino-Witten (WZW) model that resides on the boundaries \cite{Witten:1988hf}.
    However, there is a subtlety as the generic path-integral quantization scheme implements the integral over all configurations of dynamical fields (variables of path-integral) $A_{\mu}^a$ in Chern-Simons theory that may result in a non-invertible dreibein $e_{\mu}^a$.
    In quantum gravity, the gravitational path integral $Z_{\text{gravity}}=\int\mathcal{D}g_{\mu\nu}\,e^{-S_{\text{gravity}}\qty[g_{\mu\nu}]}$ is performed over geometries $g_{\mu\nu}=e_{\mu}^ae_{\nu}^b\eta_{ab}$ which are supposed to be non-degenerate.
    Thus, while Chern-Simons theory offers valuable insights, it serves as a preliminary toy model rather than a complete framework for the development of quantum gravity.
    \par In order to circumvent such a conundrum, one should select the so-called Teichm\"{u}ller component from the classical phase space of Chern-Simons theory and quantize the space of holomorphic sections of a certain line bundle over it \cite{Witten:1988hc}.
    Seminal work by H.\,Verlinde \cite{Verlinde:1989ua} addresses this quantization issue and reveals that the resulting Hilbert space is a collection of Virasoro conformal blocks of Liouville theory equipped with an abstract inner product \eqref{eq: abstract inner product}.
    This statement is refined by J.\,Teschner in his series of works \cite{Teschner:2002vx,Teschner:2003at,Teschner:2003em,Teschner:2005bz} and has been investigated extensively in mathematics literatures \cite{kashaev1998quantization, andersen2013new, Mikhaylov:2017ngi}.
    Recent work by S.\,Collier et al. \cite{Collier:2023fwi, Collier:2024mgv} advocated more tractable form of the impractical inner product \eqref{eq: abstract inner product}, whose explicit expression is provided in Subsection \ref{subsec: The Hilbert space and the crossing transformations}.
    They renamed the conventional Teichm\"uller-based TQFT with the new inner product as \textit{Virasoro TQFT} (VTQFT), for the Hilbert space associated with the 2d boundaries is a collection of Virasoro conformal blocks.
    Building on their formalism, the quantum gravity partition function is given for any $3$d hyperbolic geometry $M$ as
    \begin{align}
      Z_{\text{gravity}}(M)=\sum_{\gamma\in\MCG(\partial M)/\MCG(M)}|Z_{\text{Vir}}(\gamma\cdot M)|^2, \label{eq: gravity partition function}
    \end{align}
    even when it is uncertain whether the conventional metric approach for path-integral is readily applicable.
    $\MCG(X):=\Diff(X)/\Diff_0(X)\cong\pi_0(\Diff(X))$ is the \textit{mapping class group} of a manifold $X$.\footnote{Since a homeomorphism sends the boundary to the boundary, a bulk diffeomorphism $f$ induces the boundary diffeomorphism $f|_{\der M}$, for which we can regard $\MCG(M)$ as a subset of $\MCG(\der M)$. $\MCG(M,\partial M)=\Diff(M,\partial M)/\Diff_0(M,\partial M)\cong\pi_0(\Diff(M,\partial M))$ is a \textit{relative mapping class group}, whose elements are equivalence classes of diffeomorphism that map each boundary component to itself.}
    Notably, this rule successfully reproduces the Maloney-Witten sum for solid torus geometries \cite{Maloney:2007ud}. Another way to support its credibility is to compare the partition function to that of a putative holographic dual $2$d CFT which has been the focus of intensive investigation in recent years \cite{Cotler:2020ugk, Chandra:2022bqq, Belin:2023efa, Jafferis:2024jkb, Pelliconi:2024aqw}.
    In the literature, it is strongly suggested that the dual CFT is not a particular pure CFT but rather one characterized by an ensemble average of observables which is Gaussian to leading order supplemented with non-Gaussian corrections required by crossing symmetry.
    However, this leads to a significant enigma known as the \textit{factorization puzzle for partition functions} raised in Ref.\,\cite{Maldacena:2004rf} and studied, e.g., in Refs.\,\cite{Saad:2021uzi,Blommaert:2021fob}, which involves the non-factorization of the partition function for connected geometries with multiple boundaries.
    From the boundary point of view, the partition function should factorize $Z(M)=Z(N_1)\cdots Z(N_n)$ for disconnected boundaries $\partial M=N_1\sqcup\cdots\sqcup N_n$, whereas the connecting wormhole geometries in the bulk give rise to some entanglement between different boundaries.
    \par A potential solution to that puzzle emerges from another paradigm referred to as \textit{generalized global symmetries} \cite{Gaiotto:2014kfa}.
    In this framework, the notion of ``symmetries" is reinterpreted as invariance under the action of higher-dimensional topological operators.
    Of particular relevance to our present discussion are \textit{non-invertible symmetries} \cite{Bhardwaj:2017xup, Chang:2018iay}, forming a symmetry category rather than a symmetry group due to the absence of inverse operations.
    Non-Abelian TQFTs provide typical examples, with their topological line operators, or Wilson lines, serving as simple objects in symmetry categories possibly with non-trivial fusion rule $L_i\otimes L_j=\bigoplus_kN_{ij}^kL_k\:(\sum_kN_{ij}^k>1)$.
    For 3d TQFT, the associated category is a \textit{modular tensor category} (MTC), roughly described as a fusion category with a braiding structure.
    Anyon condensation is a non-invertible version of gauging $1$-form symmetry and similar to the invertible case, this process requires selecting an ``'t Hooft anomaly-free" object, known as a \textit{condensable anyon}, from the MTC.
    Mathematically, such an object is a \textit{connected commutative separable Frobenius algebra object} in a braided monoidal category.
    In Ref.\,\cite{Benini:2022hzx} it is shown that the anyon condensation for a general condensable anyon in non-Abelian Chern-Simons theory leads to partition function factorization in two-boundary wormhole geometry as one would expect given that global symmetries are absent in a bulk quantum gravity theory with a CFT dual \cite{Harlow:2018jwu, Harlow:2018tng}.
    They assert that the ensemble picture on the boundary emerges from the bulk generalized global symmetry formed by Wilson lines.
    The approach should also be feasible within VTQFT as pointed out in Refs.\,\cite{Collier:2023fwi,Dymarsky:2024frx}, and we show that the factorization indeed manifests in the two-boundary wormhole geometry when we consider the \textit{diagonal Lagrangian condensable anyon} despite the fact that their symmetry category is not a MTC, much less a tensor category, due to the continuum of lines.
    We still reach a meaningful result partly because VTQFT retains a well-behaved braiding structure and a number of Moore-Seiberg consistency equations.
    As a result of anyon condensation, we obtain Liouville CFT on the 2d boundaries, as expected from sandwich construction of Symmetry TFT \cite{Gaiotto:2020iye, Kaidi:2022cpf}, and we no longer require a sum over topologies as the partition function becomes entirely independent of the $3$d bulk topologies.
    To the best of our knowledge, this is the first explicit computational examples of
    gauging continuous non-invertible symmetries in the literature.
    \par This paper is organized as follows.
    In Section \ref{sec: Virasoro TQFT}, we briefly review the detailed computational foundation of Virasoro TQFT and discuss its connection to the semiclassical phenomena like Hawking-Page phase transition and Bekenstein-Hawking entropy.
    Section \ref{sec: Anyon condensation in modular tensor category} introduces anyon condensation as a general gauging method for $3$d non-Abelian TQFTs.
    We itemize various conditions for an object in symmetry category $\mathcal{C}$ to be gaugeable and define diagonal Lagrangian algebra object in the special case of TQFT with a Drinfeld center $\mathcal{Z}(\mathcal{C})=\mathcal{C}\boxtimes\overline{\mathcal{C}}$.
    Section \ref{sec: Anyon condensation in Virasoro TQFT} is devoted to the presentation of the main result.
    We establish a crucial relation involving an elementary building block called the projector, and then go on to prove that the partition function of two-boundary wormhole factorizes veritably by virtue of anyon condensation.
    Section \ref{sec: conclusions and discussions} provides concluding remarks on our result and future directions.
    Appendix \ref{append: Crossing kernels and Moore-Seiberg consistency conditions} encapsulates numerous consistency equations for crossing transformations of Virasoro conformal blocks that are ubiquitous throughout the main body.
    They are essentially a consequence of the non-rational version of the Moore-Seiberg consistency conditions.
    In Appendix \ref{append: From monoidal category to fusion category and modular tensor category}, we recap several core categorical terms needed to formulate fusion category and modular tensor category.
    The interrelations among them are visualized in a diagram which may also assist readers in consulting other relevant literature.

  \section{Virasoro TQFT} \label{sec: Virasoro TQFT}
    In this section, we provide an outline of Virasoro TQFT \cite{Collier:2023fwi} to put it into practice in Section \ref{sec: Anyon condensation in Virasoro TQFT}.
    We begin by reviewing the Hilbert space structure and the action of crossing transformations in Subsection \ref{subsec: The Hilbert space and the crossing transformations}, followed by a summary of the VTQFT path-integral rule in Subsection \ref{subsec: Heegaard splitting and path-integral on compression bodies}.
    In Subsection \ref{subsec: application to solid torus}, we provide a justification for VTQFT as a 3d gravitational theory by reproducing the Hawking-Page phase transition in 3d and the Bekenstein-Hawking entropy of the BTZ blackhole.
    Before proceeding, let us briefly explore the interplay between the theory and Chern-Simons theory, as well as the pivotal role of the quantization of Teichm\"{u}ller space.
    \par Suppose the spacetime manifold is of the form $\Sigma\times\mathbb{R}$, where $\Sigma$ is a closed Riemann surface.
    As established in Ref.\,\cite{Witten:1988hc}, the classical phase space of $SL(2,\mathbb{R})$ Chern-Simons theory on $\Sigma\times\mathbb{R}$ is the moduli space $\mathcal{M}_{\text{flat}}$ of flat $SL(2,\mathbb{R})$-bundles (principal $SL(2,\mathbb{R})$-bundles equipped with flat connections) over $\Sigma$.
    This arises because the EOMs from the Chern-Simons action require that the curvature be vanishing.
    In bundle theory \cite{tu2017differential}, each flat $SL(2,\mathbb{R})$-bundle, combined with the $2$d fundamental representation of $SL(2,\mathbb{R})$, induces the associated vector bundle of rank $2$. The Euler number of this bundle takes a specific value from $-(2g-2),-(2g-3),\cdots,2g-3,$ or $2g-2$, where the upper bound $2g-2$ coincides with the Euler number $\chi(\Sigma)$ of the tangent bundle $T\Sigma$.
    All the flat $SL(2,\mathbb{R})$-bundles are classified according to the Euler number of their associated bundles ranging from $-(2g-2)$ to $2g-2$. The moduli space $\mathcal{M}_{\text{flat}}$ has indeed $(2g-2)-(-(2g-2))+1=4g-3$ connected components labeled by the Euler number.
    On the other hand, there is a one-to-one correspondence between flat $SL(2,\mathbb{R})$-bundles and holonomy representations $\rho:\pi_1(\Sigma)\to SL(2,\mathbb{R})$ which may or may not be continuous.
    There is another match between \textit{discrete} embeddings $\pi_1(\Sigma)\to SL(2,\mathbb{R})$ and complex structures of $\Sigma$, so only a limited class of flat $SL(2,\mathbb{R})$-bundles indeed defines complex structures on $\Sigma$.
    All other flat bundles result in singular geometries which are unfavorable in gravity theory.
    Notably, the flat bundles in that class happen to be exactly those within the single connected component of $\mathcal{M}_{\text{flat}}$ with maximal Euler number $2g-2$.
    This component is isomorphic to the \bit{Teichm\"{u}ller space} $\mathcal{T}_{\Sigma}$ of $\Sigma$, the universal covering space of moduli space $\mathcal{M}_{\Sigma}$ of $\Sigma$ related by $\mathcal{M}_{\Sigma}=\mathcal{T}_{\Sigma}/\MCG(\Sigma)$ where $\MCG$ is the mapping class group of $\Sigma$ \cite{Petri:2024}.
    Thus, the quantization of $3$d gravity with a negative cosmological constant reduces to the quantization of Teichm\"{u}ller space.
    These deductions are heavily reliant on the mathematical results by M.\,F.\,Atiyah and R.\,Bott \cite{atiyah1983yang}.
    \par As noted earlier the quantization of the Teichm\"{u}ller space associated with the $n$-punctured genus $g$ Riemann surface $\Sigma_{g,n}$ gives rise to a Hilbert space $\mathcal{H}_{g,n}$ whose elements are Virasoro conformal blocks $\ket{\mathcal{F}_{g,n}^{\mathcal{C}}(\bm{p};\bm{p}_e)}$ of Liouville theory with an abstract inner product \cite{Verlinde:1989ua}
    \begin{align}
      \braket{\mathcal{F}_{g,n}^{\mathcal{C}}\qty(\bm{p};\bm{p}_e)}{\mathcal{F}_{g,n}^{\mathcal{C}}(\bm{p}^{\prime};\bm{p}_e)}=\int_{\mathcal{T}_{g,n}}d^{6g-6+2n}\bm{m}~Z_{bc}Z_{\text{TLL}}\overline{\mathcal{F}_{g,n}^{\mathcal{C}}\qty(\bm{p};\bm{p}_e|\bm{m})}\mathcal{F}_{g,n}^{\mathcal{C}}(\bm{p}^{\prime};\bm{p}_e|\bm{m}), \label{eq: abstract inner product}
    \end{align}
    where $\bm{p}\in\mathbb{R}_{\geq 0}^{3g-3+n}$ (resp.~$\bm{p}_e\in\mathbb{R}_{\geq 0}^n$) is an internal (resp.~external) Liouville momenta w.\,r.\,t.~a conformal block decomposition channel $\mathcal{C}$ and $\bm{m}$ is the  moduli coordinates on the Teichm\"{u}ller space $\mathcal{T}_{g,n}$.
    $Z_{bc}$ is the $bc$ ghost partition function for gauge fixing and $Z_{\text{TLL}}$ is the timelike Liouville partition function to cancel the Weyl anomaly just as in string theory.
    The holomorphic conformal block functions $\mathcal{F}_{g,n}^{\mathcal{C}}(\bm{p};\bm{p}_e;\bm{m})$ are obtained by taking overlap between the moduli basis $\ket{\bm{m}}~(\bm{m}\in\mathcal{T}_{g,n})$ and the conformal block basis $\ket{\mathcal{F}_{g,n}^{\mathcal{C}}(\bm{p};\bm{p}_e)}$
    \begin{align}
      \mathcal{F}_{g,n}^{\mathcal{C}}(\bm{p};\bm{p}_e;\bm{m}):=\braket{\bm{m}}{\mathcal{F}_{g,n}^{\mathcal{C}}(\bm{p};\bm{p}_e)}. \label{eq: overlap of moduli and conformal block}
    \end{align}
    In gravitational theory, we need to consider the product of the chiral and the anti-chiral sector to determine the partition function \eqref{eq: gravity partition function}.
    Hence the classical phase space of gravity is $\mathcal{T}_{g,n}\times\overline{\mathcal{T}}_{g,n}$ \cite{Scarinci:2011np}, where $\overline{\mathcal{T}}_{g,n}$ is the orientation reversal of $\mathcal{T}_{g,n}$.
    We will next explore the Hilbert space structure in detail. We only describe the chiral part $\mathcal{T}_{g,n}$ for simplicity, but the same goes for the anti-chiral part unless stated otherwise.

    \subsection{The Hilbert space and the crossing transformations} \label{subsec: The Hilbert space and the crossing transformations}
      As we see above the Hilbert space $\mathcal{H}_{g,n}$ associated with a Riemann surface $\Sigma_{g,n}$, possibly with punctures, is the space of Virasoro conformal blocks on $\Sigma_{g,n}$.
      They are parametrized by basic CFT data, namely the central charge and the spectrum (the conformal weights of the operator contents).
      \par The Liouville parameter $b$ is pertaining to the Chern-Simons level $k$ as
      \begin{align}
        b=\frac{1}{\sqrt{k-2}}, \label{eq: Liouville parameter and Chern-Simons level}
      \end{align}
      coming out of the Hamiltonian reduction of CS/WZW via the free-field Wakimoto representation.
      In terms of the back ground charge $Q:=b+\frac{1}{b}$ the central charge takes the form
      \begin{align}
        c= 1+6Q^2 = 13+6b^2+\frac{6}{b^2}, \label{eq: central charge}
      \end{align}
      which asymptotes to the Brown-Henneaux value $c=\frac{3l}{2G}$ in the semiclassical limit $G\to 0$.
      The conformal weight $h_p$ is parametrized by Liouville momentum $p$ as
      \begin{align}
        h_p= \alpha(Q-\alpha)=\frac{Q^2}{4}+p^2=\frac{c-1}{24}+p^2\qquad\Big(\alpha:= \frac{Q}{2}+ip\Big). \label{eq: conformal weight}
      \end{align}
      The state with the conformal weight $h_p$ is normalizable only when $p\in\mathbb{R}_{\geq 0}$ or equivalently when $h_p$ is above the threshold $\frac{c-1}{24}$, and it is noteworthy that the identity line $\mathbbm{1}$ ($h_{\mathbbm{1}}=0$) is unnormalizable whose Liouville momentum is $p=\pm i\frac{Q}{2}$.
      We will frequently express $p\to\mathbbm{1}$ instead of $p=\pm i\frac{Q}{2}$ to clarify that the Wilson line $p$ is mapped to the identity line $\mathbbm{1}$.
      \par In VTQFT, the abstract inner product \eqref{eq: abstract inner product} is proposed to have a simplified form\footnote{The $\mathcal{N}=1$ supersymmetric extension is presented in Ref.\,\cite{Bhattacharyya:2024vnw}.}
      \begin{gather}
        \big<\mathcal{F}_{0,3}(\bm{p}_e)\big|\mathcal{F}_{0,3}(\bm{p}_e)\big>=\frac{1}{C_0(p_1,p_2,p_3)}\qquad\qty(~\bm{p}_e=\qty(p_1,p_2,p_3)^{T}), \\
        \big<\mathcal{F}_{g,n}^{\mathcal{C}}(\bm{p}_1;\bm{p}_e)\big|\mathcal{F}_{g,n}^{\mathcal{C}}(\bm{p}_2;\bm{p}_e)\big>=\frac{\delta^{(3g-3+n)}(\bm{p}_1-\bm{p}_2)}{\rho_{g,n}^{\mathcal{C}}(\bm{p}_1)} \qquad\big((g,n)\neq(0,3)\big). \label{eq: inner product}
      \end{gather}
      Here, the subscript $e$ attached to the second argument of a conformal block $\mathcal{F}_{g,n}^{\mathcal{C}}$ is the shorthand for ``external".
      The inner product is only defined between blocks with the same external legs.
      The sphere three-point block ($g=0,n=3$) is treated differently since $\dim\mathcal{H}_{0,3}=1<\infty$ allows it to be normalizable in the usual sense while any other blocks are only delta-function normalizable due to the fact that $\dim\mathcal{H}_{g,n}=\infty$. The block $\big|\mathcal{F}_{0,3}(\bm{p}_e)\big>$ has the unique channel and no internal momenta, so there are no superscript $\mathcal{C}$ and the only arguments are external Liouville momentum $\bm{p}_e$.
      The quantity $\rho_{g,n}^{\mathcal{C}}(\bm{p}_1)$ in the denominator on the r.\,h.\,s.~of eq.\,\eqref{eq: inner product} is defined as
      \begin{align}
        \rho_{g,n}^{\mathcal{C}}(\bm{p}):=\prod_{\substack{\text{internal cuffs} \\ a}}\rho_0(p_a)\prod_{\substack{\text{trivalent junctions} \\ (i,j,k)}}C_0(p_i,p_j,p_k), \label{eq: integral measure}
      \end{align}
      where $\rho_0(p_a)$ and $C_0(p_i,p_j,p_k)$ are given by eq.\,\eqref{eq: Cardy density}, \eqref{eq: universal three point coefficient}.
      The ``internal cuffs" and the ``trivalent junctions" are solely determined by the choice of a channel $\mathcal{C}$ that specifies the way of decomposing the Riemann surface $\Sigma_{g,n}$ into $2g-2+n$ pair of pants (3-punctured sphere). There are $2g-2+n$ junctions corresponding to each pant and $(3(2g-2+n)-n)\cdot\frac{1}{2}=3g-3+n$ cuffs corresponding to each internal slice.
      $C_0(p_1,p_2,p_3)$ appears in an universal asymptotic formula for the microcanonical ensemble average of the OPE coefficients of a generic compact unitary 2d CFT with $c>1$ \cite{Collier:2019weq}, and is to some extent proportional to the Dorn-Otto-Zamolodchikov-Zamolodchikov (DOZZ) structure constant \cite{Dorn:1994xn,Zamolodchikov:1995aa}, the sphere 3-point coefficient, or equivalently the OPE coefficient, in Liouville theory\footnote{The explicit factor of proportionality is
        \begin{align}
          C_0\qty(p_1,p_2,p_3)=\uwave{\frac{\big(\pi\mu\gamma(b^2)b^{2-2b^2}\big)^{\frac{Q}{2b}}}{2^{\frac{3}{4}}\pi}\frac{\Gamma_b\qty(2Q)}{\Gamma_b\qty(Q)}}\frac{C_{\text{DOZZ}}\qty(p_1,p_2,p_3)}{\sqrt{\prod_{k=1}^3S_0\qty(p_k)\rho_0\qty(p_k)}},
        \end{align}
        where the wavy line part is independent of $p_1,p_2,p_3$.
        $S_0(p)$ is the Liouville reflection coefficient
        \begin{align}
          S_0(p):=(\pi\mu\gamma(b^2)b^{2-2b^2})^{-\frac{2ip}{b}}\frac{\Gamma_b(2ip)\Gamma_b(Q-2ip)}{\Gamma_b(Q+2ip)\Gamma_b(-2ip)}.
        \end{align}
        For a precise definition and properties of the double gamma function $\Gamma_b\qty(x)$, see e.g.~Ref.\,\cite{Ribault:2014hia, Ribault:2024rvk}.
      }
      \begin{align}
        C_0(p_1,p_2,p_3)\propto\frac{C_{\text{DOZZ}}(p_1,p_2,p_3)}{\sqrt{\prod_{k=1}^3S_0(p_k)\rho_0(p_k)}}.
      \end{align}
      $C_0(p_1,p_2,p_3)$ is symmetric under the exchange of any two arguments due to the symmetricity of $C_{\text{DOZZ}}(p_1,p_2,p_3)$:
      \begin{align}
        C_0(p_1,p_2,p_3)=C_0(p_2,p_1,p_3)=C_0(p_3,p_2,p_1).
      \end{align}
      Since both $\rho_0(p_a)$ and $C_0(p_i,p_j,p_k)$ are real, the same quantities are employed in computing the inner product in the anti-chiral part.
      For concreteness, here is an example for the inner product between two 2-punctured torus blocks, where the conformal block $\big|\mathcal{F}_{1,2}^{\mathcal{C}}(p_a,p_b;p_1,p_2)\big>$ is depicted graphically:
      \begin{align}
        \braket{
          \begin{tikzpicture}[scale = 0.5, baseline={([yshift = -0.5ex]current bounding box.center)}]
            % left torus boundary
            \draw[thick] (0, 0) ellipse (2 and 1);
            \draw[thick, bend left = 20] (1, 0.08) to (-1, 0.08);
            \draw[thick, bend right = 20] (0.78, 0) to (-0.78, 0);
            \draw[red] (0, 0) ellipse (1.2 and 0.5);
            \draw[red] (1.2, 0) -- (1.8, 0);
            \draw[red] (-1.2, 0) -- (-1.8, 0);
            % punctures
            \draw[thick] (-1.9, 0.1) -- (-1.7, -0.1);
            \draw[thick] (-1.9, -0.1) -- (-1.7, 0.1);
            \draw[thick] (1.9, 0.1) -- (1.7, -0.1);
            \draw[thick] (1.9, -0.1) -- (1.7, 0.1);
            % labels
            \node[scale = 0.6, above] at (0, 0.45) {$p_a$};
            \node[scale = 0.6, below] at (0, -0.45) {$p_b$};
            \node[scale = 0.6, above] at (-1.5, -0.05) {$p_1$};
            \node[scale = 0.6, above] at (1.5, -0.05) {$p_2$};
          \end{tikzpicture}
        \,}{\,
          \begin{tikzpicture}[scale = 0.5, baseline={([yshift = -0.5ex]current bounding box.center)}]
            % left torus boundary
            \draw[thick] (0, 0) ellipse (2 and 1);
            \draw[thick, bend left = 20] (1, 0.08) to (-1, 0.08);
            \draw[thick, bend right = 20] (0.78, 0) to (-0.78, 0);
            \draw[red] (0, 0) ellipse (1.2 and 0.5);
            \draw[red] (1.2, 0) -- (1.8, 0);
            \draw[red] (-1.2, 0) -- (-1.8, 0);
            % punctures
            \draw[thick] (-1.9, 0.1) -- (-1.7, -0.1);
            \draw[thick] (-1.9, -0.1) -- (-1.7, 0.1);
            \draw[thick] (1.9, 0.1) -- (1.7, -0.1);
            \draw[thick] (1.9, -0.1) -- (1.7, 0.1);
            % labels
            \node[scale = 0.6, above] at (0, 0.45) {$p_c$};
            \node[scale = 0.6, below] at (0, -0.45) {$p_d$};
            \node[scale = 0.6, above] at (-1.5, -0.05) {$p_1$};
            \node[scale = 0.6, above] at (1.5, -0.05) {$p_2$};
          \end{tikzpicture}
        }=\frac{\delta\qty(p_a-p_c)\delta\qty(p_b-p_d)}{\rho_0\qty(p_a)\rho_0\qty(p_b)C_0\qty(p_1,p_a,p_b)C_0\qty(p_2,p_a,p_b)}. \label{eq: inner product example}
      \end{align}
      Intuition behind the inner product \eqref{eq: inner product} is the unitarity under the action of the mapping class group $\MCG(\Sigma)$.
      The image of the projective unitary representation $U_{g,n}:\MCG(\Sigma)\ni\gamma\mapsto U_{g,n}(\gamma)\in\operatorname{End}(\mathcal{H}_{g,n})$ is generated by a non-rational version of basic crossing moves in Moore-Seiberg construction \cite{Moore:1988qv} presented below.
      \begin{enumerate}
        \renewcommand{\labelenumi}{(\roman{enumi})}
        \item fusion transformation
              \begin{align}
                \begin{tikzpicture}[scale = 0.6, baseline = {([yshift=-.5ex]current bounding box.center)}]
                  \draw[thick] (-1.6, 1.4) .. controls (-1.4, 1.3) and (-1, 0.8) .. (0, 0.8) .. controls (1, 0.8) and (1.4, 1.3) .. (1.6, 1.4);
                  \draw[thick] (-1.6, -1.4) .. controls (-1.4, -1.3) and (-1, -0.8) .. (0, -0.8) .. controls (1, -0.8) and (1.4, -1.3) .. (1.6, -1.4);
                  \draw[thick] (-1.8, 0.9) .. controls (-1.6, 0.75) and (-1, 0.4) .. (-1, 0) .. controls (-1, -0.4) and (-1.6, -0.75) .. (-1.8, -0.9);
                  \draw[thick] (1.8, 0.9) .. controls (1.6, 0.75) and (1, 0.4) .. (1, 0) .. controls (1, -0.4) and (1.6, -0.75) .. (1.8, -0.9);
                  \draw[thick, rotate around = {64:(-1.7, 1.15)}] (-1.7, 1.15) ellipse (0.27 and 0.15);
                  \draw[thick, rotate around = {-64:(-1.7, -1.15)}] (-1.7, -1.15) ellipse (0.27 and 0.15);
                  \draw[thick, rotate around = {-64:(1.7, 1.15)}] (1.7, 1.15) ellipse (0.27 and 0.15);
                  \draw[thick, rotate around = {64:(1.7, -1.15)}] (1.7, -1.15) ellipse (0.27 and 0.15);
                  \draw[red] (-1.6, 1.05) .. controls (-1.2, 0.8) and (-0.9, 0.5) .. (-0.5, 0);
                  \draw[red] (-1.6, -1.05) .. controls (-1.2, -0.8) and (-0.9, -0.5) .. (-0.5, 0);
                  \draw[red] (1.6, 1.05) .. controls (1.2, 0.8) and (0.9, 0.5) .. (0.5, 0);
                  \draw[red] (1.6, -1.05) .. controls (1.2, -0.8) and (0.9, -0.5) .. (0.5, 0);
                  \draw[red] (-0.5, 0) to (0.5, 0);
                  \fill[red] (-0.5, 0) circle [radius = 0.1];
                  \fill[red] (0.5, 0) circle [radius = 0.1];
                  \draw[bend right = 30] (0, 0.8) to (0, -0.8);
                  \draw[dashed, bend left = 30] (0, 0.8) to (0, -0.8);
                  % labels
                  \node[scale = 0.8, left] at (-1.8, 1.3) {$p_1$};
                  \node[scale = 0.8, left] at (-1.8, -1.3) {$p_2$};
                  \node[scale = 0.8, right] at (1.8, 1.3) {$p_3$};
                  \node[scale = 0.8, right] at (1.8, -1.3) {$p_4$};
                  \node[scale = 0.8, above] at (0.2, -0.1) {$p_s$};
                \end{tikzpicture}
                =\int_{0}^{\infty}dp_t\,F_{p_sp_t}
                \begin{bmatrix}
                    p_1 & p_3 \\
                    p_2 & p_4
                \end{bmatrix}
                \,
                \begin{tikzpicture}[scale = 0.6, baseline = {([yshift=-.5ex]current bounding box.center)}]
                  \draw[thick] (-1.6, 1.4) .. controls (-1.4, 1.3) and (-1, 0.8) .. (0, 0.8) .. controls (1, 0.8) and (1.4, 1.3) .. (1.6, 1.4);
                  \draw[thick] (-1.6, -1.4) .. controls (-1.4, -1.3) and (-1, -0.8) .. (0, -0.8) .. controls (1, -0.8) and (1.4, -1.3) .. (1.6, -1.4);
                  \draw[thick] (-1.8, 0.9) .. controls (-1.6, 0.75) and (-1, 0.4) .. (-1, 0) .. controls (-1, -0.4) and (-1.6, -0.75) .. (-1.8, -0.9);
                  \draw[thick] (1.8, 0.9) .. controls (1.6, 0.75) and (1, 0.4) .. (1, 0) .. controls (1, -0.4) and (1.6, -0.75) .. (1.8, -0.9);
                  \draw[thick, rotate around = {64:(-1.7, 1.15)}] (-1.7, 1.15) ellipse (0.27 and 0.15);
                  \draw[thick, rotate around = {-64:(-1.7, -1.15)}] (-1.7, -1.15) ellipse (0.27 and 0.15);
                  \draw[thick, rotate around = {-64:(1.7, 1.15)}] (1.7, 1.15) ellipse (0.27 and 0.15);
                  \draw[thick, rotate around = {64:(1.7, -1.15)}] (1.7, -1.15) ellipse (0.27 and 0.15);
                  \draw[red] (-1.6, 1.05) .. controls (-1.2, 0.8) and (-0.9, 0.6) .. (0, 0.5);
                  \draw[red] (-1.6, -1.05) .. controls (-1.2, -0.8) and (-0.9, -0.6) .. (0, -0.5);
                  \draw[red] (1.6, 1.05) .. controls (1.2, 0.8) and (0.9, 0.6) .. (0, 0.5);
                  \draw[red] (1.6, -1.05) .. controls (1.2, -0.8) and (0.9, -0.6) .. (0, -0.5);
                  \draw[red] (0, 0.5) to (0, -0.5);
                  \fill[red] (0, 0.5) circle [radius = 0.1];
                  \fill[red] (0, -0.5) circle [radius = 0.1];
                  \draw[bend right = 30] (-1, 0) to (1, 0);
                  \draw[dashed, bend left = 30] (-1, 0) to (1, 0);
                  % labels
                  \node[scale = 0.8, left] at (-1.8, 1.3) {$p_1$};
                  \node[scale = 0.8, left] at (-1.8, -1.3) {$p_2$};
                  \node[scale = 0.8, right] at (1.8, 1.3) {$p_3$};
                  \node[scale = 0.8, right] at (1.8, -1.3) {$p_4$};
                  \node[scale = 0.8, left] at (0, 0.1) {$p_t$};
                \end{tikzpicture}
                \label{eq: fusion transformation}
              \end{align}
        \item modular $S$-transformation
              \begin{align}
                \begin{tikzpicture}[scale = 0.6, baseline={([yshift = -0.5ex]current bounding box.center)}]
                  \begin{scope}
                    \clip (-2.1, 1.2) rectangle (1.1, -1.2);
                    \draw[thick] (0, 0) ellipse (2 and 1.15);
                  \end{scope}
                  \draw[thick] (1.1, 0.96) .. controls (1.3, 0.9) and (1.8, 0.6) .. (2, 0.6);
                  \draw[thick] (1.1, -0.96) .. controls (1.3, -0.9) and (1.8, -0.6) .. (2, -0.6);
                  \draw[thick] (2, 0) ellipse (0.2 and 0.595);
                  \draw[thick, bend left = 30] (1, 0.08) to (-1, 0.08);
                  \draw[thick, bend right = 30] (0.78, 0) to (-0.78, 0);
                  \draw[red] (0, 0) ellipse (1.25 and 0.7);
                  \draw[red] (1.25, 0) -- (1.8, 0);
                  \fill[red] (1.25, 0) circle [radius = 0.08];
                  % labels
                  \node[scale = 0.8, left] at (-1.25, 0) {$p$};
                  \node[scale = 0.8, above] at (1.525, 0) {$p_0$};
                \end{tikzpicture}
                \,=\int_{0}^{\infty}dp^{\prime}\,S_{pp^{\prime}}\qty[p_0]\,
                \begin{tikzpicture}[scale = 0.6, baseline={([yshift = -0.5ex]current bounding box.center)}]
                  \begin{scope}
                    \clip (-2.1, 1.2) rectangle (1.1, -1.2);
                    \draw[thick] (0, 0) ellipse (2 and 1.15);
                  \end{scope}
                  \draw[thick] (1.1, 0.96) .. controls (1.3, 0.9) and (1.8, 0.6) .. (2, 0.6);
                  \draw[thick] (1.1, -0.96) .. controls (1.3, -0.9) and (1.8, -0.6) .. (2, -0.6);
                  \draw[thick] (2, 0) ellipse (0.2 and 0.595);
                  \draw[thick, bend left = 30] (1, 0.08) to (-1, 0.08);
                  \draw[thick, bend right = 30] (0.78, 0) to (-0.78, 0);
                  \begin{scope}
                    \clip (-0.25, -0.2) rectangle (0, -1.15);
                    \draw[red] (0, -0.67) ellipse (0.2 and 0.45);
                  \end{scope}
                  \begin{scope}
                    \clip (0, -0.2) rectangle (0.25, -1.15);
                    \draw[dashed, red] (0, -0.67) ellipse (0.2 and 0.45);
                  \end{scope}
                  \draw[red] (1.8, 0) .. controls (1.5, 0) and (0.8, -0.7) .. (-0.21, -0.7);
                  \fill[red] (-0.21, -0.7) circle [radius = 0.08];
                  % labels
                  \node[scale = 0.8, left] at (-0.25, -0.67) {$p^{\prime}$};
                  \node[scale = 0.8, above] at (1.4, -0.1) {$p_0$};
                \end{tikzpicture}
              \end{align}
        \item braiding
              \begin{align}
                \begin{tikzpicture}[scale = 0.6, baseline={([yshift = -0.5ex]current bounding box.center)}]
                  \draw[thick] (0, -1.5) ellipse (0.5 and 0.2);
                  \draw[thick, rotate = 120] (0, -1.5) ellipse (0.5 and 0.2);
                  \draw[thick, rotate = 240] (0, -1.5) ellipse (0.5 and 0.2);
                  \draw[thick, bend right = 40] (-0.5, -1.5) to ({-0.75 * sqrt(3) - 0.25}, {0.75 - 0.25 * sqrt(3)});
                  \draw[thick, bend left = 40] (0.5, -1.5) to ({0.75 * sqrt(3) + 0.25}, {0.75 - 0.25 * sqrt(3)});
                  \draw[thick, bend right = 40] ({-0.75 * sqrt(3) + 0.25}, {0.75 + 0.25 * sqrt(3)}) to ({0.75 * sqrt(3) - 0.25}, {0.75 + 0.25 * sqrt(3)});
                  \draw[red] (0, 0) to (0, -1.3);
                  \draw[red] (0, 0) to ({-0.75 * sqrt(3) + 0.1 * sqrt(3)}, 0.65);
                  \draw[red] (0, 0) to ({0.75 * sqrt(3) - 0.1 * sqrt(3)}, 0.65);
                  \fill[red] (0, 0) circle [radius = 0.08];
                  % labels
                  \node[scale = 0.8, below] at (0, -1.9) {$p_3$};
                  \node[scale = 0.8, above left] at ({-0.7 * sqrt(3) - 0.2 * sqrt(3)}, 0.85) {$p_1$};
                  \node[scale = 0.8, above right] at ({0.7 * sqrt(3) + 0.2 * sqrt(3)}, 0.85) {$p_2$};
                \end{tikzpicture}
                =\,B_{p_3}^{p_1p_2}
                \begin{tikzpicture}[scale = 0.6, baseline={([yshift = -0.5ex]current bounding box.center)}]
                  \draw[thick] (0, -1.5) ellipse (0.5 and 0.2);
                  \draw[thick, rotate = 120] (0, -1.5) ellipse (0.5 and 0.2);
                  \draw[thick, rotate = 240] (0, -1.5) ellipse (0.5 and 0.2);
                  \draw[thick, bend right = 40] (-0.5, -1.5) to ({-0.75 * sqrt(3) - 0.25}, {0.75 - 0.25 * sqrt(3)});
                  \draw[thick, bend left = 40] (0.5, -1.5) to ({0.75 * sqrt(3) + 0.25}, {0.75 - 0.25 * sqrt(3)});
                  \draw[thick, bend right = 40] ({-0.75 * sqrt(3) + 0.25}, {0.75 + 0.25 * sqrt(3)}) to ({0.75 * sqrt(3) - 0.25}, {0.75 + 0.25 * sqrt(3)});
                  \draw[red] (0, 0) to (0, -1.3);
                  \draw[red] (0, 0) to ({-0.75 * sqrt(3) + 0.1 * sqrt(3)}, 0.65);
                  \draw[red, bend left = 15] (0, 0) to (-1.15, 0.13);
                  \draw[dashed, red, bend left = 30] (-1.15, 0.13) to (-0.2, 0.78);
                  \draw[red, bend right = 30] (-0.2, 0.78) to ({0.75 * sqrt(3) - 0.1 * sqrt(3)}, 0.65);
                  \fill[red] (0, 0) circle [radius = 0.08];
                  % labels
                  \node[scale = 0.8, below] at (0, -1.9) {$p_3$};
                  \node[scale = 0.8, above left] at ({-0.7 * sqrt(3) - 0.2 * sqrt(3)}, 0.85) {$p_1$};
                  \node[scale = 0.8, above right] at ({0.7 * sqrt(3) + 0.2 * sqrt(3)}, 0.85) {$p_2$};
                \end{tikzpicture}
                =\,B_{p_1p_2}^{p_3}
                \begin{tikzpicture}[scale = 0.6, baseline={([yshift = -0.5ex]current bounding box.center)}]
                  \draw[thick] (0, -1.5) ellipse (0.5 and 0.2);
                  \draw[thick, rotate = 120] (0, -1.5) ellipse (0.5 and 0.2);
                  \draw[thick, rotate = 240] (0, -1.5) ellipse (0.5 and 0.2);
                  \draw[thick, bend right = 40] (-0.5, -1.5) to ({-0.75 * sqrt(3) - 0.25}, {0.75 - 0.25 * sqrt(3)});
                  \draw[thick, bend left = 40] (0.5, -1.5) to ({0.75 * sqrt(3) + 0.25}, {0.75 - 0.25 * sqrt(3)});
                  \draw[thick, bend right = 40] ({-0.75 * sqrt(3) + 0.25}, {0.75 + 0.25 * sqrt(3)}) to ({0.75 * sqrt(3) - 0.25}, {0.75 + 0.25 * sqrt(3)});
                  \draw[red] (0, 0) to (0, -1.3);
                  \draw[red] (0, 0) to ({0.75 * sqrt(3) - 0.1 * sqrt(3)}, 0.65);
                  \draw[red, bend right = 15] (0, 0) to (1.15, 0.13);
                  \draw[dashed, red, bend right = 30] (1.15, 0.13) to (0.2, 0.78);
                  \draw[red, bend left = 30] (0.2, 0.78) to ({-0.75 * sqrt(3) + 0.1 * sqrt(3)}, 0.65);
                  \fill[red] (0, 0) circle [radius = 0.08];
                  % labels
                  \node[scale = 0.8, below] at (0, -1.9) {$p_3$};
                  \node[scale = 0.8, above left] at ({-0.7 * sqrt(3) - 0.2 * sqrt(3)}, 0.85) {$p_1$};
                  \node[scale = 0.8, above right] at ({0.7 * sqrt(3) + 0.2 * sqrt(3)}, 0.85) {$p_2$};
                \end{tikzpicture}
              \end{align}
      \end{enumerate}
      While the braiding coefficient is just a phase factor
      \begin{align}
        B_{p_3}^{p_1p_2}=e^{\pi i(h_3-h_1-h_2)},\qquad B_{p_1p_2}^{p_3}=e^{-\pi i(h_3-h_1-h_2)}, \label{eq: braiding phase}
      \end{align}
      the fusion kernel $F$ and the modular $S$-kernel $S$ have complicated forms \eqref{eq: fusion kernel}, \eqref{eq: modular S-kernel}.
      However, it is apparent without knowing such intricacy in depth that the fusion kernel $F$ is symmetric under permuting two rows or two columns
      \begin{align}
        F_{p_sp_t}
        \begin{bmatrix}
          p_1 & p_3 \\
          p_2 & p_4
        \end{bmatrix}
        =F_{p_sp_t}
        \begin{bmatrix}
          p_3 & p_1 \\
          p_4 & p_2
        \end{bmatrix}
        =F_{p_sp_t}
        \begin{bmatrix}
          p_2 & p_4 \\
          p_1 & p_3
        \end{bmatrix}
        .
      \end{align}
      All crossing kernels here are complex conjugated in calculating the anti-chiral part of VTQFT though the fusion kernel $F$ is unchanged because it is real.
      See Appendix \ref{append: Crossing kernels and Moore-Seiberg consistency conditions} for a variety of crossing equations as a result of Moore-Seiberg consistency condition, namely the pentagon, the hexagon among others.
      The set of equations are frequently referred to in Subsection \ref{subsec: Heegaard splitting and path-integral on compression bodies} and Section \ref{sec: Anyon condensation in Virasoro TQFT}.
      \par We now turn to introduce several important formulae that will be employed in the subsequent calculations.
      At the outset, let us consider the fusion transformation from the $s$-channel to the $u$-channel (see also eq.\,(3.33) in Ref.\,\cite{Collier:2024mgv})
      \begin{align}
        \begin{tikzpicture}[scale = 0.6, baseline={([yshift = -0.5ex]current bounding box.center)}]
          \draw[red] (0,0) to (4.5, 0);
          \draw[red] (1.5,0) to (1.5, 1.5);
          \draw[red] (3, 0) to (3, 1.5);
          \node[above] at (0, 0) {$p_1$};
          \node[above] at (4.5, 0) {$p_4$};
          \node[above] at (1.5, 1.5) {$p_2$};
          \node[above] at (3, 1.5) {$p_3$};
          \node[below] at (2.25, 0) {$p_s$};
          \fill[red] (1.5, 0) circle (.07);
          \fill[red] (3, 0) circle (.07);
        \end{tikzpicture}
        =\int_{0}^{\infty}dp_u\,e^{\pi i(h_s+h_u-h_1-h_4)}\,F_{p_sp_u}
        \begin{bmatrix}
          p_1 & p_3 \\
          p_2 & p_4
        \end{bmatrix}
        \begin{tikzpicture}[scale = 0.6, baseline={([yshift = -0.5ex]current bounding box.center)}]
          \draw[red] (0, 0) to (4.5, 0);
          \draw[red] (3, 0) to (1.5, 1.5);
          \fill[white] (2.25, 0.75) circle (0.125);
          \draw[red] (1.5, 0) to (3, 1.5);
          \node[above] at (0, 0) {$p_1$};
          \node[above] at (4.5, 0) {$p_4$};
          \node[above] at (1.5, 1.5) {$p_2$};
          \node[above] at (3, 1.5) {$p_3$};
          \node[below] at (2.25, 0) {$p_u$};
          \fill[red] (1.5, 0) circle (.07);
          \fill[red] (3, 0) circle (.07);
        \end{tikzpicture}\ . \label{eq: fusion kernel to u-channel}
      \end{align}
      There is an extra phase factor in the integrand compared to fusion transformation from $s$-channel to $t$-channel \eqref{eq: fusion transformation} due to the overlap of $p_2$ and $p_3$.
      To prove this equation, we apply the fusion transformation, the braiding and the fusion transformation again
      \begin{align}
        & \begin{tikzpicture}[scale = 0.6, baseline={([yshift = -0.5ex]current bounding box.center)}]
          \draw[red] (0,0) to (4.5, 0);
          \draw[red] (1.5,0) to (1.5, 1.5);
          \draw[red] (3, 0) to (3, 1.5);
          \node[above] at (0, 0) {$p_1$};
          \node[above] at (4.5, 0) {$p_4$};
          \node[above] at (1.5, 1.5) {$p_2$};
          \node[above] at (3, 1.5) {$p_3$};
          \node[below] at (2.25, 0) {$p_s$};
          \fill[red] (1.5, 0) circle (.07);
          \fill[red] (3, 0) circle (.07);
        \end{tikzpicture}
        =\int_{0}^{\infty}dp_t\,F_{p_sp_t}
        \begin{bmatrix}
          p_2 & p_3 \\
          p_1 & p_4
        \end{bmatrix}
        \begin{tikzpicture}[scale = 0.6, baseline = 10.5pt]
          \draw[red] (0,0) to (4.5, 0);
          \draw[red] (1.5, 1.5) to (2.25, 0.75);
          \draw[red] (3, 1.5) to (2.25, 0.75);
          \draw[red] (2.25, 0) to (2.25, 0.75);
          \node[above] at (0, 0) {$p_1$};
          \node[above] at (4.5, 0) {$p_4$};
          \node[above] at (1.5, 1.5) {$p_2$};
          \node[above] at (3, 1.5) {$p_3$};
          \node[above] at (1.8, 0) {$p_t$};
          \fill[red] (2.25, 0) circle (.07);
          \fill[red] (2.25, 0.75) circle (.07);
        \end{tikzpicture}
        \nonumber \\
        & \qquad\qquad\quad=\int_{0}^{\infty}dp_t\,F_{p_sp_t}
        \begin{bmatrix}
          p_2 & p_3 \\
          p_1 & p_4
        \end{bmatrix}
        e^{-\pi i(h_t-h_2-h_3)}
        \begin{tikzpicture}[scale = 0.6, baseline = 10.5pt]
          \draw[red] (0,0) to (4.5, 0);
          \draw[red] (1.5, 1.5) .. controls (2.5, 1.1) and (2.5, 0.9) .. (2.25, 0.75);
          \fill[white] (2.25, 1.13) circle (0.1);
          \draw[red] (3, 1.5) .. controls (2, 1.1) and (2, 0.9) .. (2.25, 0.75);
          \draw[red] (2.25, 0) to (2.25, 0.75);
          \node[above] at (0, 0) {$p_1$};
          \node[above] at (4.5, 0) {$p_4$};
          \node[above] at (1.5, 1.5) {$p_2$};
          \node[above] at (3, 1.5) {$p_3$};
          \node[above] at (1.8, 0) {$p_t$};
          \fill[red] (2.25, 0) circle (.07);
          \fill[red] (2.25, 0.75) circle (.07);
        \end{tikzpicture}
        \nonumber \\
        & \qquad=\int_{0}^{\infty}dp_tdp_u\,F_{p_sp_t}
        \begin{bmatrix}
          p_2 & p_3 \\
          p_1 & p_4
        \end{bmatrix}
        e^{-\pi i(h_t-h_2-h_3)}F_{p_tp_u}
        \begin{bmatrix}
          p_1 & p_3 \\
          p_4 & p_2
        \end{bmatrix}
        \begin{tikzpicture}[scale = 0.6, baseline={([yshift = -0.5ex]current bounding box.center)}]
          \draw[red] (0, 0) to (4.5, 0);
          \draw[red] (3, 0) to (1.5, 1.5);
          \fill[white] (2.25, 0.75) circle (0.125);
          \draw[red] (1.5, 0) to (3, 1.5);
          \node[above] at (0, 0) {$p_1$};
          \node[above] at (4.5, 0) {$p_4$};
          \node[above] at (1.5, 1.5) {$p_2$};
          \node[above] at (3, 1.5) {$p_3$};
          \node[below] at (2.25, 0) {$p_u$};
          \fill[red] (1.5, 0) circle (.07);
          \fill[red] (3, 0) circle (.07);
        \end{tikzpicture}
        , \label{eq: s-u crossing transform}
      \end{align}
      and at last use the hexagon identity \eqref{eq: hexagon 1}.
      It can be seen by exactly the same proof, except for the use of the other version of the hexagon identity \eqref{eq: hexagon 2}, that a similar formula holds true when $p_2$ and $p_3$ are swapped front and back
      \begin{align}
        \begin{tikzpicture}[scale = 0.6, baseline={([yshift = -0.5ex]current bounding box.center)}]
          \draw[red] (0,0) to (4.5, 0);
          \draw[red] (1.5,0) to (1.5, 1.5);
          \draw[red] (3, 0) to (3, 1.5);
          \node[above] at (0, 0) {$p_1$};
          \node[above] at (4.5, 0) {$p_4$};
          \node[above] at (1.5, 1.5) {$p_2$};
          \node[above] at (3, 1.5) {$p_3$};
          \node[below] at (2.25, 0) {$p_s$};
          \fill[red] (1.5, 0) circle (.07);
          \fill[red] (3, 0) circle (.07);
        \end{tikzpicture}
        =\int_{0}^{\infty}dp_u\,e^{\pi i(h_1+h_4-h_s-h_u)}\,F_{p_sp_u}
        \begin{bmatrix}
          p_1 & p_3 \\
          p_2 & p_4
        \end{bmatrix}
        \begin{tikzpicture}[scale = 0.6, baseline={([yshift = -0.5ex]current bounding box.center)}]
          \draw[red] (0, 0) to (4.5, 0);
          \draw[red] (1.5, 0) to (3, 1.5);
          \fill[white] (2.25, 0.75) circle (0.125);
          \draw[red] (3, 0) to (1.5, 1.5);
          \node[above] at (0, 0) {$p_1$};
          \node[above] at (4.5, 0) {$p_4$};
          \node[above] at (1.5, 1.5) {$p_2$};
          \node[above] at (3, 1.5) {$p_3$};
          \node[below] at (2.25, 0) {$p_u$};
          \fill[red] (1.5, 0) circle (.07);
          \fill[red] (3, 0) circle (.07);
        \end{tikzpicture}
        . \label{eq: fusion kernel to u-channel ver.2}
      \end{align}
      We need additional two momentous link identities regarding \bit{Wilson bubble}, \bit{Wilson triangle} and \bit{Verlinde loop}.
      The first and the last terms are borrowed from Ref.\,\cite{Post:2024itb}, while the ``Wilson triangle" is our original designation.
      The Wilson bubble is a line containing a loop and the Wilson triangle is a triangle at a trivalent junction as presented in the l.\,h.\,s.~of the following identities.
      \begin{gather}
        \begin{tikzpicture}[scale = 0.6, baseline={([yshift = -0.5ex]current bounding box.center)}]
          \draw[red] (0, 0) -- (1.5, 0);
          \draw[red] (2.5, 0) circle [radius = 1];
          \draw[red] (3.5, 0) -- (5, 0);
          \fill[red] (1.5, 0) circle[radius = 0.1];
          \fill[red] (3.5, 0) circle[radius = 0.1];
          % labels
          \node[above, scale = 0.8] at (0, 0) {$p_1$};
          \node[above, scale = 0.8] at (2.5, 1) {$p_2$};
          \node[below, scale = 0.8] at (2.5, -1) {$p_3$};
          \node[above, scale = 0.8] at (5, 0) {$p_4$};
        \end{tikzpicture}
        =\frac{\delta(p_1-p_4)}{\rho_0(p_1)C_0(p_1,p_2,p_3)}\,
        \begin{tikzpicture}[scale = 0.6, baseline = -2.7pt]
          \draw[red] (0, 0) -- (5, 0);
          \node[above, scale = 0.8] at (2.5, 0) {$p_1$};
        \end{tikzpicture}
        , \label{eq: Wilson bubble} \\
        \begin{tikzpicture}[scale = 0.5, baseline = {([yshift=-.5ex]current bounding box.center)}]
          \draw[red] (0, 1.5) to (0, 0.75);
          \draw[red] ({-0.75 * sqrt(3)}, -0.75) to ({-0.375 * sqrt(3)}, -0.375);
          \draw[red] ({0.75 * sqrt(3)}, -0.75) to ({0.375 * sqrt(3)}, -0.375);
          \draw[red] ({-0.375 * sqrt(3)}, -0.375) to ({0.375 * sqrt(3)}, -0.375);
          \draw[red] ({0.375 * sqrt(3)}, -0.375) to (0, 0.75);
          \draw[red] (0, 0.75) to ({-0.375 * sqrt(3)}, -0.375);
          \fill[red] ({-0.375 * sqrt(3)}, -0.375) circle[radius = 0.1];
          \fill[red] ({0.375 * sqrt(3)}, -0.375) circle[radius = 0.1];
          \fill[red] (0, 0.75) circle[radius = 0.1];
          % labels
          \node[scale = 0.8] at (0, 1.8) {$p_1$};
          \node[scale = 0.8] at ({-0.9 * sqrt(3)}, -0.9) {$p_2$};
          \node[scale = 0.8] at ({0.9 * sqrt(3)}, -0.9) {$p_3$};
          \node[above, scale = 0.8] at (-0.6, 0) {$p_u$};
          \node[above, scale = 0.8] at (0.6, 0) {$p_t$};
          \node[below, scale = 0.8] at (0, -0.375) {$p_s$};
        \end{tikzpicture}
        =\frac{1}{\rho_0(p_1)C_0(p_1,p_t,p_u)}F_{p_sp_1}
        \begin{bmatrix}
          p_3 & p_2 \\
          p_t & p_u
        \end{bmatrix}
        \begin{tikzpicture}[scale = 0.5, baseline = {([yshift=-.5ex]current bounding box.center)}]
          \draw[red] (0, 1.5) to (0, 0);
          \draw[red] ({-0.75 * sqrt(3)}, -0.75) to (0, 0);
          \draw[red] ({0.75 * sqrt(3)}, -0.75) to (0, 0);
          \fill[red] (0, 0) circle[radius = 0.1];
          % labels
          \node[scale = 0.8] at (0, 1.8) {$p_1$};
          \node[scale = 0.8] at ({-0.9 * sqrt(3)}, -0.9) {$p_2$};
          \node[scale = 0.8] at ({0.9 * sqrt(3)}, -0.9) {$p_3$};
        \end{tikzpicture}
        . \label{eq: Wilson triangle identity}
      \end{gather}
      These are only valid when the loop and the triangle are placed on a contractible cycle in the bulk, i.e.~they must not wrap around any genera of boundary Riemann surfaces.
      The Wilson bubble identity \eqref{eq: Wilson bubble} is indeed a corollary of the Wilson triangle identity \eqref{eq: Wilson triangle identity}, established through setting $p_s=p_t$, applying the swapping equality of the $F$-symbol (see eq.\,(3.15) and eq.\,(3.16) in Ref.\,\cite{Collier:2024mgv}) and taking $p_3\to\mathbbm{1}$ along with the reparametrization $p_u\to p_3$, $p_t\to p_2$.
      The proof of eq.\,\eqref{eq: Wilson triangle identity} requires some laborious four-boundary wormhole calculations not vital to our present analysis, so only those interested may refer to the discussion around eq.\,(3.45) in Ref.\,\cite{Collier:2023fwi}.
      The other one is the Verlinde loop, a Wilson loop that encircles a single line as illustrated below
      \begin{align}
        \begin{tikzpicture}[scale = 0.6, baseline = {([yshift=-.5ex]current bounding box.center)}]
          \draw[red] (0, 0) circle[x radius = 0.5, y radius = 1];
          \fill[white] (0.5, 0) circle[radius = 0.125];
          \draw[red] (-2.5, 0) -- (-0.625, 0);
          \draw[red] (-0.375, 0) -- (2.5, 0);
          \draw[red] (0, 1) -- (0, 2);
          \node[scale = 0.8, above] at (2, 0) {$p_1$};
          \node[scale = 0.8, left] at (0, 1.8) {$p_2$};
          \node[scale = 0.8, below left] at (-0.25, -{sqrt(0.75)}) {$p$};
        \end{tikzpicture}
        =\frac{S_{pp_1}\qty[p_2]}{S_{\mathbbm{1}p_1}\qty[\mathbbm{1}]}\,
        \begin{tikzpicture}[scale = 0.6, baseline = 13.5pt]
          \draw[red] (-2.5, 0) -- (2.5, 0);
          \draw[red] (0, 0) -- (0, 2);
          \node[scale = 0.8, above] at (2, 0) {$p_1$};
          \node[scale = 0.8, left] at (0, 1.8) {$p_2$};
        \end{tikzpicture}
        . \label{eq: Verlinde loop}
      \end{align}
      To prove the equality, one apply the general formula (A.28) in Ref.\,\cite{Post:2024itb} as follows
      \begin{align}
        \begin{tikzpicture}[scale = 0.6, baseline = {([yshift=-.5ex]current bounding box.center)}]
          \draw[red] (0, 0) circle[x radius = 0.5, y radius = 1];
          \fill[white] (0.5, 0) circle[radius = 0.125];
          \draw[red] (-2.5, 0) -- (-0.625, 0);
          \draw[red] (-0.375, 0) -- (2.5, 0);
          \draw[red] (0, 1) -- (0, 2);
          \draw[dashed, red] (0, 0) -- (0, -1);
          \node[scale = 0.8, above] at (2, 0) {$p_1$};
          \node[scale = 0.8, left] at (0, 1.8) {$p_2$};
          \node[scale = 0.8, below left] at (-0.25, -{sqrt(0.75)}) {$p$};
          \node[scale = 0.8, right] at (-0.1, -0.4) {$\mathbbm{1}$};
        \end{tikzpicture}
        \;=\, \int_0^\infty dp^{\prime}\, F_{\mathbbm{1}p^{\prime}}\qty[
          \begin{matrix}
            p~ & p \\
            p~ & p
          \end{matrix}
        ]\frac{S_{pp_1}[p^{\prime}]}{S_{\mathbbm{1}p_1}\qty[\mathbbm{1}]}\;
        \begin{tikzpicture}[scale = 0.6, baseline = 16pt]
          \draw[red] (-2.5, 0) -- (2.5, 0);
          \draw[red] (0, 0) -- (0, 0.75);
          \draw[red] (0, 1.25) circle[radius = 0.5];
          \draw[red] (0, 1.75) -- (0, 2.5);
          \node[scale = 0.8, above] at (-2.5, 0) {$p_1$};
          \node[scale = 0.8, above] at (2, 0) {$p_1$};
          \node[scale = 0.8, left] at (0, 0.4) {$p^{\prime}$};
          \node[scale = 0.8, left] at (-0.5, 1.25) {$p$};
          \node[scale = 0.8, right] at (0.5, 1.25) {$p$};
          \node[scale = 0.8, left] at (0, 2.3) {$p_2$};
        \end{tikzpicture}
        ,
      \end{align}
      and then resolve the $p$ loop by the Wilson bubble identity \eqref{eq: Wilson bubble}.

    \subsection{Heegaard splitting and path-integral on compression bodies} \label{subsec: Heegaard splitting and path-integral on compression bodies}
      We proceed to establish the procedure for computing the state $\ket{Z_{\text{Vir}}(M)}$ for a $3$-manifold $M$ with boundaries $\partial M=\bigsqcup_{i=1}^n\Sigma_{g_i,n_i}$.
      We refer to the state as the \bit{VTQFT partition function} of $M$.
      While VTQFT shares significant similarities with standard TQFTs, there are notable differences due to the presence of the continuous spectrum.
      In fact, $\ket{Z_{\text{Vir}}(M)}$ is a state in $\bigotimes_{i=1}^n\mathcal{H}_{g_i,n_i}$ or in $\mathbb{C}$ when $\der M=\emptyset$, although typically $\dim\mathcal{H}_{g_i,n_i}=\infty$.
      The significant challenge is that the surgery is not always qualified in VTQFT since the theory is only sensible for \textit{hyperbolic} manifolds. \footnote{This can partly be understood from the fact that non-hyperbolic manifolds cannot be on-shell (a solution to the classical EOM) in AdS$_3$ gravity.}
      In Chern-Simons TQFT, the partition function of $S^3$ is computed by starting with $S^1\times S^2$ containing a single Wilson line wrapping around the non-contractible cycle in the $S^1$-direction.
      One then performs Dehn surgery on the line, leading to the partition function $Z(S^3)=S_{00}$ \cite{Witten:1988hf}. However $Z_{\text{Vir}}(S^3)$ is ill-defined since $S_{\mathbbm{1}\mathbbm{1}}[\mathbbm{1}]$ is not defined in VTQFT.
      Therefore, the VTQFT path-integral can only be implemented for sufficiently complex geometry, specifically those with boundaries $\Sigma_{g,n}$ satisfying $2-2g-n<0$.
      Although an unpunctured torus does not meet this criterion---and the inner product of $\mathcal{H}_{1,0}$ is not even delta-function normalizable---it is occasionally discussed as an exception at the level of conformal blocks \cite{Collier:2023fwi,Jafferis:2024jkb}.
      \par At this point, the explicit rules are outlined as follows.
      The first rule applies to a handlebody (a $3$-manifold formed by filling the interior of a Riemann surface $\Sigma_{g,n}$) with a network of Wilson lines $\bm{p}$ inserted along a channel $\mathcal{C}$ but no extra lines
      \begin{align}
        \ket{Z_{\text{Vir}}\qty(S\Sigma_{g,n})}:=\ket{\mathcal{F}_{g,n}^{\mathcal{C}}\qty(\bm{p})}. \label{eq: path-integral on handlebodies}
      \end{align}
      In particular, the VTQFT path-integral prepares $\ket{Z_{\text{Vir}}\qty(S\Sigma_{g})}=\ket{\mathcal{F}_{g}^{\mathcal{C}}\qty(\mathbbm{1})}$ for an unpunctured Riemann surface with no extra Wilson line insertion.
      For $(g,n)=(1,0)$, the torus zero-point block $\ket{\mathcal{F}_{1,0}\qty(p)}\in\mathcal{H}_{1,0}$, or equivalently a state on a 2d boundary torus of a 3d solid torus with Wilson line $p$ wrapping around its non-contractible (longitudinal) cycle, is nothing but the Virasoro character. As such, we interchangeably denote it as $\ket{\chi_p}=\ket{\mathcal{F}_{1,0}\qty(p)}$ in what follows.
      For the wormhole geometry $\Sigma_{g,n}\times\qty[0,1]$, the VTQFT path-integral inserts a complete set of state on both boundaries diagonally
      \begin{align}
        \ket{Z_{\text{Vir}}\qty(\Sigma_{g,n}\times\qty[0,1])}:=\int d^{3g-3+n}\bm{p}~\rho_{g,n}^{\mathcal{C}}\qty(\bm{p})\ket{\mathcal{F}_{g,n}^{\mathcal{C}}\qty(\bm{p};\bm{p}_e)}\otimes\ket{\mathcal{F}_{g,n}^{\mathcal{C}}\qty(\bm{p};\bm{p}_e)}. \label{eq: wormhole path-integral}
      \end{align}
      where the explicit form of the integral measure $\rho_{g_i,n_i}^{\mathcal{C}_i}(\bm{p}_i)$'s is given in eq.\,\eqref{eq: integral measure}.
      For a generic manifold $M$, we employ the \bit{generalized Heegaard splitting} to divide $M$ into a pair of manifolds $C_{g,n}^{(1)}(g_1,n_1;\cdots;g_{m_1},n_{m_1})$ and $C_{g,n}^{(2)}(h_1,k_1;\cdots;h_{m_2},k_{m_2})$. Heegaard splitting is a generic scheme for decomposing a manifold into smaller pieces in the theory of $3$-manifolds.
      We compute the VTQFT partition function for each part and glue them together along the splitting surface, applying a twist by the $U_{g,n}(\gamma)$ for some $\gamma\in\MCG(\Sigma_{g,n})$.
      $C_{g,n}\qty(g_1,n_1;\cdots;g_m,n_m)$ is referred to as a \bit{compression body} whose boundary consists of the ``outer" segment $(\der C_{g,n})^+=\Sigma_{g,n}$ and the ``inner" segment $(\der C_{g,n})^-=\bigsqcup_{i=1}^m\Sigma_{g_i,n_i}$ (possibly $(\der C_{g,n})^-=\emptyset$).
      The simplest example of a Heegaard splitting is $M=S^3$ decomposed into two compression bodies $C_{1,0}^{(1)}=ST^2$, $C_{1,0}^{(2)}=ST^2$ (solid tori) which are glued along the boundary tori $T^2$ twisted by the modular $S$-transformation in $\MCG(T^2)\cong SL(2,\mathbb{Z})$, although the partition function of $S^3$ is not computable in VTQFT as mentioned above.
      \par The VTQFT path-integral produces a state $\big|Z_{\text{Vir}}\big(C_{g,n}\qty(g_1,n_1;\cdots;g_m,n_m)\big)\big>$ for the compression body, inserting a complete set of states $\big|\mathcal{F}_{g_i,n_i}^{\mathcal{C}_i}(\bm{p}_i)\big>$ on each inner boundary component and a certain state $\big|\Phi_{g,n}^{\mathcal{C}}(\bm{p}_1,\cdots,\bm{p}_m;\bm{q})\big>$ on the outer boundary
      \begin{align}
        & \quad\big|Z_{\text{Vir}}\big(C_{g,n}\qty(g_1,n_1;\cdots;g_m,n_m)\big)\big> \label{eq: compression body path-integral} \\
        & \qquad:=\int\prod_{i=1}^m\big(d^{3g_i-3+n_i}\bm{p}_i~\rho_{g_i,n_i}^{\mathcal{C}_i}(\bm{p}_i)\big)\big|\mathcal{F}_{g_1,n_1}^{\mathcal{C}_1}(\bm{p}_1)\big>\otimes\cdots\otimes\big|\mathcal{F}_{g_m,n_m}^{\mathcal{C}_m}(\bm{p}_m)\big>\otimes\big|\Phi_{g,n}^{\mathcal{C}}(\bm{p}_1,\cdots,\bm{p}_m:\bm{q})\big>. \nonumber
      \end{align}
      Technically, $\big|\Phi_{g,n}^{\mathcal{C}}(\bm{p}_1,\cdots,\bm{p}_m;\bm{q})\big>$ is constructed by placing the network of Wilson lines $\bm{p}_i$ on the sub-channel $\mathcal{C}_i$ of $\mathcal{C}$ corresponding to the inner boundary block $\big|\mathcal{F}_{g_i,n_i}^{\mathcal{C}_i}(\bm{p}_i)\big>$, and it may or may not contain additional network of Wilson lines $\bm{q}$ (see Figure 7 in Ref.\,\cite{Collier:2023fwi}).
      The following example illustrates the process carried out here in more detail.
      The path-integral on the wormhole \eqref{eq: wormhole path-integral} serves as an example of this compression body path-integral with $m=1$, $g_1=g$, $n_1=n$ and no additional Wilson lines $\bm{q}$.
      The two compression bodies are then glued together along their outer boundary $\Sigma_{g,n}$ twisted by $\gamma$.
      We evaluate the matrix element $\big<\Phi_{g,n}^{\mathcal{C}}(\bm{p}_1,\cdots,\bm{p}_{m_1};\bm{q}_1)\big|U_{g,n}(\gamma)\big|\Phi_{g,n}^{\mathcal{D}}(\bm{q}_1,\cdots,\bm{q}_{m_2};\bm{q}_2)\big>$ and obtain
      \begin{align}
        & \ket{Z_{\text{Vir}}\qty(M)}=\big<Z_{\text{Vir}}\big(C_{g,n}^{\qty(1)}\qty(g_1,n_1;\cdots;g_{m_1},n_{m_1})\big)\big|U_{g,n}\qty(\gamma)\big|Z_{\text{Vir}}\big(C_{g,n}^{\qty(2)}\qty(h_1,k_1;\cdots;h_{m_2},k_{m_2})\big)\big> \nonumber \\
        & \quad:=\int\prod_{i=1}^{m_1}\big(d^{3g_i-3+n_i}\bm{p}_i~\rho_{g_i,n_i}^{\mathcal{C}_i}(\bm{p}_i)\big)\int\prod_{j=1}^{m_2}\big(d^{3h_j-3+k_j}\bm{q}_j~\rho_{h_j,k_j}^{\mathcal{D}_j}(\bm{q}_j)\big) \nonumber \\
        & \quad\qquad\times\big<\Phi_{g,n}^{\mathcal{C}}(\bm{p}_1,\cdots,\bm{p}_{m_1};\bm{q}_1)\big|U_{g,n}(\gamma)\big|\Phi_{g,n}^{\mathcal{D}}(\bm{q}_1,\cdots,\bm{q}_{m_2};\bm{q}_2)\big> \\
        & \quad\qquad\times\big|\mathcal{F}_{g_1,n_1}^{\mathcal{C}_1}(\bm{p}_1)\big>\otimes\cdots\otimes\big|\mathcal{F}_{g_{m_1},n_{m_1}}^{\mathcal{C}_{m_1}}(\bm{p}_{m_1})\big>\otimes\big|\mathcal{F}_{h_1,k_1}^{\mathcal{D}_1}(\bm{q}_1)\big>\otimes\cdots\otimes\big|\mathcal{F}_{h_{m_2},k_{m_2}}^{\mathcal{D}_{m_2}}(\bm{q}_{m_2})\big>. \nonumber
      \end{align}
      In most cases, we only take into account the trivial gluing $U_{g,n}(\gamma)=\identity_{\mathcal{H}_{g,n}}$.

      \subsubsection*{Example: $\Sigma_{1,2}\times\qty[0,1]$ wormhole with non-trivial bulk linking}
      \begin{figure}[t]
        \centering
        {
          \begin{tikzpicture}[scale = 0.6]
            % left torus boundary
            \draw[thick] (-3, 0) ellipse (1 and 2);
            \draw[thick, bend left = 20] (-3.1, 1) to (-3.1, -1);
            \draw[thick, bend right = 20] (-3.02, 0.78) to (-3.02, -0.78);
            % right torus boundary
            \draw[thick] (3, 0) ellipse (1 and 2);
            \draw[thick, bend left = 20] (2.9, 1) to (2.9, -1);
            \draw[thick, bend right = 20] (2.98, 0.78) to (2.98, -0.78);
            %connecting lines
            \draw[thick, bend right = 10] (-2.97, 2) to (2.97, 2);
            \draw[thick, bend left = 10] (-2.97, -2) to (2.97, -2);
            % Wilson lines
            \draw[red] (-2.7, 1.3) .. controls (-1, 1) and (-0.4, 0.7) .. (-0.4, 0);
            \draw[red] (2.7, 1.3) .. controls (1, 1) and (0.4, 0.7) .. (0.4, 0);
            \fill[white] (-0.52, 0.42) circle [radius = 0.125];
            \fill[white] (0.52, 0.42) circle [radius = 0.125];
            \draw[red] (0, 0) ellipse (1 and 0.5);
            \fill[white] (-0.52, -0.42) circle [radius = 0.125];
            \fill[white] (0.52, -0.42) circle [radius = 0.125];
            \draw[red] (2.7, -1.3) .. controls (1, -1) and (0.4, -0.7) .. (0.4, 0);
            \draw[red] (-2.7, -1.3) .. controls (-1, -1) and (-0.4, -0.7) .. (-0.4, 0);
            % punctures
            \draw[thick] (-2.8, 1.4) to (-2.6, 1.2);
            \draw[thick] (-2.8, 1.2) to (-2.6, 1.4);
            \draw[thick] (2.8, 1.4) to (2.6, 1.2);
            \draw[thick] (2.8, 1.2) to (2.6, 1.4);
            \draw[thick] (-2.8, -1.4) to (-2.6, -1.2);
            \draw[thick] (-2.8, -1.2) to (-2.6, -1.4);
            \draw[thick] (2.8, -1.4) to (2.6, -1.2);
            \draw[thick] (2.8, -1.2) to (2.6, -1.4);
            % labels
            \node[scale = 0.8, above] at (-1.3, 1.05) {$p_1$};
            \node[scale = 0.8, above] at (1.3, 1.05) {$p_2$};
            \node[scale = 0.8, below] at (0, -0.5) {$p_3$};
          \end{tikzpicture}
        }
        \caption{$\Sigma_{1,2}\times\qty[0,1]$ wormhole with three Wilson lines tangling in the bulk}
        \label{fig: 2-punctured genus 1  wormhole}
      \end{figure}
      To provide clarity, let us do an exercise in computing the VTQFT partition function for a particular geometry that has not yet appeared in the literature, namely $\Sigma_{1,2}\times\qty[0,1]$ wormhole geometry with Wilson lines non-trivially linking in the bulk as seen in Figure \ref{fig: 2-punctured genus 1 wormhole}.
      Similar example of $\Sigma_{0,3}\times[0,1]$ can be found in Section 5 in Ref.\,\cite{deBoer:2024mqg}.
      From this point onward, we will adopt the following shorthand notation to avoid unnecessary complications,
      \begin{align}
        C_{ijk}:=C_0(p_i,p_j,p_k).
      \end{align}
      Let us first untangle the link by fusion kernel transformation and braiding
      \begin{align}
        & \lower2.1ex\hbox{\text{\scalebox{1.6}[4.5]{$|$}}}\,

        \lower2.1ex\hbox{\text{\scalebox{2}[4.5]{$\rangle$}}} \label{eq: total partition function of 2-punctured genus 1 wormhole}
      \end{align}
      We denote as $M$ the $3$-manifold with two $2$-punctured torus boundaries in the last line. Although $M$ itself is a compression body, we would better Heegaard-split it along the dashed line into two compression bodies $M_1$ and $M_2$ shown in Figure \ref{fig: 2-punctured genus 1 wormhole after Heegaard splitting}.
      \begin{figure}[t]
        \begin{minipage}[b]{0.45\linewidth}
          \centering
          %
        \end{minipage}
        \caption{Two compression bodies that constitute the boundary of the wormhole}
        \label{fig: 2-punctured genus 1  wormhole after Heegaard splitting}
      \end{figure}
      To implement the VTQFT path-integral on them, one needs to apply the general path-integral rule \eqref{eq: compression body path-integral}
      \begin{align}
        \begin{aligned}
          \ket{M_1}=\int_{0}^{\infty}dp_adp_b\,\rho_0(p_a)\rho_0(p_b)C_{ab1}^2\ket{
            %
          }.
        \end{aligned}
      \end{align}
      Both $\ket{M_1}$ and $\ket{M_2}$ are states in $\mathcal{H}_{1,2}\otimes\mathcal{H}_{1,2}$ and the blue lines $p$, $p^{\prime}$ correspond to the additional lines $\bm{q}$ in eq.\,\eqref{eq: compression body path-integral}.
      The outer boundary (colored in olive) has seemingly complicated network of Wilson lines because of $p$, $p^{\prime}$, but it can be easily resolved by fusion transformation and the Wilson triangle identity \eqref{eq: Wilson triangle identity}
      \begin{align}
        & \ket{
          \,
          %
        }.
      \end{align}
      We thus take an inner product between outer boundaries of $M_1$ and $M_2$ to glue them together into $M$ with $U_{g,n}(\gamma)=\identity_{\mathcal{H}_{g,n}}$
      \begin{align}
        \ket{M} & =\int_{0}^{\infty}dp_adp_bdp_cdp_d\,\rho_0(p_a)\rho_0(p_b)\rho_0(p_c)\rho_0(p_d)C_{ab1}^2C_{cd2}^2 \nonumber                                      \\
                & \qqquad\times\braket{
          \,
          %
        }.
      \end{align}
      where eq.\,\eqref{eq: inner product example} is used from the first to the second line.
      Substituting this into \eqref{eq: total partition function of 2-punctured genus 1 wormhole} and applying the formula \eqref{eq: reduced two-punctured torus consistency} followed by its complex conjugated version yield
      \begin{align}
         & \lower2.1ex\hbox{\text{\scalebox{1.6}[4.5]{$|$}}}\,
        %
        }. \nonumber
      \end{align}

    \subsection{Interlude: solid torus geometry in VTQFT} \label{subsec: application to solid torus}
      With the necessary tools at available, we briefly step aside to assess how VTQFT applies to a simple spacetime as a candidate for a quantum gravity model.
      Matching the VTQFT path-integral of multi-boundary wormhole geometries to ensemble CFT data has been explored extensively in the literature \cite{Collier:2024mgv, Jafferis:2024jkb, deBoer:2024mqg}, so we focus on evaluating the solid torus here.
      Solid torus is the most elementary yet momentous topology in 3d geometry as it encompasses both the thermal AdS$_3$ and the Euclidean BTZ blackhole \cite{Banados:1992wn}.
      Let us first review how to determine the torus moduli parameter $\tau$ for the thermal AdS$_3$ and the BTZ blackhole, which allows us to compute the VTQFT partition function $Z_{\text{Vir}}(ST^2;\tau)=\braket{\tau}{Z_{\text{Vir}}(ST^2)}=\chi_0(\tau)$ in the subsequent calculations.
      \par The metric for the Euclidean AdS$_3$ with temperature $T_H$ and Euclidean BTZ blackhole are given by
      \begin{align}
        g_{\text{AdS}_3} & =\qty(1+\frac{r^2}{l^2})dt_E^2+\frac{1}{1+\tfrac{r^2}{l^2}}dr^2+r^2d\theta \label{eq: metric of AdS3}, \\
        g_{\text{BTZ}} & =f(r)^2dt_E^2+\frac{1}{f(r)^2}dr^2+r^2\qty(-\frac{J_E}{2r^2}dt_E+d\theta)^2, \label{eq: metric of a BTZ blackhole}
      \end{align}
      where $f(r):=\sqrt{-M+\tfrac{r^2}{l^2}-\tfrac{J_E^2}{4r^2}}$, $J_E:=-iJ$ and $\theta\in[0,2\pi)$.
      Note that AdS$_3$ is the special case $M=-1$ and $J_E=0$ in the family of BTZ blackholes parametrized by $M$, $J_E$.
      The BTZ blackhole has two horizons corresponding to the solution of $f(r)=0$, namely
      \begin{align}
        r_{\pm}:=\sqrt{\tfrac{Ml^2\pm l\sqrt{M^2l^2+J_E^2}}{2}}=\frac{1}{2}\qty(\sqrt{l(Ml+J)}\pm\sqrt{l(Ml-J)}).
      \end{align}
      Located at the outer radius $r_+$ is the event horizon whose surface area $A$ is computed by a volume integral in terms of the induced metric $\widetilde{g}_{\text{event horizon}}:=g_{\text{BTZ}}\big|_{r=r_+,\,t_E=const.}=r_+^2d\theta^2$
      \begin{align}
        A=\int_{0}^{2\pi}d\theta\,\sqrt{\widetilde{g}_{\text{event horizon}}(\partial_{\theta},\partial_{\theta})}=2\pi r_+.
      \end{align}
      Viewed as a function of mass $M$ and angular momentum $J$ the exterior derivative $dA$ is a linear combination of $dM$ and $dJ$, and we can arrange it to
      \begin{gather}
        dM=\frac{1}{lr_+}\bigg(\frac{2}{\pi l}(r_+^2-r_-^2)dA+r_-dJ\bigg).
      \end{gather}
      This is nothing other than the first law of blackhole thermodynamics and we can read off the Hawking temperature
      \begin{align}
        T_H:=\frac{1}{\beta_H}:=\frac{r_+^2-r_-^2}{2\pi l^2r_+}=\frac{\sqrt{M^2l^2-J^2}}{\pi l(\sqrt{l(Ml+J)}+\sqrt{l(Ml-J)})}. \label{eq: Hawking temperature}
      \end{align}
      Alternatively, this result can be inferred by analyzing the near-horizon geometry and identifying the periodicity of the Euclidean time required to eliminate the conical singularities at the event horizon.\footnote{Use the coordinate $\rho^2:=r-r_+$ and approximate $f(r)$ to the second order in $\rho$. The metric \eqref{eq: metric of a BTZ blackhole} reads
      \begin{align}
        g_{\text{BTZ}}=\Big(\frac{4r_+}{l^2}-\frac{2M}{r_+}+\mathcal{O}(\rho^2)\Big)\rho^2dt_E^2+\frac{4}{\frac{4r_+}{l^2}-\frac{2M}{r_+}+\mathcal{O}(\rho^2)}d\rho^2+\cdots.
      \end{align}
      Remove the conical singularities at $\rho=0$ just as to eliminate the one at the origin of polar coordinates.}
      \par After the coordinate transformations \cite{Carlip:1994gc}
      \begin{gather}
        y_A:=\sqrt{\frac{l^2}{r^2+l^2}}\,e^{\frac{t_E}{l}},\;z_A:=\sqrt{\frac{r^2}{r^2+l^2}}\,e^{\frac{t_E}{l}-i\theta},\;\overline{z}_A:=\sqrt{\frac{r^2}{r^2+l^2}}\,e^{\frac{t_E}{l}+i\theta}, \\
        \quad y_B:=\sqrt{\frac{r_+^2-r_-^2}{r^2-r_-^2}}e^{\frac{r_+}{l}\theta-\frac{ir_-}{l^2}t_E},\;z_B:=\sqrt{\frac{r^2-r_+^2}{r^2-r_-^2}}e^{\frac{2\pi r_+}{r_++r_-}T_H(l\theta+it_E)},\;\overline{z}_B:=\sqrt{\frac{r^2-r_+^2}{r^2-r_-^2}}e^{\frac{2\pi r_+}{r_+-r_-}T_H(l\theta-it_E)}, \notag
      \end{gather}
     the metric of the thermal AdS$_3$ \eqref{eq: metric of AdS3} and a Euclidean BTZ blackhole \eqref{eq: metric of a BTZ blackhole} read in Poincar\'{e} coordinates
      \begin{align}
        g_{\text{AdS}_3}=\frac{l^2}{y_A^2}(dy_A^2+dz_Ad\overline{z}_A),\quad g_{\text{BTZ}}=\frac{l^2}{y_B^2}(dy_B^2+dz_Bd\overline{z}_B).
      \end{align}
      We mainly consider the non-rotating BTZ blackhole ($J=0$) in the subsequent discussion, where $r_-=0$.
      For the spatial asymptotic regime $y_A\to 0$ and  $y_B\to 0$ (or equivalently $r\to\infty$), their boundary coordinates $z_A$, $z_B$ are
      \begin{align}
        \lim_{r\to\infty}z_A=e^{\frac{t_E}{l}-i\theta},\quad\lim_{r\to\infty}z_B=e^{2\pi T_H(l\theta+it_E)},
      \end{align}
      that have to be identified periodically as $z_A\sim e^{\frac{\beta_H}{l}}z_A$ and $z_B\sim e^{\frac{4\pi^2l}{\beta_H}}z_B$ due to the periodicity $t_E\sim t_E+\beta_H$, $\theta\sim\theta+2\pi$ of the original coordinates.
      If we introduce the new complex coordinates
      \begin{align}
        & w_A:=\frac{i}{2\pi}\log\lim_{r\to\infty}z_A=\frac{\theta}{2\pi}+i\frac{t_E}{2\pi l}, \\
        & w_B:=\frac{i}{2\pi}\log\lim_{r\to\infty}z_B=-T_Ht_E+ilT_H\theta,
      \end{align}
      they are the coordinates on tori as expected, with periodicity being $w_A\sim w_A+1\sim w_A+\tau_{\text{AdS}_3}$ and $w_B\sim w_B+1\sim w_B+\tau_{\text{BTZ}}$ where the moduli parameters are
      \begin{align}
        \tau_{\text{AdS}_3}:=\frac{\beta_H}{2\pi l}i,\quad\tau_{\text{BTZ}}:=\frac{2\pi l}{\beta_H}i. \label{eq: torus moduli for AdS3 and BTZ}
      \end{align}
      They are related through modular $S$-transformation $\tau_{\text{BTZ}}=-\frac{1}{\tau_{\text{AdS}_3}}$.

      \subsubsection*{Hawking-Page phase transition}
        As an initial demonstration of its validity, let us analyze how VTQFT replicates the 3d version of the Hawking-Page phase transition \cite{Hawking:1982dh}.
        The derivation from the classical action is elegantly reviewed in \cite{Kurita:2004}.
        \par From eq.\,\eqref{eq: overlap of moduli and conformal block}, the torus Virasoro character is the overlap between the moduli parameter basis $\ket{\tau}$ and the torus conformal block $\ket{\chi_p}=\ket{\mathcal{F}_{1,0}(p)}$
        \begin{align}
          \chi_p\qty(\tau)=\braket{\tau}{\chi_p}.
        \end{align}
        The Virasoro character for vacuum and for the momentum above threshold $\frac{c-1}{24}$ are
        \begin{align}
          \chi_0(\tau)=\frac{q^{-\frac{c-1}{24}}(1-q)}{\eta(\tau)},\qquad\chi_p(\tau):=\frac{q^{p^2}}{\eta(\tau)}\:(p\in\mathbb{R}_{\geq 0}), \label{eq: Virasoro character}
        \end{align}
        where $q:=e^{2\pi i\tau}$ and $\eta(\tau):=q^{\frac{1}{24}}\prod_{n=1}^{\infty}(1-q^n)$ is the Dedekind eta function.
        The vacuum character has the extra factor $(1-q)$ due to the existence of a null state in the first level of its highest weight representation.
        As we see in eq.\,\eqref{eq: torus moduli for AdS3 and BTZ}, the boundary moduli for thermal AdS$_3$ and non-rotating BTZ blackhole are each given by $\tau_{\text{AdS}_3}:=\frac{\beta_H}{2\pi l}i$, $\tau_{\text{BTZ}}:=-\frac{1}{\tau_{\text{AdS}_3}}$, and by the VTQFT path-integral rule on handlebodies \eqref{eq: path-integral on handlebodies}, their VTQFT partition functions are expressed as
        \begin{gather}
          Z_{\text{Vir}}\qty(\text{thermal AdS}_3):=Z_{\text{Vir}}\qty(ST^2;\tau_{\text{AdS}_3})=\chi_0\qty(\tau_{\text{AdS}_3}), \label{eq: thermal AdS partition function} \\
          Z_{\text{Vir}}\qty(\text{BTZ}):=Z_{\text{Vir}}\qty(ST^2;-\tfrac{1}{\tau_{\text{AdS}_3}})=\chi_0\qty(-\tfrac{1}{\tau_{\text{AdS}_3}}). \label{eq: BTZ partition function}
        \end{gather}
        It is worth noting that the BTZ partition function \eqref{eq: BTZ partition function} is the vacuum Virasoro character in the dual channel ($\tau^{\prime}=-\frac{1}{\tau}$), in exact agreement with the main proposal in Subsection 2.3 of Ref.\,\cite{Mertens:2022ujr}.
        The thermodynamical free energy for each geometry is
        \begin{align}
          F_{\text{thermal AdS}_3} & :=-\frac{1}{\beta_H}\log\big|Z_{\text{Vir}}\qty(\text{thermal AdS}_3)\big|^2=-\frac{1}{\beta_H}\log\big|\chi_0\qty(\tau_{\text{AdS}_3})\big|^2, \label{eq: free energy of AdS3} \\
          F_{\text{BTZ}} & :=-\frac{1}{\beta_H}\log\big|Z_{\text{Vir}}\qty(\text{BTZ})\big|^2=-\frac{1}{\beta_H}\log\big|\chi_0\qty(-\tfrac{1}{\tau_{\text{AdS}_3}})\big|^2. \label{eq: free energy of BTZ}
        \end{align}
        It is evident that they coincide at the self-dual point
        \begin{align}
          \tau_{\text{AdS}_3}=-\frac{1}{\tau_{\text{AdS}_3}} \iff \beta_H=2\pi l, \label{eq: self-dual point}
        \end{align}
        but indeed more is true.
        Consider the difference of the free energies \eqref{eq: free energy of AdS3} and \eqref{eq: free energy of BTZ}
        \begin{align}
          f\qty(\beta_H) & :=F_{\text{BTZ}}-F_{\text{thermal AdS}_3} \nonumber \\
          & =\frac{2}{\beta_H}\log|1-\tilde{q}|+\frac{2}{\beta_H}\log|1-q|-\frac{\pi^2l\qty(c-1)}{3\beta_H^2}+\frac{c-1}{12l}\qquad(\widetilde{q}:=e^{-\frac{2\pi i}{\tau}}) \nonumber \\
          & =\frac{1}{\beta_H}\log\qty(\frac{1-e^{-\frac{\beta_H}{l}}}{1-e^{-\frac{4\pi^2l}{\beta_H}}})^2-\frac{\pi^2l(c-1)}{3\beta_H^2}+\frac{c-1}{12l}. \label{eq: free energy difference}
        \end{align}
        This function is inherently vanishing at the self-dual point \eqref{eq: self-dual point}.
        A natural question is whether there exists any other vanishing point, i.e. a phase transition point, and it turns out that the answer is negative.
        Indeed, the equation $f(\beta_H)=0$ is equivalent to $g(x)=0$ ($x:=\frac{\beta_H}{l}$), where
        \begin{align}
          g(x):=\frac{1-e^{-x}}{1-e^{-\frac{4\pi^2}{x}}}-e^{(c-1)\big(\frac{\pi^2}{6x}-\frac{x}{24}\big)},
        \end{align}
        and $g(x)$ is monotonically increasing as
        \begin{align}
          g^{\prime}(x)=\frac{e^{-x}}{1-e^{-\frac{4\pi^2}{x}}}+\frac{4\pi^2e^{-\frac{4\pi^2}{x^2}}(1-e^{-x})}{x^2\big(1-e^{-\frac{4\pi^2}{x^2}}\big)^2}+(c-1)\qty(\frac{\pi^2}{6x^2}+\frac{1}{24})e^{(c-1)\big(\frac{\pi^2}{6x}-\frac{x}{24}\big)}>0,
        \end{align}
        for any $x>0$ provided $c>1$.
        Note that $c\leq 1$ is impossible for real $b$ (see eq.\,\eqref{eq: central charge}).
        Thus, we show that within VTQFT formulation,
        the phase transition occurs precisely once not only in the semiclassical regime $c=\frac{3l}{2G}+\mathcal{O}\qty(1)~\qty(G\to 0)$, but also when the central charge $c$ has an arbitrary \textit{finite} value above $1$.

        \subsubsection*{Bekenstein-Hawking entropy}
          We next address the Bekenstein-Hawking entropy for BTZ black hole.
          This begins with expanding the VTQFT partition function of the BTZ blackhole \eqref{eq: BTZ partition function} in terms of the $S$-dual AdS$_3$ partition functions \eqref{eq: thermal AdS partition function} with a Wilson line insertion $p$ using eqs.\,\eqref{eq: Cardy density}, \eqref{eq: Virasoro character} \footnote{Similar computation without assuming VTQFT can be found in Ref.\,\cite{Mertens:2022ujr}.}
          \begin{align}
            & \left|Z_{\text{Vir}}\qty(\text{BTZ})\right|^2=\int_{0}^{\infty}dpd\overline{p}\,S_{\mathbbm{1}p}\qty[\mathbbm{1}]S_{\mathbbm{1}\overline{p}}^*\qty[\mathbbm{1}]\chi_p\qty(\tau_{\text{AdS}_3})\chi_{\overline{p}}\qty(-\overline{\tau}_{\text{AdS}_3}) \nonumber \\
            & \qquad\quad=32\int_{0}^{\infty}dpd\overline{p}\,\sinh(2\pi bp)\sinh\qty(2\pi\frac{p}{b})\sinh(2\pi b\overline{p})\sinh\qty(2\pi\frac{\overline{p}}{b})\frac{e^{-\frac{\beta}{l}(p^2+\overline{p}^2)}}{|\eta(\tau_{\text{AdS}})|^2}.
          \end{align}
          In computing the thermal entropy, the main concern at present is the semiclassical limit $b\to +0$ ($G\to +0$), where $\sinh(2\pi bp)=2\pi bp+\mathcal{O}(b^2)$, $\sinh(2\pi \frac{p}{b})=\frac{1}{2}e^{\frac{2\pi p}{b}}+\mathcal{O}(1)$. Then the partition function reduces to leading order to
          \begin{align}
            \left|Z_{\text{Vir}}\qty(\text{BTZ})\right|^2 & \approx\frac{32\pi^2b^2}{|\eta(\tau_{\text{AdS}_3})|^2}e^{\frac{2\pi^2l}{\beta b^2}}\int_{0}^{\infty}dpd\overline{p}\,p\overline{p}e^{-\frac{\beta}{l}(p-\frac{\pi l}{\beta b})^2-\frac{\beta}{l}(\overline{p}-\frac{\pi l}{\beta b})^2} \nonumber \\
            & = \frac{32\pi^2b^2}{|\eta(\tau_{\text{AdS}_3})|^2}e^{\frac{2\pi^2l}{\beta b^2}}\frac{l^2}{4\beta^2}\bigg(\sqrt{\frac{\beta}{l}\pi}\cdot\frac{\pi l}{\beta b}+\mathcal{O}(1)\bigg)^2.
          \end{align}
          From the first to the second line, we employ the integral formula
          \begin{align}
            \int dx\,xe^{-a(x-b)^2}=-\frac{1}{2a}\Big(e^{-a(x-b)^2}+\sqrt{a\pi}b\erf\big(\sqrt{a}(b-x)\big)\Big)\qquad(a,b>0),
          \end{align}
          and the fact that $\erf(x)=\mathcal{O}(1)\,(x\to\infty)$, where $\erf(x)$ is the error function $\erf(x):=\frac{2}{\sqrt{\pi}}\int_{0}^{x}dt\,e^{-t^2}$. The thermal entropy is obtained as
          \begin{align}
            & S:=\bigg(1-\beta\pdv{\beta}\bigg)\log|Z_{\text{Vir}}(\text{BTZ})|^2=\frac{4\pi^2l}{\beta b^2}+\mathcal{O}(1).
          \end{align}
          Bearing in mind that $r_-=0$ in non-rotating case, we show by eq.\,\eqref{eq: Hawking temperature} that to leading order in $b\approx\sqrt{\frac{4G}{l}}$ (see eq.\,\eqref{eq: Liouville parameter and Chern-Simons level})
          \begin{align}
            S\approx 4\pi^2l\cdot\frac{r_+}{2\pi l^2}\cdot\frac{l}{4G}=\frac{2\pi r_+}{4G}.
          \end{align}
          Therefore, VTQFT certainly provides a correct semiclassical result.

  \section{Anyon condensation in modular tensor category} \label{sec: Anyon condensation in modular tensor category}
    Let us now reaffirm that our ultimate goal is to apply ``anyon condensation" to wormhole geometries in VTQFT.
    To date, anyon condensation is mathematically well-established for categories with suitable properties, at least fusion categories, and whether it is valid in physical systems with exotic symmetry categories is not at all self-evident.
    As will be discussed later, VTQFT possesses an atypical symmetry category, making our purpose formidably challenging.
    In this section, we confine our discussion to anyon condensation in well-behaved modular tensor categories as an opening gambit for developing key intuitions.
    Section \ref{sec: Anyon condensation in Virasoro TQFT} then presents the full details of that in VTQFT.
    \par Anyon condensation \cite{kong2014anyon} is a generalization of the gauging invertible symmetries to non-invertible cases.
    For invertible symmetries, suppose $G^{(p)}$ is a group representing $p$-form symmetry of a theory, a transformation acting on $p$-dimensional (possibly non-local) operators in the theory.
    The theory is gauged by selecting a subgroup of $G^{(p)}$ that does not carry an 't Hooft anomaly and promoting the background gauge field to a dynamical variable.
    More specifically, the process involves summing over inequivalent $(p+1)$-chains if $G^{(p)}$ is discrete, or performing a path-integral over all $(p+1)$-forms modulo gauge transformations if $G^{(p)}$ is continuous.
    For general aspects of gauging invertible symmetries, see Section 3 and 4 in the excellent review \cite{Bhardwaj:2023kri}.
    In the non-invertible case, the fusion rule of codimension $(p+1)$ operators acting as $p$-form symmetries becomes non-invertible, for which the symmetry is represented not by a group but more generally by a monoidal category.
    Hence, non-invertible symmetries are often referred to as categorical symmetries.
    Anyon condensation is therefore formulated as a generalization of gauging 1-form symmetry for non-invertible symmetries.

    \subsection{3d non-Abelian TQFT and Lagrangian algebra object} \label{subsec: 3d non-Abelian TQFT and Lagrangian algebra object}
      In this subsection, we recapitulate the detailed procedures of anyon condensation, which is also reviewed in Subsection 4.3 in Ref.\,\cite{Bhardwaj:2017xup} for 2d theories with fusion category symmetries, and in Subsection 3.1 - 3.2 in Ref.\,\cite{Kaidi:2021gbs}, Subsection 4.1 in Ref.\,\cite{Benini:2022hzx}, Subsection 7.1 in Ref.\,\cite{Roumpedakis:2022aik} for 3d TQFT with modular tensor category symmetries.
      Recent developments include, for example, Refs.\,\cite{Cordova:2023jip, Cordova:2024goh}.
      \par Let $\mathcal{C}$ be a modular tensor category (MTC) associated with a $3$d TQFT $\mathcal{T}$ and $\mathcal{I}\qty(\mathcal{C})$ be the label set of simple objects in $\mathcal{C}$.
      A modular tensor category $\mathcal{C}$ is a \textit{$\mathbb{C}$-linear semisimple finite ribbon category}, where the simple objects in $\mathcal{C}$ are Wilson lines labeled by $\mathcal{I}(\mathcal{C})$ representing the fundamental excitations in $3$d TQFT $\mathcal{T}$.
      Each condition defining a modular tensor category corresponds physically to the existence of a finite family of inequivalent Wilson lines $(L_i)_{i\in\mathcal{I}(\mathcal{C})}$, the well-defined concept of product (fusion) between them, and the braiding, the exchange of different particles.
      See Appendix \ref{append: From monoidal category to fusion category and modular tensor category} for mathematical foundations and the relation among various categorical notions.
      \par In considering anyon condensation, let us first introduce a specific object named algebra object in $\mathcal{C}$, which is defined for braided tensor categories in general.
      An \bit{algebra object} in $\mathcal{C}$ is a triplet $(\mathcal{A},m,\eta)$ consisting of an object $\mathcal{A}=\bigoplus_{i\in\mathcal{I}(\mathcal{C})}Z_i^{\mathcal{A}}L_i\in\Obj\mathcal{C}$ ($Z_i^{\mathcal{A}}\in\mathbb{Z}_{\geq 0}$), the \bit{product morphism} $m:\mathcal{A}\otimes\mathcal{A}\to\mathcal{A}$ and the \bit{unit morphism} $\eta:\mathbbm{1}\to\mathcal{A}$ satisfying the \bit{associativity} and the \bit{unit axiom}
      \begin{gather}
        m\circ(m\otimes\identity_{\mathcal{A}})=m\circ(\identity_{\mathcal{A}}\otimes m)\circ\alpha_{\mathcal{A},\mathcal{A},\mathcal{A}}, \\
        m\circ(\eta\otimes\identity_{\mathcal{A}})=\identity_{\mathcal{A}}=m\circ(\identity_{\mathcal{A}}\otimes\eta),
      \end{gather}
      where $\alpha_{\mathcal{A},\mathcal{A},\mathcal{A}}:(\mathcal{A}\otimes\mathcal{A})\otimes\mathcal{A}\overset{\sim}{\to}\mathcal{A}\otimes(\mathcal{A}\otimes\mathcal{A})$ is the \bit{associator} (\textit{associativity isomorphism}) in $\mathcal{C}$.
      They are described pictorially as
      \begin{align}
        \begin{tikzpicture}[scale = 0.5, baseline = {([yshift=-.5ex]current bounding box.center)}]
          \draw[thick, red] (0, 1) to (0, 2);
          \draw[thick, red] (-2, -1) to (0, 1);
          \draw[thick, red] (2, -1) to (0, 1);
          \draw[thick, red] (0, -1) to (-1, 0);
          \node[above, scale = 0.8] at (0, 2) {$\mathcal{A}$};
          \node[below, scale = 0.8] at (-2, -1) {$\mathcal{A}$};
          \node[below, scale = 0.8] at (0, -1) {$\mathcal{A}$};
          \node[below, scale = 0.8] at (2, -1) {$\mathcal{A}$};
          \node[above left, scale = 0.8] at (-0.5, 0.5) {$\mathcal{A}$};
          \fill[red] (-1, 0) circle[radius = 0.1];
          \node[below, scale = 0.6] at (-1, -0.1) {$m$};
          \fill[red] (0, 1) circle[radius = 0.1];
          \node[below, scale = 0.6] at (0, 0.9) {$m$};
        \end{tikzpicture}
        =
        \begin{tikzpicture}[scale = 0.5, baseline = {([yshift=-.5ex]current bounding box.center)}]
          \draw[thick, red] (0, 1) to (0, 2);
          \draw[thick, red] (-2, -1) to (0, 1);
          \draw[thick, red] (2, -1) to (0, 1);
          \draw[thick, red] (0, -1) to (1, 0);
          \node[above, scale = 0.8] at (0, 2) {$\mathcal{A}$};
          \node[below, scale = 0.8] at (-2, -1) {$\mathcal{A}$};
          \node[below, scale = 0.8] at (0, -1) {$\mathcal{A}$};
          \node[below, scale = 0.8] at (2, -1) {$\mathcal{A}$};
          \node[above right, scale = 0.8] at (0.5, 0.5) {$\mathcal{A}$};
          \fill[red] (1, 0) circle[radius = 0.1];
          \node[below, scale = 0.6] at (1, -0.1) {$m$};
          \fill[red] (0, 1) circle[radius = 0.1];
          \node[below, scale = 0.6] at (0, 0.9) {$m$};
        \end{tikzpicture}
        ,\qquad
        \begin{tikzpicture}[scale = 0.5, baseline = {([yshift=-.5ex]current bounding box.center)}]
          \draw[thick, red] (0, 0.5) to (0, 2);
          \draw[thick, red] (-0.75, -0.25) to (0, 0.5);
          \draw[dashed] (-1.5, -1) to (-0.75, -0.25);
          \draw[thick, red] (1.5, -1) to (0, 0.5);
          \node[above, scale = 0.8] at (0, 2) {$\mathcal{A}$};
          \node[below, scale = 0.8] at (-1.5, -1) {$\mathbbm{1}$};
          \node[below, scale = 0.8] at (1.5, -1) {$\mathcal{A}$};
          \fill[red!30!white] (-0.75, -0.25) circle[radius = 0.1];
          \node[below, scale = 0.6] at (-0.75, -0.35) {$\eta$};
          \fill[red] (0, 0.5) circle[radius = 0.1];
          \node[below, scale = 0.6] at (0, 0.4) {$m$};
        \end{tikzpicture}
        =\;\;
        \begin{tikzpicture}[scale = 0.5, baseline = {([yshift=-.5ex]current bounding box.center)}]
          \draw[thick, red] (0, -1) to (0, 2);
          \node[above, scale = 0.8] at (0, 2) {$\mathcal{A}$};
          \node[below, scale = 0.8] at (0, -1) {$\mathcal{A}$};
        \end{tikzpicture}
        \;\;=
        \begin{tikzpicture}[scale = 0.5, baseline = {([yshift=-.5ex]current bounding box.center)}]
          \draw[thick, red] (0, 0.5) to (0, 2);
          \draw[thick, red] (-1.5, -1) to (0, 0.5);
          \draw[thick, red] (0.75, -0.25) to (0, 0.5);
          \draw[dashed] (1.5, -1) to (0.75, -0.25);
          \node[above, scale = 0.8] at (0, 2) {$\mathcal{A}$};
          \node[below, scale = 0.8] at (-1.5, -1) {$\mathcal{A}$};
          \node[below, scale = 0.8] at (1.5, -1) {$\mathbbm{1}$};
          \fill[red!30!white] (0.75, -0.25) circle[radius = 0.1];
          \node[below, scale = 0.6] at (0.75, -0.35) {$\eta$};
          \fill[red] (0, 0.5) circle[radius = 0.1];
          \node[below, scale = 0.6] at (0, 0.4) {$m$};
        \end{tikzpicture}
        . \label{eq: the associativity and the unit axiom (pic)}
      \end{align}
      In an analogous fashion, a \bit{co-algebra object} in $\mathcal{C}$ is a triplet $(\mathcal{A},\Delta,\iota)$ where $\mathcal{A}\in\Obj\mathcal{C}$ is an object, $\Delta:\mathcal{A}\to\mathcal{A}\otimes\mathcal{A}$ is the \bit{co-product morphism} and $\iota:\mathcal{A}\to\mathbbm{1}$ is the \bit{co-unit morphism}, complying the \bit{co-associativity} and the \bit{co-unit axiom}
      \begin{gather}
        \alpha_{\mathcal{A},\mathcal{A},\mathcal{A}}\circ(\Delta\otimes\identity_{\mathcal{A}})\circ\Delta=(\identity_{\mathcal{A}}\otimes\Delta)\circ\Delta, \\
        (\iota\otimes\identity_{\mathcal{A}})\circ\Delta=\identity_{\mathcal{A}}=(\identity_{\mathcal{A}}\otimes\iota)\circ\Delta,
      \end{gather}
      expressed as the same diagram as eq.\,\eqref{eq: the associativity and the unit axiom (pic)} but with their orientation reversed vertically
      \begin{align}
        \begin{tikzpicture}[scale = 0.5, baseline = {([yshift=-.5ex]current bounding box.center)}]
          \draw[thick, red] (0, -1) to (0, -2);
          \draw[thick, red] (-2, 1) to (0, -1);
          \draw[thick, red] (2, 1) to (0, -1);
          \draw[thick, red] (0, 1) to (-1, 0);
          \node[below, scale = 0.8] at (0, -2) {$\mathcal{A}$};
          \node[above, scale = 0.8] at (-2, 1) {$\mathcal{A}$};
          \node[above, scale = 0.8] at (0, 1) {$\mathcal{A}$};
          \node[above, scale = 0.8] at (2, 1) {$\mathcal{A}$};
          \node[below left, scale = 0.8] at (-0.5, -0.5) {$\mathcal{A}$};
          \fill[red] (-1, 0) circle[radius = 0.1];
          \node[above, scale = 0.6] at (-1, 0.1) {$\Delta$};
          \fill[red] (0, -1) circle[radius = 0.1];
          \node[above, scale = 0.6] at (0, -0.9) {$\Delta$};
        \end{tikzpicture}
        =
        \begin{tikzpicture}[scale = 0.5, baseline = {([yshift=-.5ex]current bounding box.center)}]
          \draw[thick, red] (0, -1) to (0, -2);
          \draw[thick, red] (-2, 1) to (0, -1);
          \draw[thick, red] (2, 1) to (0, -1);
          \draw[thick, red] (0, 1) to (1, 0);
          \node[below, scale = 0.8] at (0, -2) {$\mathcal{A}$};
          \node[above, scale = 0.8] at (-2, 1) {$\mathcal{A}$};
          \node[above, scale = 0.8] at (0, 1) {$\mathcal{A}$};
          \node[above, scale = 0.8] at (2, 1) {$\mathcal{A}$};
          \node[below left, scale = 0.8] at (-0.5, -0.5) {$\mathcal{A}$};
          \fill[red] (1, 0) circle[radius = 0.1];
          \node[above, scale = 0.6] at (1, 0.1) {$\Delta$};
          \fill[red] (0, -1) circle[radius = 0.1];
          \node[above, scale = 0.6] at (0, -0.9) {$\Delta$};
        \end{tikzpicture}
        ,\qquad
        \begin{tikzpicture}[scale = 0.5, baseline = {([yshift=-.5ex]current bounding box.center)}]
          \draw[thick, red] (0, -0.5) to (0, -2);
          \draw[thick, red] (-0.75, 0.25) to (0, -0.5);
          \draw[dashed] (-1.5, 1) to (-0.75, 0.25);
          \draw[thick, red] (1.5, 1) to (0, -0.5);
          \node[below, scale = 0.8] at (0, -2) {$\mathcal{A}$};
          \node[above, scale = 0.8] at (-1.5, 1) {$\mathbbm{1}$};
          \node[above, scale = 0.8] at (1.5, 1) {$\mathcal{A}$};
          \fill[red!30!white] (-0.75, 0.25) circle[radius = 0.1];
          \node[above, scale = 0.6] at (-0.75, 0.35) {$\iota$};
          \fill[red] (0, -0.5) circle[radius = 0.1];
          \node[above, scale = 0.6] at (0, -0.4) {$\Delta$};
        \end{tikzpicture}
        =\;\;
        \begin{tikzpicture}[scale = 0.5, baseline = {([yshift=-.5ex]current bounding box.center)}]
          \draw[thick, red] (0, 1) to (0, -2);
          \node[below, scale = 0.8] at (0, -2) {$\mathcal{A}$};
          \node[above, scale = 0.8] at (0, 1) {$\mathcal{A}$};
        \end{tikzpicture}
        \;\;=
        \begin{tikzpicture}[scale = 0.5, baseline = {([yshift=-.5ex]current bounding box.center)}]
          \draw[thick, red] (0, -0.5) to (0, -2);
          \draw[thick, red] (-1.5, 1) to (0, -0.5);
          \draw[thick, red] (0.75, 0.25) to (0, -0.5);
          \draw[dashed] (1.5, 1) to (0.75, 0.25);
          \node[below, scale = 0.8] at (0, -2) {$\mathcal{A}$};
          \node[above, scale = 0.8] at (-1.5, 1) {$\mathcal{A}$};
          \node[above, scale = 0.8] at (1.5, 1) {$\mathbbm{1}$};
          \fill[red!30!white] (0.75, 0.25) circle[radius = 0.1];
          \node[above, scale = 0.6] at (0.75, 0.35) {$\iota$};
          \fill[red] (0, -0.5) circle[radius = 0.1];
          \node[above, scale = 0.6] at (0, -0.4) {$\Delta$};
        \end{tikzpicture}
        .
      \end{align}
      If $\mathcal{A}$ is both an algebra object and a co-algebra object in $\mathcal{C}$, it must further fulfill the following conditions to be condensable.
      These conditions are imposed to ensure that the result of anyon condensation is independent of the choice of the triangulation of the spacetime manifold, and the theory still retain the unit object $\mathbbm{1}$ even after condensation.
      We will shortly see how triangulation is involved in the process of anyon condensation.
      The first condition is the \bit{separability}
      \begin{align}
        m\circ\Delta=c\cdot\identity_{\mathcal{A}}\;(\exists c\in\mathbb{Z}_{\geq 0}),
      \end{align}
      which states that fusing after branching amounts to doing nothing up to a constant factor:
      \begin{align}
        \begin{tikzpicture}[scale = 0.5, baseline = {([yshift=-.5ex]current bounding box.center)}]
          \draw[thick, red] (0, -1.25) to (0, -2);
          \draw[thick, red] (0, -0.5) circle[radius = 0.75];
          \draw[thick, red] (0, 0.25) to (0, 1);
          \node[below, scale = 0.8] at (0, -2) {$\mathcal{A}$};
          \node[above, scale = 0.8] at (0, 1) {$\mathcal{A}$};
          \node[left, scale = 0.8] at (-0.75, -0.5) {$\mathcal{A}$};
          \node[right, scale = 0.8] at (0.75, -0.5) {$\mathcal{A}$};
          \fill[red] (0, -1.25) circle[radius = 0.1];
          \node[above, scale = 0.6] at (0, -1.15) {$\Delta$};
          \fill[red] (0, 0.25) circle[radius = 0.1];
          \node[below, scale = 0.6] at (0, 0.15) {$m$};
        \end{tikzpicture}
        \;\propto\;\;
        \begin{tikzpicture}[scale = 0.5, baseline = {([yshift=-.5ex]current bounding box.center)}]
          \draw[thick, red] (0, 1) to (0, -2);
          \node[below, scale = 0.8] at (0, -2) {$\mathcal{A}$};
          \node[above, scale = 0.8] at (0, 1) {$\mathcal{A}$};
        \end{tikzpicture}
        .
      \end{align}
      A separable algebra object is \bit{connected} (or \bit{haploid}) if
      \begin{align}
        \dim\Hom_{\mathcal{C}}(\mathbbm{1},\mathcal{A})=1,
      \end{align}
      that is, $\mathcal{A}$ has only single identity object $\mathbbm{1}$ in it.
      More complex and more paramount is the \bit{Frobenius condition}
      \begin{align}
        (\identity_{\mathcal{A}}\otimes m)\circ(\Delta\otimes\identity_{\mathcal{A}})=\Delta\circ m=(m\otimes\identity_{\mathcal{A}})\circ(\identity_{\mathcal{A}}\otimes\Delta),
      \end{align}
      requiring the consistency under crossing:
      \begin{align}
        \begin{tikzpicture}[scale = 0.5, baseline = {([yshift=-.5ex]current bounding box.center)}]
          \draw[thick, red] (-1.5, -1.5) to (1.5, 1.5);
          \draw[thick, red] (-0.5, -0.5) to (-1.5, 1.5);
          \draw[thick, red] (0.5, 0.5) to (1.5, -1.5);
          \node[below, scale = 0.8] at (-1.5, -1.5) {$\mathcal{A}$};
          \node[below, scale = 0.8] at (1.5, -1.5) {$\mathcal{A}$};
          \node[above, scale = 0.8] at (-1.5, 1.5) {$\mathcal{A}$};
          \node[above, scale = 0.8] at (1.5, 1.5) {$\mathcal{A}$};
          \fill[red] (-0.5, -0.5) circle[radius = 0.1];
          \node[above, scale = 0.6] at (-0.45, -0.3) {$\Delta$};
          \fill[red] (0.5, 0.5) circle[radius = 0.1];
          \node[below, scale = 0.6] at (0.45, 0.3) {$m$};
        \end{tikzpicture}
        =
        \begin{tikzpicture}[scale = 0.5, baseline = {([yshift=-.5ex]current bounding box.center)}]
          \draw[thick, red] (-1, -1.5) to (0, -0.5);
          \draw[thick, red] (1, -1.5) to (0, -0.5);
          \draw[thick, red] (0, -0.5) to (0, 0.5);
          \draw[thick, red] (-1, 1.5) to (0, 0.5);
          \draw[thick, red] (1, 1.5) to (0, 0.5);
          \node[below, scale = 0.8] at (-1, -1.5) {$\mathcal{A}$};
          \node[below, scale = 0.8] at (1, -1.5) {$\mathcal{A}$};
          \node[above, scale = 0.8] at (-1, 1.5) {$\mathcal{A}$};
          \node[above, scale = 0.8] at (1, 1.5) {$\mathcal{A}$};
          \node[left, scale = 0.8] at (0, 0) {$\mathcal{A}$};
          \fill[red] (0, -0.5) circle[radius = 0.1];
          \node[below, scale = 0.6] at (0, -0.6) {$m$};
          \fill[red] (0, 0.5) circle[radius = 0.1];
          \node[above, scale = 0.6] at (0, 0.6) {$\Delta$};
        \end{tikzpicture}
        =
        \begin{tikzpicture}[scale = 0.5, baseline = {([yshift=-.5ex]current bounding box.center)}]
          \draw[thick, red] (1.5, -1.5) to (-1.5, 1.5);
          \draw[thick, red] (0.5, -0.5) to (1.5, 1.5);
          \draw[thick, red] (-0.5, 0.5) to (-1.5, -1.5);
          \node[below, scale = 0.8] at (-1.5, -1.5) {$\mathcal{A}$};
          \node[below, scale = 0.8] at (1.5, -1.5) {$\mathcal{A}$};
          \node[above, scale = 0.8] at (-1.5, 1.5) {$\mathcal{A}$};
          \node[above, scale = 0.8] at (1.5, 1.5) {$\mathcal{A}$};
          \fill[red] (0.5, -0.5) circle[radius = 0.1];
          \node[above, scale = 0.6] at (0.45, -0.3) {$\Delta$};
          \fill[red] (-0.5, 0.5) circle[radius = 0.1];
          \node[below, scale = 0.6] at (-0.45, 0.3) {$m$};
        \end{tikzpicture}
        .
      \end{align}
      An algebra object $\mathcal{A}$ is \bit{commutative} if
      \begin{align}
        m=m\circ c_{\mathcal{A},\mathcal{A}}, \label{eq: commutative condition}
      \end{align}
      where $c_{\mathcal{A},\mathcal{A}}:\mathcal{A}\otimes\mathcal{A}\overset{\sim}{\to}\mathcal{A}\otimes\mathcal{A}$ is the \bit{braiding isomorphism} between two $\mathcal{A}$'s in $\mathcal{C}$.
      Graphically, it guarantees that $\mathcal{A}$ acquires no non-trivial phase after braiding
      \begin{align}
        \begin{tikzpicture}[scale = 0.5, baseline = {([yshift=-.5ex]current bounding box.center)}]
          \draw[thick, red] (0, 0.5) to (0, 2);
          \draw[thick, red] (-1.5, -1) .. controls (0.7, 0) and (0.7, 0.4) .. (0, 0.5);
          \fill[white] (0, -0.19) circle[radius = 0.125];
          \draw[thick, red] (1.5, -1) .. controls (-0.7, 0) and (-0.7, 0.4) .. (0, 0.5);
          \node[above, scale = 0.8] at (0, 2) {$\mathcal{A}$};
          \node[below, scale = 0.8] at (-1.5, -1) {$\mathcal{A}$};
          \node[below, scale = 0.8] at (1.5, -1) {$\mathcal{A}$};
          \fill[red] (0, 0.5) circle[radius = 0.1];
          \node[below, scale = 0.6] at (0, 0.45) {$m$};
        \end{tikzpicture}
        =
        \begin{tikzpicture}[scale = 0.5, baseline = {([yshift=-.5ex]current bounding box.center)}]
          \draw[thick, red] (0, 0.5) to (0, 2);
          \draw[thick, red] (-1.5, -1) to (0, 0.5);
          \draw[thick, red] (1.5, -1) to (0, 0.5);
          \node[above, scale = 0.8] at (0, 2) {$\mathcal{A}$};
          \node[below, scale = 0.8] at (-1.5, -1) {$\mathcal{A}$};
          \node[below, scale = 0.8] at (1.5, -1) {$\mathcal{A}$};
          \fill[red] (0, 0.5) circle[radius = 0.1];
          \node[below, scale = 0.6] at (0, 0.4) {$m$};
        \end{tikzpicture}
        .
      \end{align}
      Combining the above conditions, a \bit{condensable anyon} in 3d TQFT is \textit{a connected commutative separable Frobenius algebra object} $\mathcal{A}=\bigoplus_{i\in\mathcal{I}(\mathcal{C})}Z_i^{\mathcal{A}}L_i\in\Obj\mathcal{C}$.\footnote{A condensable anyon in 2d, on the other hand, is a connected \textit{symmetric} separable Frobenius algebra object \cite{Bhardwaj:2017xup}.}
      A condensable anyon is \bit{Lagrangian} if its quantum dimension (Frobenius-Perron dimension\footnote{Frobenius-Perron dimension coincides with quantum dimension in general spherical fusion categories.}) $\dim\mathcal{A}$ squares to the total quantum dimension $\mathcal{D}:=\sqrt{\sum_{i\in\mathcal{I}(\mathcal{C})}(\dim L_i)^2}$ of the MTC $\mathcal{C}$.
      This is equivalent to asserting that $\mathcal{A}$ is the ``maximal" object that can be condensed in the symmetry category $\mathcal{C}$, in a sense that adding any other object in $\mathcal{C}$ to $\mathcal{A}$ breaks at least one of its defining properties (typically by yielding a non-trivial braiding phase).
      In general, there are multiple inequivalent choices of Lagrangian condensable anyons in $\mathcal{C}$.
      \par The anyon condensation in question is achieved by placing the condensable anyons on a fine mesh over the spacetime manifold. For a general $d$-manifold ($d>2$) $M$, a \bit{fine mesh} refers to the graph that is dual to the $1$-skeleton formed by the vertices and edges of the $d$-simplices obtained from triangulating $M$.
      The above conditions guarantee that the condensable anyon on a fine mesh remains equivalent under the action of \textit{Pachner 1-4} and/or \textit{2-3 moves}.
      All the triangulations are related to each other by a certain number of consecutive basic Pachner moves, so they indeed assure that the result of anyon condensation on a given spacetime is independent of the choice of triangulations.
      Specific examples, such as the case of handlebodies, are discussed in Subsection 3.2 of Ref.\,\cite{Benini:2022hzx}, and our main focus---the factorization---is demonstrated based on that within the context of Chern-Simons theory.
      Just as gauging a Lagrangian subgroup of invertible symmetries makes the partition function completely independent of the bulk geometry, condensing a Lagrangian condensable anyon produces the partition function solely bulk geometry.
      \begin{figure}[t]
        \centering
        \begin{minipage}[b]{0.49\columnwidth}
            \centering
            \tikzset{every picture/.style={line width=0.75pt}} %set default line width to 0.75pt
            \begin{tikzpicture}[x=0.75pt,y=0.75pt,yscale=-1,xscale=1]
              %uncomment if require: \path (0,300); %set diagram left start at 0, and has height of 300
              %Curve Lines [id:da12895505533240426]
              \draw[color = green!40!black, thin] (99,120) .. controls (125,119) and (162,105) .. (207,133);
              %Straight Lines [id:da7655080975619653]
              \draw[dash pattern={on 4.5pt off 4.5pt}] (128.98,151) -- (249.98,151.01);
              %Shape: Parallelogram [id:dp022902354394234492]
              \draw  [color = white][fill = red, fill opacity=0.15 ] (189.97,99) -- (189.97,190) -- (249.99,151) -- (249.99,60.01) -- cycle;
              %Straight Lines [id:da9216676379984665]
              \draw (189.97,190) -- (310.97,190);
              %Straight Lines [id:da5887415195216888]
              \draw (249.99,60.01) -- (371,60.01);
              %Straight Lines [id:da1562599462901203]
              \draw[dash pattern={on 4.5pt off 4.5pt}] (249.98,151.01) -- (370.98,151.01) ;
              %Curve Lines [id:da255076435282537]
              \draw [color = red, thin] (239.77,123) .. controls (279.77,93) and (326,127) .. (360,120);
              \fill[red] (239.77,123) circle[radius = 2];
              %Straight Lines [id:da77348936154095]
              \draw[dash pattern={on 4.5pt off 4.5pt}] (249.99,60.01) -- (249.98,151.01);
              %Straight Lines [id:da3056597639949409]
              \draw (249.99,60.01) -- (189.99,99);
              %Straight Lines [id:da13023815734331923]
              \draw (189.99,99) -- (189.97,190);
              %Straight Lines [id:da6802233948596155]
              \draw[dash pattern={on 4.5pt off 4.5pt}] (189.97,190) -- (249.98,151.01);
              %Curve Lines [id:da4430775090194363]
              \draw [color = blue, thin] (199.97,170.67) .. controls (249.97,129.67) and (197.97,107.67) .. (243.88,78.43);
              %Straight Lines [id:da20374047093024583]
              \draw (189.99,99) -- (310.99,99);
              %Curve Lines [id:da7510254113297262]
              \draw[color = green!40!black, thin] (207,133) .. controls (242,161) and (283,91) .. (328,147);
              \fill[green!40!black] (207,133) circle[radius = 2];
              %Straight Lines [id:da41193149599606116]
              \draw (68.97,189.99) -- (189.97,190);
              %Straight Lines [id:da09248951234926572]
              \draw (68.99,98.99) -- (189.99,99);
              %Straight Lines [id:da6158813772393403]
              \draw (128.99,60) -- (249.99,60.01);
              \draw[color = blue, thin] (243.88,78.43) .. controls (292.78,74) and (312.78,93) .. (354.78,73);
              \fill[blue] (243.88,78.43) circle[radius = 2];
              % Text Node
              \draw (257,161) node [anchor=north west][inner sep=0.75pt]  [font=\normalsize] [align=left] {original phase};
              % Text Node
              \draw (56,160) node [anchor=north west][inner sep=0.75pt]   [align=left] {condensed phase};
            \end{tikzpicture}
            \subcaption{partial condensation and topological interface}
            \label{fig: a}
        \end{minipage}
        \begin{minipage}[b]{0.49\columnwidth}
            \centering
            \tikzset{every picture/.style={line width=0.75pt}} %set default line width to 0.75pt
            \begin{tikzpicture}[x=0.75pt,y=0.75pt,yscale=-1,xscale=1]
              %Straight Lines [id:da7655080975619653]
              %Shape: Parallelogram [id:dp022902354394234492]
              \draw  [color = white][fill = red, fill opacity=0.15 ] (189.97,99) -- (189.97,190) -- (249.99,151) -- (249.99,60.01) -- cycle;
              %Straight Lines [id:da9216676379984665]
              \draw (189.97,190) -- (310.97,190);
              %Straight Lines [id:da5887415195216888]
              \draw (249.99,60.01) -- (371,60.01);
              %Straight Lines [id:da1562599462901203]
              \draw[dash pattern={on 4.5pt off 4.5pt}] (249.98,151.01) -- (370.98,151.01) ;
              %Curve Lines [id:da255076435282537]
              \draw [color = red, thin] (239.77,123) .. controls (279.77,93) and (326,127) .. (360,120);
              \fill[red] (239.77,123) circle[radius = 2];
              %Straight Lines [id:da77348936154095]
              \draw[dash pattern={on 4.5pt off 4.5pt}] (249.99,60.01) -- (249.98,151.01);
              %Straight Lines [id:da3056597639949409]
              \draw (249.99,60.01) -- (189.99,99);
              %Straight Lines [id:da13023815734331923]
              \draw (189.99,99) -- (189.97,190);
              %Straight Lines [id:da6802233948596155]
              \draw[dash pattern={on 4.5pt off 4.5pt}] (189.97,190) -- (249.98,151.01);
              %Curve Lines [id:da4430775090194363]
              \draw[color = blue, thin] (199.97,170.67) .. controls (249.97,129.67) and (197.97,107.67) .. (243.88,78.43);
              %Straight Lines [id:da20374047093024583]
              \draw (189.99,99) -- (310.99,99);
              %Straight Lines [id:da6158813772393403]
              \draw[color = blue, thin] (243.88,78.43) .. controls (292.78,74) and (312.78,93) .. (354.78,73);
              \fill[blue] (243.88,78.43) circle[radius = 2];
              % Text Node
              \draw (257,161) node [anchor=north west][inner sep=0.75pt]  [font=\normalsize] [align=left] {original phase};
            \end{tikzpicture}
            \subcaption{full condensation and topological boundary}
            \label{fig: b}
        \end{minipage}
        \caption{The red wall is the topological interface induced from a condensable anyon $\mathcal{A}$.
        \textbf{(a)} The red line is contained in $\mathcal{A}$ and therefore ``condensed", meaning that it can end on the topological interface, leaving a point-like operator on the boundary theory.
        The blue line is outside $\mathcal{A}$ and braids nontrivially with $\mathcal{A}$ so that it is ``confined", meaning it is restricted on the topological boundary.
        Only those colored in green can braid trivially with $\mathcal{A}$ and pass through the topological interface.
        \textbf{(b)} When the condensable anyon is Lagrangian, the condensed phase (gauged theory) is trivial: no lines can traverse the interface, so the surface is aptly termed the ``topological boundary" rather than the ``topological interface".}
      \end{figure}
      \par Let us remark that the factorization is quite natural for general $3$d TQFT $\mathcal{T}$, given that condensable anyons admit an equivalent representation as what is called the \textit{topological interface}.
      This correspondence becomes evident upon continuously deforming, or, say, ``fattening", the condensable anyon lines $\mathcal{A}$ to form interfaces as elucidated in Subsection 3.1 in Ref.\,\cite{Kaidi:2021gbs}.
      The topological interface stemming from $\mathcal{A}$ finds its mathematical characterization in the right $\mathcal{A}$-module category $\mathcal{M}_{\mathcal{A}}$, and it bridges the original theory $\mathcal{C}$ on the right and the condensed (gauged) theory on the left (see Fig.\,\ref{fig: a}).
      In the presence of a topological interface $\mathcal{M}_{\mathcal{A}}$, any anyon lines contained in the algebra object $\mathcal{A}$ can terminate on it, leaving point-like operators, and never seep out to the other side (red line).
      Anyon lines outside $\mathcal{A}$ braid either trivially with all the anyon lines in $\mathcal{A}$ or non-trivially with at least one of them.
      The former lines pass through the topological interface (green line), while the latter are \textit{confined} on it, meaning they create extra excitations on the interface once they touch it (blue line).
      If $\mathcal{A}$ is Lagrangian, all the anyon lines outside $\mathcal{A}$ braid non-trivially with at least one member of $\mathcal{A}$, so the resulting condensed phase is the trivial one (See Fig.\,\ref{fig: b}).
      In this case, the topological interface is referred to as the \textit{topological boundary}.
      Inserting a Lagrangian condensable anyon $\mathcal{A}$ on the fine mesh of the wormhole $\Sigma\times[0,1]$ is equivalent to partitioning the geometry into two disconnected parts $\text{\textbf{Slab}}_1$, $\text{\textbf{Slab}}_2$, with the trivial phase intervening them (see Fig.\,\ref{fig: splitting by trivial phase}).
      This qualitative speculation concludes that the wormhole partition function after anyon condensation is naturally factorized into the product of partition functions on the two slabs
      \begin{align}
        Z_{\mathcal{T}}(\Sigma\times[0,1];\mathcal{A})=Z_{\mathcal{T}}(\text{\textbf{Slab}}_1)Z_{\mathcal{T}}(\text{\textbf{Slab}}_2)
      \end{align}
      For each slabs on the two side we can think of the black boundary as a physical boundary and the red boundary as a topological boundary.
      Then the sandwich construction of symmetry TFT \cite{Gaiotto:2020iye, Kaidi:2022cpf} allows us to identify the $3$d TQFT $\mathcal{T}$ on the slab with 2d QFT $\mathcal{B}$ on the physical boundary constrained by the topological boundary, for which it holds
      \begin{align}
        Z_{\mathcal{T}}(\Sigma\times[0,1];\mathcal{A})=Z_{\mathcal{B}}(\Sigma)Z_{\mathcal{B}}(\Sigma).
      \end{align}
      Typical examples include the case where $\mathcal{T}$ is toric code and $\mathcal{B}$ is $2$d untwisted $\mathcal{Z}_2$ Dijkgraaf-Witten theory, and where $\mathcal{T}$ is $SU(2)_k$ Chern-Simons and $\mathcal{B}$ is $SU(2)$ level $k$ WZW CFT.
      \begin{figure}[t]
        \centering
        \tikzset{every picture/.style={line width=0.75pt}} %set default line width to 0.75pt
        \begin{tikzpicture}[x=0.75pt,y=0.75pt,yscale=-1,xscale=1]
          \draw   (241,39) .. controls (254.81,39) and (266,63.62) .. (266,94) .. controls (266,124.38) and (254.81,149) .. (241,149) .. controls (227.19,149) and (216,124.38) .. (216,94) .. controls (216,63.62) and (227.19,39) .. (241,39) -- cycle ;
          \draw  [draw opacity=0] (243.09,121.65) .. controls (238.21,121.65) and (234.25,109.34) .. (234.25,94.15) .. controls (234.25,78.96) and (238.21,66.65) .. (243.09,66.65) -- (243.09,94.15) -- cycle ; \draw   (243.09,121.65) .. controls (238.21,121.65) and (234.25,109.34) .. (234.25,94.15) .. controls (234.25,78.96) and (238.21,66.65) .. (243.09,66.65) ;
          \draw  [draw opacity=0] (240.93,68.67) .. controls (240.96,68.66) and (240.98,68.66) .. (241,68.66) .. controls (244.43,68.66) and (247.22,80.01) .. (247.22,94) .. controls (247.22,107.99) and (244.43,119.34) .. (241,119.34) .. controls (240.71,119.34) and (240.42,119.25) .. (240.14,119.1) -- (241,94) -- cycle ; \draw   (240.93,68.67) .. controls (240.96,68.66) and (240.98,68.66) .. (241,68.66) .. controls (244.43,68.66) and (247.22,80.01) .. (247.22,94) .. controls (247.22,107.99) and (244.43,119.34) .. (241,119.34) .. controls (240.71,119.34) and (240.42,119.25) .. (240.14,119.1) ;
          \draw   (429,39) .. controls (442.81,39) and (454,63.62) .. (454,94) .. controls (454,124.38) and (442.81,149) .. (429,149) .. controls (415.19,149) and (404,124.38) .. (404,94) .. controls (404,63.62) and (415.19,39) .. (429,39) -- cycle ;
          \draw  [draw opacity=0] (431.09,121.65) .. controls (426.21,121.65) and (422.25,109.34) .. (422.25,94.15) .. controls (422.25,78.96) and (426.21,66.65) .. (431.09,66.65) -- (431.09,94.15) -- cycle ; \draw   (431.09,121.65) .. controls (426.21,121.65) and (422.25,109.34) .. (422.25,94.15) .. controls (422.25,78.96) and (426.21,66.65) .. (431.09,66.65) ;
          \draw  [draw opacity=0] (428.93,68.67) .. controls (428.96,68.66) and (428.98,68.66) .. (429,68.66) .. controls (432.43,68.66) and (435.22,80.01) .. (435.22,94) .. controls (435.22,107.99) and (432.43,119.34) .. (429,119.34) .. controls (428.71,119.34) and (428.42,119.25) .. (428.14,119.1) -- (429,94) -- cycle ; \draw   (428.93,68.67) .. controls (428.96,68.66) and (428.98,68.66) .. (429,68.66) .. controls (432.43,68.66) and (435.22,80.01) .. (435.22,94) .. controls (435.22,107.99) and (432.43,119.34) .. (429,119.34) .. controls (428.71,119.34) and (428.42,119.25) .. (428.14,119.1) ;
          \draw  [color={rgb, 255:red, 255; green, 0; blue, 0 }  ,draw opacity=1 ] (303.97,43) .. controls (316.1,43) and (325.93,65.16) .. (325.93,92.5) .. controls (325.93,119.84) and (316.1,142) .. (303.97,142) .. controls (291.83,142) and (282,119.84) .. (282,92.5) .. controls (282,65.16) and (291.83,43) .. (303.97,43) -- cycle ;
          \draw  [draw opacity=0] (305.8,117.38) .. controls (305.8,117.38) and (305.8,117.38) .. (305.8,117.38) .. controls (301.51,117.38) and (298.04,106.3) .. (298.04,92.63) .. controls (298.04,78.97) and (301.51,67.88) .. (305.8,67.88) -- (305.8,92.63) -- cycle ; \draw  [color={rgb, 255:red, 255; green, 0; blue, 0 }  ,draw opacity=1 ] (305.8,117.38) .. controls (305.8,117.38) and (305.8,117.38) .. (305.8,117.38) .. controls (301.51,117.38) and (298.04,106.3) .. (298.04,92.63) .. controls (298.04,78.97) and (301.51,67.88) .. (305.8,67.88) ;
          \draw  [draw opacity=0] (303.91,69.7) .. controls (303.93,69.7) and (303.95,69.7) .. (303.97,69.7) .. controls (306.98,69.7) and (309.43,79.91) .. (309.43,92.5) .. controls (309.43,105.09) and (306.98,115.3) .. (303.97,115.3) .. controls (303.7,115.3) and (303.45,115.23) .. (303.19,115.08) -- (303.97,92.5) -- cycle ; \draw  [color={rgb, 255:red, 255; green, 0; blue, 0 }  ,draw opacity=1 ] (303.91,69.7) .. controls (303.93,69.7) and (303.95,69.7) .. (303.97,69.7) .. controls (306.98,69.7) and (309.43,79.91) .. (309.43,92.5) .. controls (309.43,105.09) and (306.98,115.3) .. (303.97,115.3) .. controls (303.7,115.3) and (303.45,115.23) .. (303.19,115.08) ;
          \draw  [color={rgb, 255:red, 255; green, 0; blue, 0 }  ,draw opacity=1 ] (362.97,42) .. controls (375.1,42) and (384.93,64.16) .. (384.93,91.5) .. controls (384.93,118.84) and (375.1,141) .. (362.97,141) .. controls (350.83,141) and (341,118.84) .. (341,91.5) .. controls (341,64.16) and (350.83,42) .. (362.97,42) -- cycle ;
          \draw  [draw opacity=0] (364.8,116.38) .. controls (364.8,116.38) and (364.8,116.38) .. (364.8,116.38) .. controls (360.51,116.38) and (357.04,105.3) .. (357.04,91.63) .. controls (357.04,77.97) and (360.51,66.88) .. (364.8,66.88) -- (364.8,91.63) -- cycle ; \draw  [color={rgb, 255:red, 255; green, 0; blue, 0 }  ,draw opacity=1 ] (364.8,116.38) .. controls (364.8,116.38) and (364.8,116.38) .. (364.8,116.38) .. controls (360.51,116.38) and (357.04,105.3) .. (357.04,91.63) .. controls (357.04,77.97) and (360.51,66.88) .. (364.8,66.88) ;
          \draw  [draw opacity=0] (362.91,68.7) .. controls (362.93,68.7) and (362.95,68.7) .. (362.97,68.7) .. controls (365.98,68.7) and (368.43,78.91) .. (368.43,91.5) .. controls (368.43,104.09) and (365.98,114.3) .. (362.97,114.3) .. controls (362.7,114.3) and (362.45,114.23) .. (362.19,114.08) -- (362.97,91.5) -- cycle ; \draw  [color={rgb, 255:red, 255; green, 0; blue, 0 }  ,draw opacity=1 ] (362.91,68.7) .. controls (362.93,68.7) and (362.95,68.7) .. (362.97,68.7) .. controls (365.98,68.7) and (368.43,78.91) .. (368.43,91.5) .. controls (368.43,104.09) and (365.98,114.3) .. (362.97,114.3) .. controls (362.7,114.3) and (362.45,114.23) .. (362.19,114.08) ;
          \draw    (241,39) .. controls (270,41) and (288,43) .. (303.97,43) ;
          \draw    (241,149) .. controls (266,145) and (288,142) .. (303.97,142) ;
          \draw    (429,39) .. controls (408.03,41) and (389.03,42) .. (362.97,42) ;
          \draw    (429,149) .. controls (410.03,146) and (389.03,143) .. (362.97,141) ;
          \draw [color={rgb, 255:red, 255; green, 0; blue, 0 }  ,draw opacity=1 ]   (303.97,43) .. controls (328,44) and (343,44) .. (362.97,42) ;
          \draw [color={rgb, 255:red, 255; green, 0; blue, 0 }  ,draw opacity=1 ]   (303.97,142) .. controls (327,140) and (343,140) .. (362.97,141) ;
          \draw [color={rgb, 255:red, 255; green, 0; blue, 0 }  ,draw opacity=1 ]   (360,28) .. controls (341,27) and (328,48) .. (334,72) ;
          \draw   (307,35.82) .. controls (307,31.15) and (304.67,28.82) .. (300,28.82) -- (279.5,28.82) .. controls (272.83,28.82) and (269.5,26.49) .. (269.5,21.82) .. controls (269.5,26.49) and (266.17,28.82) .. (259.5,28.82)(262.5,28.82) -- (239,28.82) .. controls (234.33,28.82) and (232,31.15) .. (232,35.82) ;
          %Shape: Brace [id:dp6089236618439431]
          \draw   (359,153) .. controls (359,157.67) and (361.33,160) .. (366,160) -- (386.5,160) .. controls (393.17,160) and (396.5,162.33) .. (396.5,167) .. controls (396.5,162.33) and (399.83,160) .. (406.5,160)(403.5,160) -- (427,160) .. controls (431.67,160) and (434,157.67) .. (434,153) ;
          % Text Node
          \draw (362,16) node [anchor=north west][inner sep=0.75pt]   [align=left] {\textcolor[rgb]{1,0,0}{trivial phase}};
          \draw (252,7) node [anchor=north west][inner sep=0.75pt]   [align=left] {\textbf{Slab}$_{1}$};
          \draw (380,169) node [anchor=north west][inner sep=0.75pt]   [align=left] {\textbf{Slab}$_{2}$};
          \draw (190,82) node [anchor=north west, scale=1.6][inner sep=0.75pt]   [align=left] {$\Sigma$};
          \draw (457,82) node [anchor=north west, scale=1.6][inner sep=0.75pt]   [align=left] {$\Sigma$};
        \end{tikzpicture}
        \caption{The condenable anyon on a fine mesh splits the wormhole geometry into two parts $\text{\textbf{Slab}}_1$, $\text{\textbf{Slab}}_2$.}
        \label{fig: splitting by trivial phase}
      \end{figure}

    \subsection{Diagonal Lagrangian condensable anyon}  \label{subsec: diagonal Lagrangian condensable anyon}
      Consider the theory with the Drinfeld center symmetry $\mathcal{Z}(\mathcal{C})=\mathcal{C}\boxtimes\overline{\mathcal{C}}$, where $\overline{\mathcal{C}}$ corresponds to the orientation-reversal of the original 3d TQFT. In simple terms, all the crossing operations of $\overline{\mathcal{C}}$ are given by the complex conjugates of those of $\mathcal{C}$.
      In this subsection, we introduce the canonical condensable anyon associated with $\mathcal{C}\boxtimes\overline{\mathcal{C}}$, referred to as the diagonal Lagrangian condensable anyon. The symbol $\boxtimes$ here denotes the tensor product of multiple categories, known as \bit{Deligne's tensor product} (see Section 1.11 in Ref.\,\cite{etingof2015tensor}), and is distinct from the symbol $\otimes$ used to fuse objects within a single category $\mathcal{C}$.
      \par The \bit{diagonal Lagrangian condensable anyon} is
      \begin{align}
        \mathcal{A}:=\bigoplus_{i\in\mathcal{I}\qty(\mathcal{C})}L_i\boxtimes\overline{L}_i. \label{eq: diagonal Lagrangian condensable anyon}
      \end{align}
      The algebra object $\mathcal{A}$ is manifestly commutative as the braiding phases from the chiral part and the anti-chiral part cancel out.
      Given the fusion rules among Wilson lines
      \begin{align}
        L_i\otimes L_j=\bigoplus_kN_{ij}^kL_k\qquad(N_{ij}^k\in\mathbb{Z}_{\geq 0}),
      \end{align}
      the product morphism $m:\mathcal{A}\otimes\mathcal{A}\to\mathcal{A}$ is defined by
      \begin{align}
        m:=\bigoplus_{i,j,k,\alpha}m_{ij,\alpha}^k\boxtimes\overline{m}_{ij,\alpha}^k, \label{eq: product}
      \end{align}
      where $m_{ij,\alpha}^k\in\Hom_{\mathcal{C}}\qty(L_i\otimes L_j,L_k)$, $\overline{m}_{ij,\alpha}^k\in\Hom_{\overline{\mathcal{C}}}\qty(\overline{L}_i\otimes \overline{L}_j,\overline{L}_k)$ ($\alpha=1,\cdots,N_{ij}^k$) is the basis.
      The co-product morphism $\Delta:\mathcal{A}\to\mathcal{A}\otimes\mathcal{A}$ is given by
      \begin{align}
        \Delta\Big(\bigoplus_{i\in\mathcal{I}(\mathcal{C})}L_i\boxtimes\overline{L}_i\Big):=\bigoplus_{i,j,k}N_{jk}^i\overline{N}_{jk}^i\qty(L_j\boxtimes\overline{L}_j)\otimes\qty(L_k\boxtimes\overline{L}_k). \label{eq: co-product}
      \end{align}
      The fact that the algebra object $\qty(\mathcal{A},m,\eta,\Delta,\iota)$ defined by eqs.\,\eqref{eq: product}, \eqref{eq: co-product} is indeed condensable (technically a Frobenius algebra object) is guaranteed by Proposition 7.20.1 in Ref.\,\cite{etingof2015tensor}.\footnote{Readers may refer the definition of dual morphisms like $m^*$ and $e^*$ appearing in Proposition 7.20.1 to eq.\,(2.47) and (2.48) (p.41) in this reference.} The co-associativity and the co-unit axiom follow automatically once the Frobenius condition is satisfied.\footnote{We thank K.\,Ohmori for telling us this point.}

      \paragraph{Example: $SU(2)_2$ Chern-Simons theory} ~ \\
        Let us confirm that separability holds in the simplest non-Abelian case $SU(2)_2$.
        The symmetry category of the chiral half of $SU(2)_k$ Chern-Simons theory is the category of representations $\mathcal{C}=\operatorname{Rep}\big(\,\mathcal{U}_q(\mathfrak{sl}(2,\mathbb{C}))\big)$ $(q=e^{\frac{\pi i}{k+2}})$ of the quantum group $\mathcal{U}_q(\mathfrak{sl}(2,\mathbb{C}))$, and the Wilson lines, i.e.\ simple objects in $\mathcal{C}$, are labeled by spin variables $j=0,\frac{1}{2},1,\cdots,\frac{k}{2}$ of integrable representations, obeying the fusion rule
        \begin{align}
          L_i\otimes L_j=\bigoplus_{l=|i-j|}^{\min\{i+j,\,k-(i+j)\}}L_l.
        \end{align}
        $l$ runs by increments of 1 in the sum.
        In $k=2$ case, it is explicitly written down as
        \begin{gather}
          L_0\otimes L_0=L_0,\;\;L_0\otimes L_{\frac{1}{2}}=L_{\frac{1}{2}},\;\;L_0\otimes L_1=L_1, \\
          L_{\frac{1}{2}}\otimes L_{\frac{1}{2}}=L_0\oplus L_1,\;\;L_{\frac{1}{2}}\otimes L_1=L_{\frac{1}{2}},\;\;L_1\otimes L_1=L_0.
        \end{gather}
        This TQFT shares the same fusion rule as the modular Ising category ($\operatorname{TY}(\mathbb{Z}_2,\xi,\tau)$ with a sutable braiding structure) if we identify $L_{\frac{1}{2}}$ and $L_1$ with the Ising anyon $\sigma$ and $\psi$, but their braiding data are different.
        The diagonal Lagrangian algebra in $\mathcal{C}\boxtimes\overline{\mathcal{C}}$ is
        \begin{align}
          \mathcal{A}=L_0\boxtimes\overline{L}_0\oplus L_{\frac{1}{2}}\boxtimes\overline{L}_{\frac{1}{2}}\oplus L_1\boxtimes\overline{L}_1
        \end{align}
        and the co-product \eqref{eq: co-product} maps $\mathcal{A}$ to
        \begin{align}
          \Delta(\mathcal{A}) & =(L_0\boxtimes\overline{L}_0)\otimes(L_0\boxtimes\overline{L}_0)\oplus 2(L_0\boxtimes\overline{L}_0)\otimes(L_{\frac{1}{2}}\boxtimes\overline{L}_{\frac{1}{2}})\oplus 2(L_0\boxtimes\overline{L}_0)\otimes(L_1\boxtimes\overline{L}_1) \nonumber \\
          & \oplus 2(L_{\frac{1}{2}}\boxtimes\overline{L}_{\frac{1}{2}})\otimes(L_{\frac{1}{2}}\boxtimes\overline{L}_{\frac{1}{2}})\oplus 2(L_{\frac{1}{2}}\boxtimes\overline{L}_{\frac{1}{2}})\otimes(L_1\boxtimes\overline{L}_1)\oplus(L_1\boxtimes\overline{L}_1)\otimes(L_1\boxtimes\overline{L}_1)
        \end{align}
        so we conclude that
        \begin{align}
          m\circ\Delta\qty(\mathcal{A})=4\mathcal{A}
        \end{align}
    \par Parenthetically, we note the link between general, possibly non-diagonal, Lagrangian condensable anyons and the topological boundary conditions.
    If we feed the diagonal Lagrangian condensable anyon \eqref{eq: diagonal Lagrangian condensable anyon} into the topological boundary picture, the resulting boundary theory is diagonal since all the bulk diagonal lines can terminate, leaving point-like operators on the boundary, while all the bulk non-diagonal lines are confined (see Fig.\,\ref{fig: b}).
    For $SU(2)_k$ Chern-Simons theory, the boundary theory is the diagonal $SU(2)$ level $k$ WZW model.
    If we instead select a non-diagonal Lagrangian condensable anyon as a topological boundary, we can easily infer that the corresponding boundary theory is a non-diagonal WZW model.
    Since the consistent WZW models are modular invariant and obey ADE classification \cite{Cappelli:1986hf, Cappelli:1987xt}, there should be a trinity relation described in Fig.\,\ref{fig: trinity relation} (There are subtleties regarding the Morita equivalence among Lagrangian algebra objects, but we do not need the details here.).
    \begin{figure}[t]
      \centering
      \tikzset{every picture/.style={line width=0.75pt}} %set default line width to 0.75pt
      \begin{tikzpicture}[x=0.75pt,y=0.75pt,yscale=-1,xscale=1]
        %Shape: Rectangle [id:dp18083371792458136]
        \draw   (89.82,20) -- (259.82,20) -- (259.82,80) -- (89.82,80) -- cycle ;
        %Shape: Rectangle [id:dp2008754324110591]
        \draw   (410,19.88) -- (580,19.88) -- (580,80) -- (410,80) -- cycle ;
        %Shape: Rectangle [id:dp525058660329054]
        \draw   (250,130) -- (420,130) -- (420,190.53) -- (250,190.53) -- cycle ;
        %Straight Lines [id:da9021263603425564]
        \draw    (169.7,80) -- (249.88,160.88) ;
        %Straight Lines [id:da6762824251552344]
        \draw    (419.92,160.35) -- (500.88,80) ;
        %Straight Lines [id:da7663302362487242]
        \draw    (259.68,50) -- (409.92,49.88) ;
        % Text Node
        \draw (97,35) node [anchor=north west][inner sep=0.75pt]   [align=left] {\begin{minipage}[lt]{114.53pt}\setlength\topsep{0pt}
        \begin{center}
          Lagrangian condensable\\anyons in $\displaystyle SU( 2)_{k}$ CS
        \end{center}
        \end{minipage}};
          % Text Node
        \draw (274,145) node [anchor=north west][inner sep=0.75pt]   [align=left] {\begin{minipage}[lt]{90pt}\setlength\topsep{0pt}
        \begin{center}
          modular invariant \\ $\displaystyle SU(2)_k$ WZW
        \end{center}
        \end{minipage}};
        % Text Node
        \draw (422,36) node [anchor=north west][inner sep=0.75pt]   [align=left] {\begin{minipage}[lt]{110pt}\setlength\topsep{0pt}
        \begin{center}
          topological boundary \\ conditions
        \end{center}
        \end{minipage}};
      \end{tikzpicture}
      \caption{The trinity relation among Lagrangian condensable anyons, topological boundaries, and boundary modular invariant WZW model in $SU(2)_k$ Chern-Simons theory.
      Each of them obeys the ADE classification.}
      \label{fig: trinity relation}
    \end{figure}
    This is consistent with the fact that all Lagrangian condensable anyons in $\mathcal{C}=\operatorname{Rep}\big(\,\mathcal{U}_q(\mathfrak{sl}(2,\mathbb{C}))\big)$ obey the ADE classification \cite{kirillov2002q}.

  \section{Anyon condensation in Virasoro TQFT} \label{sec: Anyon condensation in Virasoro TQFT}
    We extend anyon condensation to VTQFT with associated category denoted as $\mathcal{C}$.
    $\mathcal{C}$ is a ribbon category, that is $\mathbb{C}$-linear Abelian rigid braided monoidal category with a ribbon structure (twist).
    However, it is not semisimple nor locally finite due to the infinite number of Wilson lines (simple objects).
    In this way $\mathcal{C}$ is a non-semisimple non-locally finite $\mathbb{C}$-linear Abelian ribbon category, whose relative position is indicated by the green area in Figure \ref{fig: inclusion relation} in Appendix \ref{append: From monoidal category to fusion category and modular tensor category}.
    In spite of these problems, we will define
    \begin{align}
      \mathcal{A}:=\int_{\mathbb{R}_{\geq 0}}^{\oplus}dp\,L_p\boxtimes\overline{L}_p \label{eq: diagonal Lagrangian algebra object in VTQFT}
    \end{align}
    and still call it the \bit{diagonal condensable anyon}.
    The modifier ``Lagrangian" is being removed because $\mathcal{A}$ does not contain the identity line $\mathbbm{1}\boxtimes\mathbbm{1}$.
    Although the continuous direct sums is not guaranteed in a mathematically rigorous manner, we leave it for the discussion in Section \ref{sec: conclusions and discussions} for now.
    As stated in the introduction, VTQFT as a gravitational theory suffers from the factorization puzzle for partition function, the non-factorization of the partition function for geometries with multiple boundaries.
    We solve the paradox by condensing the VTQFT diagonal condensable anyon \eqref{eq: diagonal Lagrangian algebra object in VTQFT} for two-boundary wormhole geometries in line with the method described in Ref.\,\cite{Benini:2022hzx}.

    \subsection{Projector}
      In this subsection, we introduce a special link of Wilson lines that frequently appears in the anyon condensation process when the diagonal condensable anyon is placed along a fine mesh of a manifold.
      The link drastically simplify the computation of VTQFT path-integral.
      A \bit{projector} is an object indicated in the l.\,h.\,s.~of
      \begin{align}
        \begin{tikzpicture}[scale = 0.5, baseline = {([yshift=-.5ex]current bounding box.center)}]
          \draw[thick, red] (0, -2.5) -- (0, 2.5);
          \draw[thick, red] (0, 1.5) .. controls (2.5, 0.9) and (2.5, -0.9) .. (0, -1.5);
          \fill[white] (1.5, 0.775) circle[radius = 0.125];
          \draw (1.5, 2.5) -- (1.5, -0.65);
          \draw (1.5, -0.9) -- (1.5, -2.5);
          \node[scale = 0.8, above] at (0, 2.5) {$\mathcal{A}$};
          \node[scale = 0.8, left] at (0, 0) {$\mathcal{A}$};
          \node[scale = 0.8, right] at (1.8, 0) {$\mathcal{A}$};
          \node[scale = 0.8, below] at (0, -2.5) {$\mathcal{A}$};
          \node[scale = 0.8, above] at (2, 2.5) {$L_{p_x}\boxtimes\overline{L}_{\overline{p}_x}$};
        \end{tikzpicture}
        \propto
        \begin{tikzpicture}[scale = 0.5, baseline = {([yshift=-.5ex]current bounding box.center)}]
          \draw[thick, red] (0, 2.5) -- (0.75, 1);
          \draw (1.5, 2.5) -- (0.75, 1);
          \draw[thick, red] (0.75, 1) -- (0.75, -1);
          \draw[thick, red] (0.75, -1) -- (0, -2.5);
          \draw (0.75, -1) -- (1.5, -2.5);
          \node[scale = 0.8, above] at (-0.5, 2.5) {$\mathcal{A}$};
          \node[scale = 0.8, below] at (-0.5, -2.5) {$\mathcal{A}$};
          \node[scale = 0.8, above] at (2, 2.5) {$L_{p_x}\boxtimes\overline{L}_{p_x}$};
          \node[scale = 0.8, below] at (2, -2.5) {$L_{p_x}\boxtimes\overline{L}_{p_x}$};
          \node[scale = 0.8, left] at (0.75, 0) {$\mathcal{A}$};
        \end{tikzpicture}
        . \label{eq: projector proportionality}
      \end{align}
      In contrast to Section \ref{sec: Virasoro TQFT}, we use a single link diagram or single ket $\ket{\:\cdot\:}$ to represent the tensor product of the chiral part and the anti-chiral part, which is why we use the notation $L_{p_x}\boxtimes\overline{L}_{\overline{p}_x}$ in the above equation.
      From this point forward, we will adhere to this rule.
      The condensable anyon $\mathcal{A}$ (the red line) shown in this figure represents a superposition of its component.
      The concept of projector first appeared in eq.\,(5.34) of Ref.\,\cite{Fuchs:2002cm} in the context of modular tensor category.
      The central point to be confirmed is that the projector is proportional to the r.\,h.\,s.~of eq.\,\eqref{eq: projector proportionality}, as briefly explained in the last paragraph of Subsection 4.1 in Ref.\,\cite{Benini:2022hzx} in the case of Chern-Simons theory.
      To begin with, let us clarify that in this paper inserting the diagonal condensable anyon \eqref{eq: diagonal Lagrangian algebra object in VTQFT} means a superposition of each component weighted by $\rho_0(p_i)$ for each internal line $p_i$ and $C_{ijk}$ for each trivalent junctions $
      \begin{tikzpicture}[scale = 0.12, baseline = -1.5pt]
        \draw (0, 0) to (0, 2);
        \draw (0, 0) to (-1.5, -1);
        \draw (0, 0) to (1.5, -1);
        \fill (0, 0) circle[radius = 0.25];
        \node[scale = 0.6, left] at (0.4, 1.6) {$i$};
        \node[scale = 0.5, above] at (-1.5, -1.15) {$j$};
        \node[scale = 0.5, above] at (1.5, -1) {$k$};
      \end{tikzpicture}
      $. Under this rule, the projector is presented as
      \begin{align}
        \begin{tikzpicture}[scale = 0.5, baseline = {([yshift=-.5ex]current bounding box.center)}]
          \draw[thick, red] (0, -2.5) -- (0, 2.5);
          \draw[thick, red] (0, 1.5) .. controls (2.5, 0.9) and (2.5, -0.9) .. (0, -1.5);
          \fill[white] (1.5, 0.775) circle[radius = 0.125];
          \draw (1.5, 2.5) -- (1.5, -0.65);
          \draw (1.5, -0.9) -- (1.5, -2.5);
          \node[scale = 0.8, above] at (0, 2.5) {$\mathcal{A}$};
          \node[scale = 0.8, left] at (0, 0) {$\mathcal{A}$};
          \node[scale = 0.8, right] at (1.8, 0) {$\mathcal{A}$};
          \node[scale = 0.8, below] at (0, -2.5) {$\mathcal{A}$};
          \node[scale = 0.8, above] at (2, 2.5) {$L_{p_x}\boxtimes\overline{L}_{\overline{p}_x}$};
        \end{tikzpicture}
        =\int_{0}^{\infty}dp_1dp_2dp_3dp_4\,\rho_0\qty(p_3)\rho_0\qty(p_4)C_{134}C_{234}
        \begin{tikzpicture}[scale = 0.5, baseline = {([yshift=-.5ex]current bounding box.center)}]
          \draw (0, -2.5) -- (0, 2.5);
          \draw (0, 1.5) .. controls (2.5, 0.9) and (2.5, -0.9) .. (0, -1.5);
          \fill[white] (1.5, 0.775) circle[radius = 0.125];
          \draw (1.5, 2.5) -- (1.5, -0.65);
          \draw (1.5, -0.9) -- (1.5, -2.5);
          \node[scale = 0.8, above] at (-0.5, 2.5) {$L_{p_1}\boxtimes\overline{L}_{p_1}$};
          \node[scale = 0.8, left] at (0, 0) {$L_{p_3}\boxtimes\overline{L}_{p_3}$};
          \node[scale = 0.8, below] at (-0.5, -2.5) {$L_{p_2}\boxtimes\overline{L}_{p_2}$};
          \node[scale = 0.8, above] at (2, 2.5) {$L_{p_x}\boxtimes\overline{L}_{\overline{p}_x}$};
          \node[scale = 0.8, below right] at (1.7, 0.2) {$L_{p_4}\boxtimes\overline{L}_{p_4}$};
        \end{tikzpicture}
        . \label{eq: projector}
      \end{align}
      In this diagram, $p_x$ and $\overline{p}_x$ are a different variables so that $L_{p_x}\boxtimes\overline{L}_{\overline{p}_x}$ is a general Wilson line in $\mathcal{C}\boxtimes\overline{\mathcal{C}}$ while $L_{p_i}\boxtimes\overline{L}_{p_i}$ are restricted to diagonal lines contained in $\mathcal{A}$.
      We then implement fusion transformation with $L_{p_3}\boxtimes\overline{L}_{p_3}$ as the internal line to create the Verlinde loop
      \begin{align}
        \begin{tikzpicture}[scale = 0.5, baseline = {([yshift=-.5ex]current bounding box.center)}]
          \draw (0, -2.5) -- (0, 2.5);
          \draw (0, 1.5) .. controls (2.5, 0.9) and (2.5, -0.9) .. (0, -1.5);
          \fill[white] (1.5, 0.775) circle[radius = 0.125];
          \draw (1.5, 2.5) -- (1.5, -0.65);
          \draw (1.5, -0.9) -- (1.5, -2.5);
          \node[scale = 0.8, above] at (-0.5, 2.5) {$L_{p_1}\boxtimes\overline{L}_{p_1}$};
          \node[scale = 0.8, left] at (0, 0) {$L_{p_3}\boxtimes\overline{L}_{p_3}$};
          \node[scale = 0.8, below] at (-0.5, -2.5) {$L_{p_2}\boxtimes\overline{L}_{p_2}$};
          \node[scale = 0.8, above] at (2, 2.5) {$L_{p_x}\boxtimes\overline{L}_{\overline{p}_x}$};
          \node[scale = 0.8, below right] at (1.7, 0.2) {$L_{p_4}\boxtimes\overline{L}_{p_4}$};
        \end{tikzpicture}
        =\int_{0}^{\infty}dp_ad\overline{p}_a\,F_{p_3p_a}
        \begin{bmatrix}
          p_1 & p_2 \\
          p_4 & p_4
        \end{bmatrix}
        F_{p_3\overline{p}_a}
        \begin{bmatrix}
          p_1 & p_2 \\
          p_4 & p_4
        \end{bmatrix}
        \begin{tikzpicture}[scale = 0.5, baseline = {([yshift=-.5ex]current bounding box.center)}]
          \draw (0, -2.5) -- (0, 2.5);
          \draw (0, 0) -- (0.7, 0);
          \draw (1.5, 0) ellipse (0.8 and 0.4);
          \fill[white] (1.5, 0.4) circle[radius = 0.125];
          \draw (1.5, 2.5) -- (1.5, -0.275);
          \draw (1.5, -0.525) -- (1.5, -2.5);
          \draw[->, > = stealth, bend left = 20] (2, -1.4) to (0.35, -0.2);
          \node[scale = 0.8, above] at (-0.5, 2.5) {$L_{p_1}\boxtimes\overline{L}_{p_1}$};
          \node[scale = 0.8, below] at (-0.5, -2.5) {$L_{p_2}\boxtimes\overline{L}_{p_2}$};
          \node[scale = 0.8, above] at (2, 2.5) {$L_{p_x}\boxtimes\overline{L}_{\overline{p}_x}$};
          \node[scale = 0.8, above right] at (1.9, 0.1) {$L_{p_4}\boxtimes\overline{L}_{p_4}$};
          \node[scale = 0.8, right] at (1.9, -1.4) {$L_{p_a}\boxtimes\overline{L}_{\overline{p}_a}$};
        \end{tikzpicture}
        .
      \end{align}
      Performing the integration w.\,r.\,t.~$p_3$ in eq.\,\eqref{eq: projector}, utilizing eq.\,\eqref{eq: version of invertibility}, and resolving the Verlinde loop by eq.\,\eqref{eq: Verlinde loop} yield
      \begin{align}
        \begin{tikzpicture}[scale = 0.5, baseline = {([yshift=-.5ex]current bounding box.center)}]
          \draw[thick, red] (0, -2.5) -- (0, 2.5);
          \draw[thick, red] (0, 1.5) .. controls (2.5, 0.9) and (2.5, -0.9) .. (0, -1.5);
          \fill[white] (1.5, 0.775) circle[radius = 0.125];
          \draw (1.5, 2.5) -- (1.5, -0.65);
          \draw (1.5, -0.9) -- (1.5, -2.5);
          \node[scale = 0.8, above] at (0, 2.5) {$\mathcal{A}$};
          \node[scale = 0.8, left] at (0, 0) {$\mathcal{A}$};
          \node[scale = 0.8, right] at (1.8, 0) {$\mathcal{A}$};
          \node[scale = 0.8, below] at (0, -2.5) {$\mathcal{A}$};
          \node[scale = 0.8, above] at (2, 2.5) {$L_{p_x}\boxtimes\overline{L}_{\overline{p}_x}$};
        \end{tikzpicture}
        & =\int_{0}^{\infty}dp_1dp_2dp_4dp_a\,\rho_0\qty(p_4)\rho_0(p_a)C_{44a}C_{12a}
        \begin{tikzpicture}[scale = 0.5, baseline = {([yshift=-.5ex]current bounding box.center)}]
          \draw (0, -2.5) -- (0, 2.5);
          \draw (0, 0) -- (0.7, 0);
          \draw (1.5, 0) ellipse (0.8 and 0.4);
          \fill[white] (1.5, 0.4) circle[radius = 0.125];
          \draw (1.5, 2.5) -- (1.5, -0.275);
          \draw (1.5, -0.525) -- (1.5, -2.5);
          \draw[->, > = stealth, bend left = 20] (2, -1.4) to (0.35, -0.2);
          \node[scale = 0.8, above] at (-0.5, 2.5) {$L_{p_1}\boxtimes\overline{L}_{p_1}$};
          \node[scale = 0.8, below] at (-0.5, -2.5) {$L_{p_2}\boxtimes\overline{L}_{p_2}$};
          \node[scale = 0.8, above] at (2, 2.5) {$L_{p_x}\boxtimes\overline{L}_{\overline{p}_x}$};
          \node[scale = 0.8, above right] at (1.9, 0.1) {$L_{p_4}\boxtimes\overline{L}_{p_4}$};
          \node[scale = 0.8, right] at (1.9, -1.4) {$L_{p_a}\boxtimes\overline{L}_{p_a}$};
        \end{tikzpicture}
        \nonumber \\
        = & \int_{0}^{\infty}dp_1dp_2dp_4dp_a\,\rho_0\qty(p_4)\rho_0(p_a)C_{44a}C_{12a}\frac{S_{p_4p_x}[p_a]S_{p_4\overline{p}_x}^*[p_a]}{S_{\mathbbm{1}p_x}\qty[\mathbbm{1}]S_{\mathbbm{1}\overline{p}_x}\qty[\mathbbm{1}]}
        \begin{tikzpicture}[scale = 0.5, baseline = {([yshift=-.5ex]current bounding box.center)}]
          \draw (0, -2.5) -- (0, 2.5);
          \draw (0, 0) -- (1.5, 0);
          \draw (1.5, 2.5) -- (1.5, -2.5);
          \draw[->, > = stealth, bend left = 20] (2, -1.4) to (0.75, -0.2);
          \node[scale = 0.8, above] at (-0.5, 2.5) {$L_{p_1}\boxtimes\overline{L}_{p_1}$};
          \node[scale = 0.8, below] at (-0.5, -2.5) {$L_{p_2}\boxtimes\overline{L}_{p_2}$};
          \node[scale = 0.8, above] at (2, 2.5) {$L_{p_x}\boxtimes\overline{L}_{\overline{p}_x}$};
          \node[scale = 0.8, below] at (2, -2.5) {$L_{p_x}\boxtimes\overline{L}_{\overline{p}_x}$};
          \node[scale = 0.8, right] at (1.9, -1.4) {$L_{p_a}\boxtimes\overline{L}_{p_a}$};
        \end{tikzpicture}
        .
      \end{align}
      The crucial point is that eq.\,\eqref{eq: version of invertibility} applies in this case because the first subscript and the four arguments of the two fusion kernels are identical.
      The chiral part Liouville momentum $p$ and the anti-chiral part Liouville momentum $\overline{p}$ are different in general, but they coincide due to the diagonality of the condensable anyon $\mathcal{A}$.
      We then employ eq.\,\eqref{eq: S-kernel lael exchange} to exchange the subscripts of $S_{p_4p_x}$ and perform the integration w.\,r.\,t.~$p_4$ by eq.\,\eqref{eq: unitarity of S-kernel}
      \begin{align}
        \begin{tikzpicture}[scale = 0.5, baseline = {([yshift=-.5ex]current bounding box.center)}]
          \draw[thick, red] (0, -2.5) -- (0, 2.5);
          \draw[thick, red] (0, 1.5) .. controls (2.5, 0.9) and (2.5, -0.9) .. (0, -1.5);
          \fill[white] (1.5, 0.775) circle[radius = 0.125];
          \draw (1.5, 2.5) -- (1.5, -0.65);
          \draw (1.5, -0.9) -- (1.5, -2.5);
          \node[scale = 0.8, above] at (0, 2.5) {$\mathcal{A}$};
          \node[scale = 0.8, left] at (0, 0) {$\mathcal{A}$};
          \node[scale = 0.8, right] at (1.8, 0) {$\mathcal{A}$};
          \node[scale = 0.8, below] at (0, -2.5) {$\mathcal{A}$};
          \node[scale = 0.8, above] at (2, 2.5) {$L_{p_x}\boxtimes\overline{L}_{\overline{p}_x}$};
        \end{tikzpicture}
          =\int_{0}^{\infty}dp_1dp_2dp_a\,\rho_0(p_a)C_{xxa}C_{12a}\frac{\delta(p_x-\overline{p}_x)}{S_{\mathbbm{1}p_x}\qty[\mathbbm{1}]}
        \begin{tikzpicture}[scale = 0.5, baseline = {([yshift=-.5ex]current bounding box.center)}]
          \draw (0, -2.5) -- (0, 2.5);
          \draw (0, 0) -- (1.5, 0);
          \draw (1.5, 2.5) -- (1.5, -2.5);
          \draw[->, > = stealth, bend left = 20] (2, -1.4) to (0.75, -0.2);
          \node[scale = 0.8, above] at (-0.5, 2.5) {$L_{p_1}\boxtimes\overline{L}_{p_1}$};
          \node[scale = 0.8, below] at (-0.5, -2.5) {$L_{p_2}\boxtimes\overline{L}_{p_2}$};
          \node[scale = 0.8, above] at (2, 2.5) {$L_{p_x}\boxtimes\overline{L}_{p_x}$};
          \node[scale = 0.8, below] at (2, -2.5) {$L_{p_x}\boxtimes\overline{L}_{p_x}$};
          \node[scale = 0.8, right] at (1.9, -1.4) {$L_{p_a}\boxtimes\overline{L}_{p_a}$};
        \end{tikzpicture}
        .
      \end{align}
      The graph in the integrand is further modified by the fusion transformation as
      \begin{align}
        \begin{tikzpicture}[scale = 0.5, baseline = {([yshift=-.5ex]current bounding box.center)}]
          \draw (0, -2.5) -- (0, 2.5);
          \draw (0, 0) -- (1.5, 0);
          \draw (1.5, 2.5) -- (1.5, -2.5);
          \draw[->, > = stealth, bend left = 20] (2, -1.4) to (0.75, -0.2);
          \node[scale = 0.8, above] at (-0.5, 2.5) {$L_{p_1}\boxtimes\overline{L}_{p_1}$};
          \node[scale = 0.8, below] at (-0.5, -2.5) {$L_{p_2}\boxtimes\overline{L}_{p_2}$};
          \node[scale = 0.8, above] at (2, 2.5) {$L_{p_x}\boxtimes\overline{L}_{p_x}$};
          \node[scale = 0.8, below] at (2, -2.5) {$L_{p_x}\boxtimes\overline{L}_{p_x}$};
          \node[scale = 0.8, right] at (1.9, -1.4) {$L_{p_a}\boxtimes\overline{L}_{p_a}$};
        \end{tikzpicture}
        =\int_{0}^{\infty}dp_bd\overline{p}_bF_{p_ap_b}
        \begin{bmatrix}
          p_1 & p_x \\
          p_2 & p_x
        \end{bmatrix}
        F_{p_a\overline{p}_b}
        \begin{bmatrix}
          p_1 & p_x \\
          p_2 & p_x
        \end{bmatrix}
        \begin{tikzpicture}[scale = 0.5, baseline = {([yshift=-.5ex]current bounding box.center)}]
          \draw (0, 2.5) -- (0.75, 1);
          \draw (1.5, 2.5) -- (0.75, 1);
          \draw (0.75, 1) -- (0.75, -1);
          \draw (0.75, -1) -- (0, -2.5);
          \draw (0.75, -1) -- (1.5, -2.5);
          \node[scale = 0.8, above] at (-0.5, 2.5) {$L_{p_1}\boxtimes\overline{L}_{p_1}$};
          \node[scale = 0.8, below] at (-0.5, -2.5) {$L_{p_2}\boxtimes\overline{L}_{p_2}$};
          \node[scale = 0.8, above] at (2, 2.5) {$L_{p_x}\boxtimes\overline{L}_{p_x}$};
          \node[scale = 0.8, below] at (2, -2.5) {$L_{p_x}\boxtimes\overline{L}_{p_x}$};
          \node[scale = 0.8, right] at (0.75, 0) {$L_{p_b}\boxtimes\overline{L}_{\overline{p}_b}$};
        \end{tikzpicture}
        .
      \end{align}
      We ultimately establish the relation \eqref{eq: projector proportionality} by performing the integration w.\,r.\,t.~$p_a$ with the assistance of eq.\,\eqref{eq: version of invertibility}
      \begin{align}
        \begin{tikzpicture}[scale = 0.5, baseline = {([yshift=-.5ex]current bounding box.center)}]
          \draw[thick, red] (0, -2.5) -- (0, 2.5);
          \draw[thick, red] (0, 1.5) .. controls (2.5, 0.9) and (2.5, -0.9) .. (0, -1.5);
          \fill[white] (1.5, 0.775) circle[radius = 0.125];
          \draw (1.5, 2.5) -- (1.5, -0.65);
          \draw (1.5, -0.9) -- (1.5, -2.5);
          \node[scale = 0.8, above] at (0, 2.5) {$\mathcal{A}$};
          \node[scale = 0.8, left] at (0, 0) {$\mathcal{A}$};
          \node[scale = 0.8, right] at (1.8, 0) {$\mathcal{A}$};
          \node[scale = 0.8, below] at (0, -2.5) {$\mathcal{A}$};
          \node[scale = 0.8, above] at (2, 2.5) {$L_{p_x}\boxtimes\overline{L}_{\overline{p}_x}$};
        \end{tikzpicture}
        & =\int_{0}^{\infty}dp_1dp_2dp_b\,\rho_0(p_b)C_{1xb}C_{2xb}\frac{\delta(p_x-\overline{p}_x)}{\rho_0(p_x)}
        \begin{tikzpicture}[scale = 0.5, baseline = {([yshift=-.5ex]current bounding box.center)}]
          \draw (0, 2.5) -- (0.75, 1);
          \draw (1.5, 2.5) -- (0.75, 1);
          \draw (0.75, 1) -- (0.75, -1);
          \draw (0.75, -1) -- (0, -2.5);
          \draw (0.75, -1) -- (1.5, -2.5);
          \node[scale = 0.8, above] at (-0.5, 2.5) {$L_{p_1}\boxtimes\overline{L}_{p_1}$};
          \node[scale = 0.8, below] at (-0.5, -2.5) {$L_{p_2}\boxtimes\overline{L}_{p_2}$};
          \node[scale = 0.8, above] at (2, 2.5) {$L_{p_x}\boxtimes\overline{L}_{p_x}$};
          \node[scale = 0.8, below] at (2, -2.5) {$L_{p_x}\boxtimes\overline{L}_{p_x}$};
          \node[scale = 0.8, right] at (0.75, 0) {$L_{p_b}\boxtimes\overline{L}_{p_b}$};
        \end{tikzpicture}
        \nonumber \\
        & =\frac{\delta(p_x-\overline{p}_x)}{\rho_0(p_x)}
        \begin{tikzpicture}[scale = 0.5, baseline = {([yshift=-.5ex]current bounding box.center)}]
          \draw[thick, red] (0, 2.5) -- (0.75, 1);
          \draw (1.5, 2.5) -- (0.75, 1);
          \draw[thick, red] (0.75, 1) -- (0.75, -1);
          \draw[thick, red] (0.75, -1) -- (0, -2.5);
          \draw (0.75, -1) -- (1.5, -2.5);
          \node[scale = 0.8, above] at (-0.5, 2.5) {$\mathcal{A}$};
          \node[scale = 0.8, below] at (-0.5, -2.5) {$\mathcal{A}$};
          \node[scale = 0.8, above] at (2, 2.5) {$L_{p_x}\boxtimes\overline{L}_{p_x}$};
          \node[scale = 0.8, below] at (2, -2.5) {$L_{p_x}\boxtimes\overline{L}_{p_x}$};
          \node[scale = 0.8, left] at (0.75, 0) {$\mathcal{A}$};
        \end{tikzpicture}
        . \label{eq: projector formula}
      \end{align}

      \subsection{Factorization}
      Now that everything is in place, we will exemplify the factorization of the partition function of a wormhole geometry for two simple cases, the torus wormhole $T^2\times\qty[0,1]$ and the genus two wormhole $\Sigma_{2,0}\times\qty[0,1]$ coupled to matter fields.
      The technical reason for introducing matter fields will be made clear momentarily.

      \subsubsection{Torus wormhole}
        \begin{figure}[t]
          \centering
          \begin{tikzpicture}[scale = 0.6, baseline={([yshift = -0.5ex]current bounding box.center)}]
            \draw[thick] (0, 0) ellipse (3.6 and 2);
            \draw[thick, bend left = 20] (1, 0) to (-1, 0);
            \draw[thick, bend right = 30] (0.78, -0.08) to (-0.78, -0.08);
            \draw[dashed] (0, 0) ellipse (1.95 and 0.85);
            \draw[dashed] (0, 0) ellipse (2.75 and 1.35);
            % Wilson lines
            \draw[thick, red] (0, 0) ellipse (1.65 and 0.62);
            \draw[thick, red] (-1.25, 0.41) .. controls (-2.4, 0.38) and (-2.9, 0.2) .. (-2.9, 0);
            \draw[thick, red] (-1.25, -0.41) .. controls (-2.4, -0.38) and (-2.9, -0.2) .. (-2.9, 0);
            \fill[white, rotate around = {-15:(-2.555, -0.26)}] (-2.63, -0.21) rectangle (-2.48, -0.31);
            \fill[white, rotate around = {-10:(-2.245, -0.33)}] (-2.32, -0.28) rectangle (-2.17, -0.38);
            \fill[white, rotate around = {-6:(-1.945, -0.37)}] (-2.02, -0.32) rectangle (-1.87, -0.42);
            \draw (2.5, 0) -- (3.35, 0);
            % punctures
            \draw[thick] (2.6, 0.1) -- (2.4, -0.1);
            \draw[thick] (2.6, -0.1) -- (2.4, 0.1);
            \draw[thick] (3.45, 0.1) -- (3.25, -0.1);
            \draw[thick] (3.45, -0.1) -- (3.25, 0.1);
            % labels
            \node[scale = 0.8] at (-3.3, 0) {$\mathcal{A}$};
            \node[scale = 0.8] at (-1.3, 0) {$\mathcal{A}$};
            \node[scale = 0.8, above] at (-0.5, 0.6) {$\mathcal{A}$};
            \node[scale = 0.8, above] at (2, 0.55) {$L_p\boxtimes\overline{L}_{\overline{p}}$};
            \draw[->, > = stealth, bend left = 20] (2.75, 0.7) to (3, 0.1);
          \end{tikzpicture}
          \caption{The torus wormhole with the Wilson line insertion connecting the two boundaries. The dashed lines are the inner boundary $\Sigma_{1,1}$.}
          \label{fig: torus wormhole}
        \end{figure}
        We avoid the pure torus wormhole $T^2\times\qty[0,1]$ and instead add the Wilson line $L_p\boxtimes\overline{L}_{\overline{p}}$ ($p,\overline{p}\in\mathbb{R}_{\geq 0}$) that connects the two boundaries, because the torus wormhole $\Sigma_{1,0}\times\qty[0,1]$ is non-hyperbolic for which the VTQFT partition function is possibly ill-behaved. \footnote{The wormhole geometry $\Sigma_{g,0}\times[0,1]$ is non-hyperbolic for $g=1$ and hyperbolic for $g\geq 2$ in a sense that we can find a hyperbolic metric that makes the boundaries $\Sigma_{g,0}\times\{0\}$, $\Sigma_{g,0}\times\{1\}$ convex.
        This is guaranteed by the Alfors-Bers theory elucidated, e.g.\,in chapter 3 in the all-encompassing textbook \cite{marden2016hyperbolic}. We thank Prof.\,Ken'ichi Ohshika for correspondence.}
        Indeed, one will find that there appear multiple divergent constants when computing the anyon condensation in the same way as what is done in the following for pure $T^2\times\qty[0,1]$.
        \par The target geometry is given in Figure \ref{fig: torus wormhole} where the red lines represent the diagonal condensable anyons $\mathcal{A}$ placed on the fine mesh.
        The VTQFT path-integral \eqref{eq: compression body path-integral} assigns
        \begin{align}
          \lower2.1ex\hbox{\text{\scalebox{1.6}[4.5]{$|$}}}\,
          \begin{tikzpicture}[scale = 0.4, baseline={([yshift = -0.5ex]current bounding box.center)}]
            \draw[thick] (0, 0) ellipse (3.6 and 2);
            \draw[thick, bend left = 20] (1, 0) to (-1, 0);
            \draw[thick, bend right = 30] (0.78, -0.08) to (-0.78, -0.08);
            \draw[dashed] (0, 0) ellipse (1.95 and 0.85);
            \draw[dashed] (0, 0) ellipse (2.75 and 1.35);
            % Wilson lines
            \draw[thick, red] (0, 0) ellipse (1.65 and 0.62);
            \draw[thick, red] (-1.25, 0.41) .. controls (-2.4, 0.38) and (-2.9, 0.2) .. (-2.9, 0);
            \draw[thick, red] (-1.25, -0.41) .. controls (-2.4, -0.38) and (-2.9, -0.2) .. (-2.9, 0);
            \fill[white, rotate around = {-15:(-2.555, -0.26)}] (-2.63, -0.21) rectangle (-2.48, -0.31);
            \fill[white, rotate around = {-10:(-2.245, -0.33)}] (-2.32, -0.28) rectangle (-2.17, -0.38);
            \fill[white, rotate around = {-6:(-1.945, -0.37)}] (-2.02, -0.32) rectangle (-1.87, -0.42);
            \draw (2.5, 0) -- (3.35, 0);
            % punctures
            \draw[thick] (2.6, 0.1) -- (2.4, -0.1);
            \draw[thick] (2.6, -0.1) -- (2.4, 0.1);
            \draw[thick] (3.45, 0.1) -- (3.25, -0.1);
            \draw[thick] (3.45, -0.1) -- (3.25, 0.1);
            % labels
            \node[scale = 0.8, above] at (2, 0.55) {$L_p\boxtimes\overline{L}_{\overline{p}}$};
            \draw[->, > = stealth, bend left = 20] (2.75, 0.7) to (3, 0.1);
          \end{tikzpicture}
          \lower2.1ex\hbox{\text{\scalebox{2}[4.5]{$\rangle$}}} & =\int_0^{\infty}dp_xd\overline{p}_x\,\rho_0(p_x)\rho_0(\overline{p}_x)C_{xxp}C_{\overline{x}\overline{x}\overline{p}} \nonumber \\
          & \qqquad\times\lower2.1ex\hbox{\text{\scalebox{1.6}[4.5]{$|$}}}\,
          \begin{tikzpicture}[scale = 0.4, baseline={([yshift = -0.5ex]current bounding box.center)}]
            \draw[thick] (0, 0) ellipse (3.6 and 2);
            \draw[thick, bend left = 20] (1, 0) to (-1, 0);
            \draw[thick, bend right = 30] (0.78, -0.08) to (-0.78, -0.08);
            % Wilson lines
            \draw[green!60!black] (0, 0) ellipse (2.3 and 1.2);
            \draw (2.3, 0) -- (3.35, 0);
            \fill (2.3, 0) circle[radius = 0.1];
            % punctures
            \draw[thick] (3.45, 0.1) -- (3.25, -0.1);
            \draw[thick] (3.45, -0.1) -- (3.25, 0.1);
            % labels
            \node[scale = 0.8, above] at (2.35, 0.55) {$L_p\boxtimes\overline{L}_{\overline{p}}$};
            \draw[->, > = stealth, bend left = 15] (2.6, 0.9) to (2.85, 0.1);
            \node[scale = 0.8] at (-2.1, 0) {$L_{p_x}\boxtimes\overline{L}_{\overline{p}_x}$};
          \end{tikzpicture}
          \lower2.1ex\hbox{\text{\scalebox{2}[4.5]{$\rangle$}}}\otimes\lower2.1ex\hbox{\text{\scalebox{1.6}[4.5]{$|$}}}\,
          \begin{tikzpicture}[scale = 0.4, baseline={([yshift = -0.5ex]current bounding box.center)}]
            \draw[thick] (0, 0) ellipse (3.6 and 2);
            \draw[thick, bend left = 20] (1, 0) to (-1, 0);
            \draw[thick, bend right = 30] (0.78, -0.08) to (-0.78, -0.08);
            % Wilson lines
            \draw[thick, red] (0, 0) ellipse (1.65 and 0.62);
            \draw[thick, red] (-1.25, -0.41) .. controls (-2.4, -0.38) and (-2.9, -0.2) .. (-2.9, 0);
            \fill[white] (-2.19, -0.34) circle[radius = 0.125];
            \draw[green!60!black] (0, 0) ellipse (2.3 and 1.2);
            \fill[white] (-2.19, 0.34) circle[radius = 0.125];
            \draw[thick, red] (-1.25, 0.41) .. controls (-2.4, 0.38) and (-2.9, 0.2) .. (-2.9, 0);
            \draw (2.3, 0) -- (3.35, 0);
            \fill (2.3, 0) circle[radius = 0.1];
            % punctures
            \draw[thick] (3.45, 0.1) -- (3.25, -0.1);
            \draw[thick] (3.45, -0.1) -- (3.25, 0.1);
            % labels
            \node[scale = 0.8, above] at (2.35, 0.55) {$L_p\boxtimes\overline{L}_{\overline{p}}$};
            \draw[->, > = stealth, bend left = 15] (2.6, 0.9) to (2.85, 0.1);
            \node[scale = 0.8, above] at (-1.6, 0.5) {$L_{p_x}\boxtimes\overline{L}_{\overline{p}_x}$};
          \end{tikzpicture}
          \lower2.1ex\hbox{\text{\scalebox{2}[4.5]{$\rangle$}}}. \label{eq: torus wormhole partition function}
        \end{align}
        Recall that the ket $\ket{\:\cdot\:}$ actually represents the tensor product of the chiral and the anti-chiral state, which accounts for the presence of the multiple integral by $p_x,\,\overline{p}_x$.
        We have to simplify the outer boundary state that contains the complicated network of condensable anyons $\mathcal{A}$.
        As mentioned earlier, we interpret the insertion of condensable anyons $\mathcal{A}$ as a superposition of its simple components together with $\rho_0(p_i)$ and $C_{ijk}$ attached to each internal momentum and trivalent junction
        \begin{align}
          \lower2.1ex\hbox{\text{\scalebox{1.6}[4.5]{$|$}}}\,
          \begin{tikzpicture}[scale = 0.4, baseline={([yshift = -0.5ex]current bounding box.center)}]
            \draw[thick] (0, 0) ellipse (3.6 and 2);
            \draw[thick, bend left = 20] (1, 0) to (-1, 0);
            \draw[thick, bend right = 30] (0.78, -0.08) to (-0.78, -0.08);
            % Wilson lines
            \draw[thick, red] (0, 0) ellipse (1.65 and 0.62);
            \draw[thick, red] (-1.25, -0.41) .. controls (-2.4, -0.38) and (-2.9, -0.2) .. (-2.9, 0);
            \fill[white] (-2.19, -0.34) circle[radius = 0.125];
            \draw[green!60!black] (0, 0) ellipse (2.3 and 1.2);
            \fill[white] (-2.19, 0.34) circle[radius = 0.125];
            \draw[thick, red] (-1.25, 0.41) .. controls (-2.4, 0.38) and (-2.9, 0.2) .. (-2.9, 0);
            \draw (2.3, 0) -- (3.35, 0);
            \fill (2.3, 0) circle[radius = 0.1];
            % punctures
            \draw[thick] (3.45, 0.1) -- (3.25, -0.1);
            \draw[thick] (3.45, -0.1) -- (3.25, 0.1);
          \end{tikzpicture}
          \lower2.1ex\hbox{\text{\scalebox{2}[4.5]{$\rangle$}}}=\prod_{i=1}^3\bigg(\int_0^{\infty}dp_i\,\rho_0(p_i)\bigg)C_{123}^2\,\lower2.1ex\hbox{\text{\scalebox{1.6}[4.5]{$|$}}}\,
          \begin{tikzpicture}[scale = 0.4, baseline={([yshift = -0.5ex]current bounding box.center)}]
            \draw[thick] (0, 0) ellipse (3.6 and 2);
            \draw[thick, bend left = 20] (1, 0) to (-1, 0);
            \draw[thick, bend right = 30] (0.78, -0.08) to (-0.78, -0.08);
            % Wilson lines
            \draw (0, 0) ellipse (1.65 and 0.62);
            \draw (-1.25, -0.41) .. controls (-2.4, -0.38) and (-2.9, -0.2) .. (-2.9, 0);
            \fill[white] (-2.19, -0.34) circle[radius = 0.125];
            \draw[green!60!black] (0, 0) ellipse (2.3 and 1.2);
            \fill[white] (-2.19, 0.34) circle[radius = 0.125];
            \draw (-1.25, 0.41) .. controls (-2.4, 0.38) and (-2.9, 0.2) .. (-2.9, 0);
            \draw (2.3, 0) -- (3.35, 0);
            \fill (2.3, 0) circle[radius = 0.1];
            \fill (-1.25, 0.41) circle[radius = 0.1];
            \fill (-1.25, -0.41) circle[radius = 0.1];
            % punctures
            \draw[thick] (3.45, 0.1) -- (3.25, -0.1);
            \draw[thick] (3.45, -0.1) -- (3.25, 0.1);
            % labels
            \node[scale = 0.8, above] at (2.4, 0.8) {$L_{p_1}\boxtimes\overline{L}_{p_1}$};
            \draw[->, > = stealth, bend left = 15] (2.4, 1.1) to (1.65, 0.25);
            \node[scale = 0.8, above] at (-2, 0.6) {$L_{p_3}\boxtimes\overline{L}_{p_3}$};
            \draw[->, > = stealth] (-2, 1) to (-2, 0.45);
            \node[scale = 0.8, below] at (-2, -0.8) {$L_{p_2}\boxtimes\overline{L}_{p_2}$};
            \draw[->, > = stealth] (-1.05, -1.4) ..controls (-0.5, -1) and (-0.4, -0.25) .. (-1.55, 0.03);
          \end{tikzpicture}
          \lower2.1ex\hbox{\text{\scalebox{2}[4.5]{$\rangle$}}}. \label{eq: torus wormhole outer boundary state}
        \end{align}
        The projector formula \eqref{eq: projector formula} enables us to resolve the link on the left side of the genus, alternating $p_i\:(i=2,3)$ integrals to $p_{\alpha}$ integrals as
        \begin{align}
          & \prod_{i\in\qty{2,3}}\bigg(\int_0^{\infty}dp_i\,\rho_0(p_i)\bigg)C_{123}^2\,\lower2.1ex\hbox{\text{\scalebox{1.6}[4.5]{$|$}}}\,
          \begin{tikzpicture}[scale = 0.4, baseline={([yshift = -0.5ex]current bounding box.center)}]
            \draw[thick] (0, 0) ellipse (3.6 and 2);
            \draw[thick, bend left = 20] (1, 0) to (-1, 0);
            \draw[thick, bend right = 30] (0.78, -0.08) to (-0.78, -0.08);
            % Wilson lines
            \draw (0, 0) ellipse (1.65 and 0.62);
            \draw (-1.25, -0.41) .. controls (-2.4, -0.38) and (-2.9, -0.2) .. (-2.9, 0);
            \fill[white] (-2.19, -0.34) circle[radius = 0.125];
            \draw[green!60!black] (0, 0) ellipse (2.3 and 1.2);
            \fill[white] (-2.19, 0.34) circle[radius = 0.125];
            \draw (-1.25, 0.41) .. controls (-2.4, 0.38) and (-2.9, 0.2) .. (-2.9, 0);
            \draw (2.3, 0) -- (3.35, 0);
            \fill (2.3, 0) circle[radius = 0.1];
            \fill (-1.25, 0.41) circle[radius = 0.1];
            \fill (-1.25, -0.41) circle[radius = 0.1];
            % punctures
            \draw[thick] (3.45, 0.1) -- (3.25, -0.1);
            \draw[thick] (3.45, -0.1) -- (3.25, 0.1);
          \end{tikzpicture}
          \lower2.1ex\hbox{\text{\scalebox{2}[4.5]{$\rangle$}}} \nonumber \\
          & \qqquad=\frac{\delta(p_x-\overline{p}_x)}{\rho_0(p_x)}\int_0^{\infty}dp_{\alpha}\,\rho_0(p_{\alpha})C_{1x\alpha}^2\,\lower2.1ex\hbox{\text{\scalebox{1.6}[4.5]{$|$}}}
          \begin{tikzpicture}[scale = 0.4, baseline={([yshift = -0.5ex]current bounding box.center)}]
            \draw[thick] (0, 0) ellipse (3.6 and 2);
            \draw[thick, bend left = 20] (1, 0) to (-1, 0);
            \draw[thick, bend right = 30] (0.78, -0.08) to (-0.78, -0.08);
            % Wilson lines
            \begin{scope}
              \clip (0, 0.65) rectangle (1.7, -0.65);
              \draw (0, 0) ellipse (1.65 and 0.62);
            \end{scope}
            \begin{scope}
              \clip (0, 1.25) rectangle (2.35, -1.25);
              \draw[green!60!black] (0, 0) ellipse (2.3 and 1.2);
            \end{scope}
            \begin{scope}
              \clip (-0.7, 1.25) rectangle (0.02, 0.9);
              \draw[green!60!black] (0, 0.91) ellipse (0.65 and 0.29);
            \end{scope}
            \begin{scope}
              \clip (-0.7, 0.93) rectangle (0.02, 0.6);
              \draw (0, 0.91) ellipse (0.65 and 0.29);
            \end{scope}
            \begin{scope}
              \clip (-0.7, -1.25) rectangle (0.02, -0.9);
              \draw[green!60!black] (0, -0.91) ellipse (0.65 and 0.29);
            \end{scope}
            \begin{scope}
              \clip (-0.7, -0.93) rectangle (0.02, -0.6);
              \draw (0, -0.91) ellipse (0.65 and 0.29);
            \end{scope}
            \begin{scope}
              \clip (-2.5, 0.93) rectangle (-0.64, -0.93);
              \draw (-0.65, 0) ellipse (1.8 and 0.91);
            \end{scope}
            \draw (2.3, 0) -- (3.35, 0);
            \fill (2.3, 0) circle[radius = 0.1];
            \fill (-0.65, 0.91) circle[radius = 0.1];
            \fill (-0.65, -0.91) circle[radius = 0.1];
            % punctures
            \draw[thick] (3.45, 0.1) -- (3.25, -0.1);
            \draw[thick] (3.45, -0.1) -- (3.25, 0.1);
            % labels
            \node[scale = 0.8, above] at (1.2, 0.85) {$L_{p_x}\boxtimes\overline{L}_{p_x}$};
            \node[scale = 0.8, below] at (1.8, -0.85) {$L_{p_x}\boxtimes\overline{L}_{p_x}$};
            \node[scale = 0.8] at (-3.15, 0.3) {$L_{p_{\alpha}}\boxtimes\overline{L}_{p_{\alpha}}$};
            \node[scale = 0.8, below] at (-2.2, -0.9) {$L_{p_1}\boxtimes\overline{L}_{p_1}$};
            \draw[->, > = stealth] (-2.2, -1.1) ..controls (-1.5, -0.3) and (-0.4, -0.2) .. (0.3, -0.5);
            \node[scale = 0.8, above] at (3.5, -0.2)
            {$L_{p}\boxtimes\overline{L}_{\overline{p}}$};
            \draw[->, > = stealth] (-2.2, -1.1) ..controls (-1.5, -0.3) and (-0.4, -0.2) .. (0.3, -0.5);
          \end{tikzpicture}
          \lower2.1ex\hbox{\text{\scalebox{2}[4.5]{$\rangle$}}} \\
          & \qquad=\frac{\delta(p_x-\overline{p}_x)}{\rho_0(p_x)}\int_0^{\infty}dp_{\alpha}\,\frac{\rho_0(p_{\alpha})C_{1x\alpha}^2}{\rho_0(p)C_{pxx}\rho_0(\overline{p})C_{\overline{p}xx}}F_{p_1p}
          \begin{bmatrix}
            p_{\alpha} & p_{\alpha} \\
            p_x & p_x
          \end{bmatrix}
          F_{p_1\overline{p}}
          \begin{bmatrix}
            p_{\alpha} & p_{\alpha} \\
            p_x & p_x
          \end{bmatrix}
          \,\lower2.1ex\hbox{\text{\scalebox{1.6}[4.5]{$|$}}}\,
          \begin{tikzpicture}[scale = 0.4, baseline={([yshift = -0.5ex]current bounding box.center)}]
            \draw[thick] (0, 0) ellipse (3.6 and 2);
            \draw[thick, bend left = 20] (1, 0) to (-1, 0);
            \draw[thick, bend right = 30] (0.78, -0.08) to (-0.78, -0.08);
            % Wilson lines
            \draw (0, 0) ellipse (2.3 and 1.2);
            \draw (2.3, 0) -- (3.35, 0);
            \fill (2.3, 0) circle[radius = 0.1];
            % punctures
            \draw[thick] (3.45, 0.1) -- (3.25, -0.1);
            \draw[thick] (3.45, -0.1) -- (3.25, 0.1);
            % labels
            \node[scale = 0.8, above] at (2.35, 0.55) {$L_p\boxtimes\overline{L}_{\overline{p}}$};
            \draw[->, > = stealth, bend left = 15] (2.6, 0.9) to (2.85, 0.1);
            \node[scale = 0.8] at (-2.1, 0) {$L_{p_{\alpha}}\boxtimes\overline{L}_{p_{\alpha}}$};
          \end{tikzpicture}
          \lower2.1ex\hbox{\text{\scalebox{2}[4.5]{$\rangle$}}}. \nonumber
        \end{align}
        The $p_1$ integrals remain unchanged under this operation.
        It should be noted that in the second line the Wilson line $L_{p_x}\boxtimes\overline{L}_{\overline{p}_x}$ is projected onto the component $L_{p_x}\boxtimes\overline{L}_{p_x}$ of the diagonal condensable anyon \eqref{eq: diagonal Lagrangian algebra object in VTQFT} by virtue of the projectors.
        From the second to the third line, we apply the Wilson triangle identity \eqref{eq: Wilson triangle identity}.
        Using the formula \eqref{eq: version of invertibility} to the $dp_1$ integral, the outer boundary state \eqref{eq: torus wormhole outer boundary state} reads
        \begin{align}
          \lower2.1ex\hbox{\text{\scalebox{1.6}[4.5]{$|$}}}\,
          \begin{tikzpicture}[scale = 0.4, baseline={([yshift = -0.5ex]current bounding box.center)}]
            \draw[thick] (0, 0) ellipse (3.6 and 2);
            \draw[thick, bend left = 20] (1, 0) to (-1, 0);
            \draw[thick, bend right = 30] (0.78, -0.08) to (-0.78, -0.08);
            % Wilson lines
            \draw[thick, red] (0, 0) ellipse (1.65 and 0.62);
            \draw[thick, red] (-1.25, -0.41) .. controls (-2.4, -0.38) and (-2.9, -0.2) .. (-2.9, 0);
            \fill[white] (-2.19, -0.34) circle[radius = 0.125];
            \draw[green!60!black] (0, 0) ellipse (2.3 and 1.2);
            \fill[white] (-2.19, 0.34) circle[radius = 0.125];
            \draw[thick, red] (-1.25, 0.41) .. controls (-2.4, 0.38) and (-2.9, 0.2) .. (-2.9, 0);
            \draw (2.3, 0) -- (3.35, 0);
            \fill (2.3, 0) circle[radius = 0.1];
            % punctures
            \draw[thick] (3.45, 0.1) -- (3.25, -0.1);
            \draw[thick] (3.45, -0.1) -- (3.25, 0.1);
          \end{tikzpicture}
          \lower2.1ex\hbox{\text{\scalebox{2}[4.5]{$\rangle$}}}=\frac{\delta(p_x-\overline{p}_x)\delta(p-\overline{p})}{\rho_0(p_x)\rho_0(p)C_{pxx}}\int_0^{\infty}dp_{\alpha}\,\rho_0(p_{\alpha})C_{\alpha\alpha p}\,\lower2.1ex\hbox{\text{\scalebox{1.6}[4.5]{$|$}}}\,
          \begin{tikzpicture}[scale = 0.4, baseline={([yshift = -0.5ex]current bounding box.center)}]
            \draw[thick] (0, 0) ellipse (3.6 and 2);
            \draw[thick, bend left = 20] (1, 0) to (-1, 0);
            \draw[thick, bend right = 30] (0.78, -0.08) to (-0.78, -0.08);
            % Wilson lines
            \draw (0, 0) ellipse (2.3 and 1.2);
            \draw (2.3, 0) -- (3.35, 0);
            \fill (2.3, 0) circle[radius = 0.1];
            % punctures
            \draw[thick] (3.45, 0.1) -- (3.25, -0.1);
            \draw[thick] (3.45, -0.1) -- (3.25, 0.1);
            % labels
            \node[scale = 0.8, above] at (2.35, 0.55) {$L_p\boxtimes\overline{L}_{\overline{p}}$};
            \draw[->, > = stealth, bend left = 15] (2.6, 0.9) to (2.85, 0.1);
            \node[scale = 0.8] at (-2.1, 0) {$L_{p_{\alpha}}\boxtimes\overline{L}_{\overline{p}_{\alpha}}$};
          \end{tikzpicture}
          \lower2.1ex\hbox{\text{\scalebox{2}[4.5]{$\rangle$}}}.
        \end{align}
        By substituting this result to eq.\,\eqref{eq: torus wormhole partition function}, we finally conclude that the partition function actually factorizes
        \begin{align}
          \lower2.1ex\hbox{\text{\scalebox{1.6}[4.5]{$|$}}}\,
          \begin{tikzpicture}[scale = 0.4, baseline={([yshift = -0.5ex]current bounding box.center)}]
            \draw[thick] (0, 0) ellipse (3.6 and 2);
            \draw[thick, bend left = 20] (1, 0) to (-1, 0);
            \draw[thick, bend right = 30] (0.78, -0.08) to (-0.78, -0.08);
            \draw[dashed] (0, 0) ellipse (1.95 and 0.85);
            \draw[dashed] (0, 0) ellipse (2.75 and 1.35);
            % Wilson lines
            \draw[thick, red] (0, 0) ellipse (1.65 and 0.62);
            \draw[thick, red] (-1.25, 0.41) .. controls (-2.4, 0.38) and (-2.9, 0.2) .. (-2.9, 0);
            \draw[thick, red] (-1.25, -0.41) .. controls (-2.4, -0.38) and (-2.9, -0.2) .. (-2.9, 0);
            \fill[white, rotate around = {-15:(-2.555, -0.26)}] (-2.63, -0.21) rectangle (-2.48, -0.31);
            \fill[white, rotate around = {-10:(-2.245, -0.33)}] (-2.32, -0.28) rectangle (-2.17, -0.38);
            \fill[white, rotate around = {-6:(-1.945, -0.37)}] (-2.02, -0.32) rectangle (-1.87, -0.42);
            \draw (2.5, 0) -- (3.35, 0);
            % punctures
            \draw[thick] (2.6, 0.1) -- (2.4, -0.1);
            \draw[thick] (2.6, -0.1) -- (2.4, 0.1);
            \draw[thick] (3.45, 0.1) -- (3.25, -0.1);
            \draw[thick] (3.45, -0.1) -- (3.25, 0.1);
            % labels
            \node[scale = 0.8, above] at (2, 0.55) {$L_p\boxtimes\overline{L}_{\overline{p}}$};
            \draw[->, > = stealth, bend left = 20] (2.75, 0.7) to (3, 0.1);
          \end{tikzpicture}
          \lower2.1ex\hbox{\text{\scalebox{2}[4.5]{$\rangle$}}} & =\frac{\delta(p-\overline{p})}{\rho_0(p)}\int_0^{\infty}dp_xdp_{\alpha}\,\rho_0(p_x)\rho_0(p_{\alpha})C_{xxp}C_{\alpha\alpha p} \nonumber \\
          & \qqquad\times\lower2.1ex\hbox{\text{\scalebox{1.6}[4.5]{$|$}}}\,
          \begin{tikzpicture}[scale = 0.4, baseline={([yshift = -0.5ex]current bounding box.center)}]
            \draw[thick] (0, 0) ellipse (3.6 and 2);
            \draw[thick, bend left = 20] (1, 0) to (-1, 0);
            \draw[thick, bend right = 30] (0.78, -0.08) to (-0.78, -0.08);
            % Wilson lines
            \draw[green!60!black] (0, 0) ellipse (2.3 and 1.2);
            \draw (2.3, 0) -- (3.35, 0);
            \fill (2.3, 0) circle[radius = 0.1];
            % punctures
            \draw[thick] (3.45, 0.1) -- (3.25, -0.1);
            \draw[thick] (3.45, -0.1) -- (3.25, 0.1);
            % labels
            \node[scale = 0.8, above] at (2.35, 0.55) {$L_p\boxtimes\overline{L}_{\overline{p}}$};
            \draw[->, > = stealth, bend left = 15] (2.6, 0.9) to (2.85, 0.1);
            \node[scale = 0.8] at (-2.1, 0) {$L_{p_x}\boxtimes\overline{L}_{\overline{p}_x}$};
          \end{tikzpicture}
          \lower2.1ex\hbox{\text{\scalebox{2}[4.5]{$\rangle$}}}\otimes\lower2.1ex\hbox{\text{\scalebox{1.6}[4.5]{$|$}}}\,
          \begin{tikzpicture}[scale = 0.4, baseline={([yshift = -0.5ex]current bounding box.center)}]
            \draw[thick] (0, 0) ellipse (3.6 and 2);
            \draw[thick, bend left = 20] (1, 0) to (-1, 0);
            \draw[thick, bend right = 30] (0.78, -0.08) to (-0.78, -0.08);
            % Wilson lines
            \draw (0, 0) ellipse (2.3 and 1.2);
            \draw (2.3, 0) -- (3.35, 0);
            \fill (2.3, 0) circle[radius = 0.1];
            % punctures
            \draw[thick] (3.45, 0.1) -- (3.25, -0.1);
            \draw[thick] (3.45, -0.1) -- (3.25, 0.1);
            % labels
            \node[scale = 0.8, above] at (2.35, 0.55) {$L_p\boxtimes\overline{L}_{\overline{p}}$};
            \draw[->, > = stealth, bend left = 15] (2.6, 0.9) to (2.85, 0.1);
            \node[scale = 0.8] at (-2.1, 0) {$L_{p_{\alpha}}\boxtimes\overline{L}_{\overline{p}_{\alpha}}$};
          \end{tikzpicture}
          \lower2.1ex\hbox{\text{\scalebox{2}[4.5]{$\rangle$}}} \nonumber \\
          & = \frac{\delta(p-\overline{p})}{\rho_0(p)}\,\lower2.1ex\hbox{\text{\scalebox{1.6}[4.5]{$|$}}}\,
          \begin{tikzpicture}[scale = 0.4, baseline={([yshift = -0.5ex]current bounding box.center)}]
            \draw[thick] (0, 0) ellipse (3.6 and 2);
            \draw[thick, bend left = 20] (1, 0) to (-1, 0);
            \draw[thick, bend right = 30] (0.78, -0.08) to (-0.78, -0.08);
            % Wilson lines
            \draw[thick, red] (0, 0) ellipse (2.3 and 1.2);
            \draw (2.3, 0) -- (3.35, 0);
            \fill (2.3, 0) circle[radius = 0.1];
            % punctures
            \draw[thick] (3.45, 0.1) -- (3.25, -0.1);
            \draw[thick] (3.45, -0.1) -- (3.25, 0.1);
            % labels
            \node[scale = 0.8, above] at (2.35, 0.55) {$L_p\boxtimes\overline{L}_{\overline{p}}$};
            \draw[->, > = stealth, bend left = 15] (2.6, 0.9) to (2.85, 0.1);
            \node[scale = 0.8] at (-2.8, 0) {$\mathcal{A}$};
          \end{tikzpicture}
          \lower2.1ex\hbox{\text{\scalebox{2}[4.5]{$\rangle$}}}\,\otimes\lower2.1ex\hbox{\text{\scalebox{1.6}[4.5]{$|$}}}\,
          \begin{tikzpicture}[scale = 0.4, baseline={([yshift = -0.5ex]current bounding box.center)}]
            \draw[thick] (0, 0) ellipse (3.6 and 2);
            \draw[thick, bend left = 20] (1, 0) to (-1, 0);
            \draw[thick, bend right = 30] (0.78, -0.08) to (-0.78, -0.08);
            % Wilson lines
            \draw[thick, red] (0, 0) ellipse (2.3 and 1.2);
            \draw (2.3, 0) -- (3.35, 0);
            \fill (2.3, 0) circle[radius = 0.1];
            % punctures
            \draw[thick] (3.45, 0.1) -- (3.25, -0.1);
            \draw[thick] (3.45, -0.1) -- (3.25, 0.1);
            % labels
            \node[scale = 0.8, above] at (2.35, 0.55) {$L_p\boxtimes\overline{L}_{\overline{p}}$};
            \draw[->, > = stealth, bend left = 15] (2.6, 0.9) to (2.85, 0.1);
            \node[scale = 0.8] at (-2.8, 0) {$\mathcal{A}$};
          \end{tikzpicture}
          \lower2.1ex\hbox{\text{\scalebox{2}[4.5]{$\rangle$}}}. \label{eq: torus whormhole factorization}
        \end{align}
        As repeated reminder, the insertion of condensable anyons is a superposition of the simple components with $\rho_0(p_i)$ and $C_{ijk}$ assigned to each internal momentum and trivalent junction.
        \par A few comments are in order on the last equation.
        First of all, the prefactor $\delta(p-\overline{p})$ functions as a constraint that enforces diagonality on the Wilson line $L_p\boxtimes\overline{L}_{\overline{p}}$.
        Physical interpretation is quite clear as $L_p\boxtimes\overline{L}_{\overline{p}}$ touches the condensable anyon $\mathcal{A}$, which serves as a topological boundary so that only the diagonal line $L_p\boxtimes\overline{L}_p$ can terminate on it, as explained in Section \ref{sec: Anyon condensation in modular tensor category}.
        Eq.\,\eqref{eq: torus whormhole factorization} as a whole represents the factorization of the state assigned to the wormhole by path-integral, and taking overlap with moduli coordinate basis $\ket{\tau_1}$, $\ket{\tau_2}$ of the two boundaries as in eq.\,\eqref{eq: overlap of moduli and conformal block} produces the factorization \textit{at the level of the partition function}:
        \begin{align}
          \big\langle\tau_1,\tau_2\big|\Sigma_{1,1}\times[0,1],\mathcal{A}\big\rangle=\frac{\delta(p-\overline{p})}{\rho_0(p)}\big\langle\tau_1\big|\Sigma_{1,1},\mathcal{A}\big\rangle\big\langle\tau_2\big|\Sigma_{1,1},\mathcal{A}\big\rangle.
        \end{align}
        Here the ket states are written in abbreviated form, where $\mathcal{A}$ signifies the insertion of the diagonal condensable anyon on each state.
        The sandwich construction identifies $\langle\tau_i\big|\Sigma_{1,1},\mathcal{A}\big\rangle$ ($i=1,2$) with the Liouville CFT one-point function on the torus, if we regard $\mathcal{A}$ as a topological boundary and the torus boundary as a physical boundary.
        Since the diagonal condensable anyon dos not contain the identity $\mathbbm{1}\boxtimes\mathbbm{1}$, the resulting theory will not possess the vacuum, which is consistent with the fact that the vacuum is unnormalizable in Liouville theory.
        It also merits emphasis that the simultaneous limit $p,\overline{p}\to\mathbbm{1}$ is ill-behaved since there appear multiple divergent ``constants" $\delta(0)$, $\int_{0}^{\infty}dp=\infty$ due to the relation \eqref{eq: structure constant limit} even when we ignore the factor $\displaystyle\frac{\delta(p-\overline{p})}{\rho_0(p)}$:
        \begin{align}
          \lim_{p,\overline{p}\to\mathbbm{1}}\lower2.1ex\hbox{\text{\scalebox{1.6}[4.5]{$|$}}}\,
          \begin{tikzpicture}[scale = 0.4, baseline={([yshift = -0.5ex]current bounding box.center)}]
            \draw[thick] (0, 0) ellipse (3.6 and 2);
            \draw[thick, bend left = 20] (1, 0) to (-1, 0);
            \draw[thick, bend right = 30] (0.78, -0.08) to (-0.78, -0.08);
            \draw[dashed] (0, 0) ellipse (1.95 and 0.85);
            \draw[dashed] (0, 0) ellipse (2.75 and 1.35);
            % Wilson lines
            \draw[thick, red] (0, 0) ellipse (1.65 and 0.62);
            \draw[thick, red] (-1.25, 0.41) .. controls (-2.4, 0.38) and (-2.9, 0.2) .. (-2.9, 0);
            \draw[thick, red] (-1.25, -0.41) .. controls (-2.4, -0.38) and (-2.9, -0.2) .. (-2.9, 0);
            \fill[white, rotate around = {-15:(-2.555, -0.26)}] (-2.63, -0.21) rectangle (-2.48, -0.31);
            \fill[white, rotate around = {-10:(-2.245, -0.33)}] (-2.32, -0.28) rectangle (-2.17, -0.38);
            \fill[white, rotate around = {-6:(-1.945, -0.37)}] (-2.02, -0.32) rectangle (-1.87, -0.42);
            \draw (2.5, 0) -- (3.35, 0);
            % punctures
            \draw[thick] (2.6, 0.1) -- (2.4, -0.1);
            \draw[thick] (2.6, -0.1) -- (2.4, 0.1);
            \draw[thick] (3.45, 0.1) -- (3.25, -0.1);
            \draw[thick] (3.45, -0.1) -- (3.25, 0.1);
            % labels
            \node[scale = 0.8, above] at (2, 0.55) {$L_p\boxtimes\overline{L}_{\overline{p}}$};
            \draw[->, > = stealth, bend left = 20] (2.75, 0.7) to (3, 0.1);
          \end{tikzpicture}
          \lower2.1ex\hbox{\text{\scalebox{2}[4.5]{$\rangle$}}} & \propto\int_0^{\infty}dp_xdp_{\alpha}\,\adunderbrace{\delta(p_x-p_x)}{=\delta(0)}\adunderbrace{\delta(p_{\alpha}-p_{\alpha})}{=\delta(0)} \nonumber \\
          & \qqquad\times\lower2.1ex\hbox{\text{\scalebox{1.6}[4.5]{$|$}}}\,
          \begin{tikzpicture}[scale = 0.4, baseline={([yshift = -0.5ex]current bounding box.center)}]
            \draw[thick] (0, 0) ellipse (3.6 and 2);
            \draw[thick, bend left = 20] (1, 0) to (-1, 0);
            \draw[thick, bend right = 30] (0.78, -0.08) to (-0.78, -0.08);
            % Wilson lines
            \draw[green!60!black] (0, 0) ellipse (2.3 and 1.2);
            % labels
            \node[scale = 0.8] at (-2.1, 0) {$L_{p_x}\boxtimes\overline{L}_{\overline{p}_x}$};
          \end{tikzpicture}
          \lower2.1ex\hbox{\text{\scalebox{2}[4.5]{$\rangle$}}}\otimes\lower2.1ex\hbox{\text{\scalebox{1.6}[4.5]{$|$}}}\,
          \begin{tikzpicture}[scale = 0.4, baseline={([yshift = -0.5ex]current bounding box.center)}]
            \draw[thick] (0, 0) ellipse (3.6 and 2);
            \draw[thick, bend left = 20] (1, 0) to (-1, 0);
            \draw[thick, bend right = 30] (0.78, -0.08) to (-0.78, -0.08);
            % Wilson lines
            \draw (0, 0) ellipse (2.3 and 1.2);
            % labels
            \node[scale = 0.8] at (-2.1, 0) {$L_{p_{\alpha}}\boxtimes\overline{L}_{\overline{p}_{\alpha}}$};
          \end{tikzpicture}
          \lower2.1ex\hbox{\text{\scalebox{2}[4.5]{$\rangle$}}} \nonumber \\
          & \not\propto\lower2.1ex\hbox{\text{\scalebox{1.6}[4.5]{$|$}}}\,
          \begin{tikzpicture}[scale = 0.4, baseline={([yshift = -0.5ex]current bounding box.center)}]
            \draw[thick] (0, 0) ellipse (3.6 and 2);
            \draw[thick, bend left = 20] (1, 0) to (-1, 0);
            \draw[thick, bend right = 30] (0.78, -0.08) to (-0.78, -0.08);
            % Wilson lines
            \draw[thick, red] (0, 0) ellipse (2.3 and 1.2);
            % labels
            \node[scale = 0.8] at (-2.8, 0) {$\mathcal{A}$};
          \end{tikzpicture}
          \lower2.1ex\hbox{\text{\scalebox{2}[4.5]{$\rangle$}}}\,\otimes\lower2.1ex\hbox{\text{\scalebox{1.6}[4.5]{$|$}}}\,
          \begin{tikzpicture}[scale = 0.4, baseline={([yshift = -0.5ex]current bounding box.center)}]
            \draw[thick] (0, 0) ellipse (3.6 and 2);
            \draw[thick, bend left = 20] (1, 0) to (-1, 0);
            \draw[thick, bend right = 30] (0.78, -0.08) to (-0.78, -0.08);
            % Wilson lines
            \draw[thick, red] (0, 0) ellipse (2.3 and 1.2);
            % labels
            \node[scale = 0.8] at (-2.8, 0) {$\mathcal{A}$};
          \end{tikzpicture}
          \lower2.1ex\hbox{\text{\scalebox{2}[4.5]{$\rangle$}}}.
          \label{eq: factorization mismatch in torus wormhole}
        \end{align}
        We further observe that, even after discarding $\delta(0)$, the second line of eq.\,\eqref{eq: factorization mismatch in torus wormhole} fails to be proportional to the third line, as the integrals fall short of the factors $\rho_0(p_x)$, $\rho_0(p_{\alpha})$.\footnote{
            More precisely, the second line of eq.\,\eqref{eq: factorization mismatch in torus wormhole} is factorizing up to an overall divergent constant but involves integrating the scalar Virasoro characters \textit{with a flat measure} regarding the Liouville momentum.
        }
        This is consistent with the absence of a systematic treatment of non-hyperbolic geometries such as $T^2\times[0,1]$ in VTQFT, and with the fact that Liouville CFT is not well-defined on the unpunctured torus.

      \subsubsection{Genus two wormhole}
      The wormhole geometry $\Sigma_{2,0}\times\qty[0,1]$ can be represented as a handlebody $S\Sigma_{2,0}$ with a smaller handlebody $S\Sigma_{2,0}^{\prime}$ curved out from its interior.
      We first consider $\Sigma_{2,0}\times\qty[0,1]$ with a Wilson line (matter field) insertion as a probe instead of the pure $\Sigma_{2,0}\times\qty[0,1]$ even though $\Sigma_{2,0}\times\qty[0,1]$ is hyperbolic.
      This is because there emerges multiple divergent ``constants" $\delta(0)$ if starting from $\Sigma_{2,0}\times[0,1]$ without any boundary connecting Wilson line.
      However, we can take sensible $p,\overline{p}\to\mathbbm{1}$ limit once we arrive at the final result given in eq.\,\eqref{eq: genus two with Wilson line wormhole factorization}, where we can bypass the divergence and find the favorable result.
      The Wilson line insertion and the fine mesh of $\Sigma_{2,0}\times\qty[0,1]$ are illustrated in Figure \ref{fig: genus 2  wormhole}.
      \begin{figure}[t]
        \centering
        {
          \begin{tikzpicture}[scale = 0.7]
            \draw[thick] (0, 1.35) .. controls (-0.8, 1.35) and (-1.5, 2) .. (-2.5, 2) .. controls (-4, 2) and (-4.7, 0.8) .. (-4.7, 0) .. controls (-4.7, -0.8) and (-4, -2) .. (-2.5, -2) .. controls (-1.5, -2) and (-0.8, -1.35) .. (0, -1.35) .. controls (0.8, -1.35) and (1.5, -2) .. (2.5, -2) .. controls (4, -2) and (4.7, -0.8) .. (4.7, 0) .. controls (4.7, 0.8) and (4, 2) .. (2.5, 2) .. controls (1.5, 2) and (0.8, 1.35) .. (0, 1.35);
            % 2 genus
            \draw[thick, bend right = 30] (-3.1, 0) to (-1.7, 0);
            \draw[thick, bend left = 30] (-3, -0.05) to (-1.8, -0.05);
            \draw[thick, bend left = 30] (3.1, 0) to (1.7, 0);
            \draw[thick, bend right = 30] (3, -0.05) to (1.8, -0.05);

            %inner boundary
            \draw[thick, dashed] (0, 0.75) .. controls (-0.8, 0.75) and (-1.5, 1.45) .. (-2.5, 1.45) .. controls (-3.7, 1.45) and (-4.25, 0.5) .. (-4.25, 0) .. controls (-4.25, -0.5) and (-3.7, -1.45) .. (-2.5, -1.45) .. controls (-1.5, -1.45) and (-0.8, -0.75) .. (0, -0.75) .. controls (0.8, -0.75) and (1.5, -1.45) .. (2.5, -1.45) .. controls (3.7, -1.45) and (4.25, -0.5) .. (4.25, 0) .. controls (4.25, 0.5) and (3.7, 1.45) .. (2.5, 1.45) .. controls (1.5, 1.45) and (0.8, 0.75) .. (0, 0.75);
            \draw[thick, dashed] (-2.4, 0) ellipse (1.3 and 0.9);
            \draw[thick, dashed] (2.4, 0) ellipse (1.3 and 0.9);
            % finemesh
            \draw[thick, red] (-2.4, 0) ellipse (1.1 and 0.65);
            \draw[thick, red] (2.4, 0) ellipse (1.1 and 0.65);
            \draw[thick, red] (-1.3, 0) to (-1.1, 0);
            \draw[dashed, thick, red] (-1.1, 0) to (1.1, 0);
            \draw[thick, red] (1.1, 0) to (1.3, 0);
            \draw[thick, red] (-3.25, 0.41) .. controls (-4, 0.38) and (-4.4, 0.1) .. (-4.4, 0);
            \draw[thick, red] (-4.4, 0) .. controls (-4.4, -0.07) and (-4.25, -0.14) .. (-4.23, -0.155);
            \draw[dashed, thick, red] (-4.23, -0.155) .. controls (-3.95, -0.315) and (-3.7, -0.345) .. (-3.58, -0.367);
            \draw[thick, red, bend left = 8] (-3.58, -0.367) .. controls (-3.51, -0.39) and (-3.28, -0.4) .. (-3.25, -0.41);
            \draw[thick, red] (3.25, 0.41) .. controls (4, 0.38) and (4.4, 0.1) .. (4.4, 0);
            \draw[thick, red] (4.4, 0) .. controls (4.4, -0.07) and (4.25, -0.14) .. (4.23, -0.155);
            \draw[dashed, thick, red] (4.23, -0.155) .. controls (3.95, -0.315) and (3.7, -0.345) .. (3.58, -0.367);
            \draw[thick, red, bend left = 8] (3.58, -0.367) .. controls (3.51, -0.39) and (3.28, -0.4) .. (3.25, -0.41);
            % Wilson line
            \draw (0, 0.3) to (0, 1.1);
            % punctures
            \draw[thick] (-0.1, 0.4) to (0.1, 0.2);
            \draw[thick] (-0.1, 0.2) to (0.1, 0.4);
            \draw[thick] (-0.1, 1.2) to (0.1, 1);
            \draw[thick] (-0.1, 1) to (0.1, 1.2);
            % labels
            \node[scale = 0.8, right] at (0, 0.7) {$L_p\boxtimes\overline{L}_{\overline{p}}$};
            \node[scale = 0.8, above] at (-2.4, 0.6) {$\mathcal{A}$};
            \node[scale = 0.8, above] at (2.4, 0.6) {$\mathcal{A}$};
            \node[scale = 0.8, above] at (-4.4, 0.2) {$\mathcal{A}$};
            \node[scale = 0.8, above] at (4.4, 0.2) {$\mathcal{A}$};
            \node[scale = 0.8, right] at (-3.55, 0) {$\mathcal{A}$};
            \node[scale = 0.8, left] at (3.55, 0) {$\mathcal{A}$};
          \end{tikzpicture}
        }
        \caption{$\Sigma_{2,0}\times\qty[0,1]$ wormhole with $\mathcal{A}$ inserted along its fine mesh. The dashed lines are the inner boundary $\Sigma_{2,0}$.}
        \label{fig: genus 2  wormhole}
      \end{figure}
      The path-integral rule on a compression body \eqref{eq: compression body path-integral} prepares a state
      \begin{align}
         & \lower2.1ex\hbox{\text{\scalebox{1.6}[4.5]{$|$}}}\,
        \begin{tikzpicture}[scale = 0.4, baseline = {([yshift=-.5ex]current bounding box.center)}]
          \draw[thick] (0, 1.35) .. controls (-0.8, 1.35) and (-1.5, 2) .. (-2.5, 2) .. controls (-4, 2) and (-4.7, 0.8) .. (-4.7, 0) .. controls (-4.7, -0.8) and (-4, -2) .. (-2.5, -2) .. controls (-1.5, -2) and (-0.8, -1.35) .. (0, -1.35) .. controls (0.8, -1.35) and (1.5, -2) .. (2.5, -2) .. controls (4, -2) and (4.7, -0.8) .. (4.7, 0) .. controls (4.7, 0.8) and (4, 2) .. (2.5, 2) .. controls (1.5, 2) and (0.8, 1.35) .. (0, 1.35);
          % 2 genus
          \draw[thick, bend right = 30] (-3.1, 0) to (-1.7, 0);
          \draw[thick, bend left = 30] (-3, -0.05) to (-1.8, -0.05);
          \draw[thick, bend left = 30] (3.1, 0) to (1.7, 0);
          \draw[thick, bend right = 30] (3, -0.05) to (1.8, -0.05);
          %inner boundary
          \draw[thick, dashed] (0, 0.75) .. controls (-0.8, 0.75) and (-1.5, 1.45) .. (-2.5, 1.45) .. controls (-3.7, 1.45) and (-4.25, 0.5) .. (-4.25, 0) .. controls (-4.25, -0.5) and (-3.7, -1.45) .. (-2.5, -1.45) .. controls (-1.5, -1.45) and (-0.8, -0.75) .. (0, -0.75) .. controls (0.8, -0.75) and (1.5, -1.45) .. (2.5, -1.45) .. controls (3.7, -1.45) and (4.25, -0.5) .. (4.25, 0) .. controls (4.25, 0.5) and (3.7, 1.45) .. (2.5, 1.45) .. controls (1.5, 1.45) and (0.8, 0.75) .. (0, 0.75);
          \draw[thick, dashed] (-2.4, 0) ellipse (1.3 and 0.9);
          \draw[thick, dashed] (2.4, 0) ellipse (1.3 and 0.9);
          % finemesh
          \draw[thick, red] (-2.4, 0) ellipse (1.1 and 0.65);
          \draw[thick, red] (2.4, 0) ellipse (1.1 and 0.65);
          \draw[thick, red] (-1.3, 0) to (-1.1, 0);
          \draw[dashed, thick, red] (-1.1, 0) to (1.1, 0);
          \draw[thick, red] (1.1, 0) to (1.3, 0);
          \draw[thick, red] (-3.25, 0.41) .. controls (-4, 0.38) and (-4.4, 0.1) .. (-4.4, 0);
          \draw[thick, red] (-4.4, 0) .. controls (-4.4, -0.07) and (-4.25, -0.14) .. (-4.23, -0.155);
          \draw[dashed, thick, red] (-4.23, -0.155) .. controls (-3.95, -0.315) and (-3.7, -0.345) .. (-3.58, -0.367);
          \draw[thick, red, bend left = 8] (-3.58, -0.367) .. controls (-3.51, -0.39) and (-3.28, -0.4) .. (-3.25, -0.41);
          \draw[thick, red] (3.25, 0.41) .. controls (4, 0.38) and (4.4, 0.1) .. (4.4, 0);
          \draw[thick, red] (4.4, 0) .. controls (4.4, -0.07) and (4.25, -0.14) .. (4.23, -0.155);
          \draw[dashed, thick, red] (4.23, -0.155) .. controls (3.95, -0.315) and (3.7, -0.345) .. (3.58, -0.367);
          \draw[thick, red, bend left = 8] (3.58, -0.367) .. controls (3.51, -0.39) and (3.28, -0.4) .. (3.25, -0.41);
          % Wilson line
          \draw (0, 0.3) to (0, 1.1);
          % punctures
          \draw[thick] (-0.1, 0.4) to (0.1, 0.2);
          \draw[thick] (-0.1, 0.2) to (0.1, 0.4);
          \draw[thick] (-0.1, 1.2) to (0.1, 1);
          \draw[thick] (-0.1, 1) to (0.1, 1.2);
        \end{tikzpicture}
        \lower2.1ex\hbox{\text{\scalebox{2}[4.5]{$\rangle$}}}\,=\int_{0}^{\infty}dp_xd\overline{p}_xdp_yd\overline{p}_ydp_zd\overline{p}_zdp_wd\overline{p}_w\,\rho_0(p_x)\rho_0(\overline{p}_x)\rho_0(p_y)\rho_0(\overline{p}_y) \nonumber               \\
         & \qqqquad\qquad\times\rho_0(p_z)\rho_0(\overline{p}_z)\rho_0(p_w)\rho_0(\overline{p}_w)C_{xxz}C_{\overline{x}\overline{x}\overline{z}}C_{yyw}C_{\overline{y}\overline{y}\overline{zw}}C_{pzw}C_{\overline{p}\overline{z}\overline{w}}  \nonumber \\
         & \qqquad\qqquad\times\lower2.1ex\hbox{\text{\scalebox{1.6}[4.5]{$|$}}}\,
        \begin{tikzpicture}[scale = 0.4, baseline = {([yshift=-.5ex]current bounding box.center)}]
          \draw[thick] (0, 1.35) .. controls (-0.8, 1.35) and (-1.5, 2) .. (-2.5, 2) .. controls (-4, 2) and (-4.7, 0.8) .. (-4.7, 0) .. controls (-4.7, -0.8) and (-4, -2) .. (-2.5, -2) .. controls (-1.5, -2) and (-0.8, -1.35) .. (0, -1.35) .. controls (0.8, -1.35) and (1.5, -2) .. (2.5, -2) .. controls (4, -2) and (4.7, -0.8) .. (4.7, 0) .. controls (4.7, 0.8) and (4, 2) .. (2.5, 2) .. controls (1.5, 2) and (0.8, 1.35) .. (0, 1.35);
          % 2 genus
          \draw[thick, bend right = 30] (-3.1, 0) to (-1.7, 0);
          \draw[thick, bend left = 30] (-3, -0.05) to (-1.8, -0.05);
          \draw[thick, bend left = 30] (3.1, 0) to (1.7, 0);
          \draw[thick, bend right = 30] (3, -0.05) to (1.8, -0.05);
          % complete set
          \draw[green!60!black] (-2.4, 0) ellipse (1.57 and 1.1);
          \draw[green!60!black] (2.4, 0) ellipse (1.57 and 1.1);
          \draw[green!60!black] (-0.83, 0) to (0.83, 0);
          % Wilson line
          \draw (0, 0) to (0, 1.1);
          \fill[green!60!black] (-0.83, 0) circle[radius = 0.1];
          \fill[green!60!black] (0.83, 0) circle[radius = 0.1];
          \fill (0, 0) circle[radius = 0.1];
          % punctures
          \draw[thick] (-0.1, 1.2) to (0.1, 1);
          \draw[thick] (-0.1, 1) to (0.1, 1.2);
          % labels
          \node[scale = 0.8, above] at (-2.4, 0.8) {$L_{p_x}\boxtimes\overline{L}_{\overline{p}_x}$};
          \node[scale = 0.8, above] at (2.4, 0.8) {$L_{p_y}\boxtimes\overline{L}_{\overline{p}_y}$};
          \node[scale = 0.8, below] at (-1.5, -0.8) {$L_{p_z}\boxtimes\overline{L}_{\overline{p}_z}$};
          \draw[->, > = stealth, bend right = 15] (-0.6, -1) to (-0.4, -0.1);
          \node[scale = 0.8, below] at (2, -0.8) {$L_{p_w}\boxtimes\overline{L}_{\overline{p}_w}$};
          \draw[->, > = stealth, bend left = 15] (0.6, -1) to (0.4, -0.1);
        \end{tikzpicture}
        \lower2.1ex\hbox{\text{\scalebox{2}[4.5]{$\rangle$}}}\otimes\,\lower2.1ex\hbox{\text{\scalebox{1.6}[4.5]{$|$}}}\,
        \begin{tikzpicture}[scale = 0.4, baseline = {([yshift=-.5ex]current bounding box.center)}]
          \draw[thick] (0, 1.35) .. controls (-0.8, 1.35) and (-1.5, 2) .. (-2.5, 2) .. controls (-4, 2) and (-4.7, 0.8) .. (-4.7, 0) .. controls (-4.7, -0.8) and (-4, -2) .. (-2.5, -2) .. controls (-1.5, -2) and (-0.8, -1.35) .. (0, -1.35) .. controls (0.8, -1.35) and (1.5, -2) .. (2.5, -2) .. controls (4, -2) and (4.7, -0.8) .. (4.7, 0) .. controls (4.7, 0.8) and (4, 2) .. (2.5, 2) .. controls (1.5, 2) and (0.8, 1.35) .. (0, 1.35);
          % 2 genus
          \draw[thick, bend right = 30] (-3.1, 0) to (-1.7, 0);
          \draw[thick, bend left = 30] (-3, -0.05) to (-1.8, -0.05);
          \draw[thick, bend left = 30] (3.1, 0) to (1.7, 0);
          \draw[thick, bend right = 30] (3, -0.05) to (1.8, -0.05);
          % fine mesh
          \draw[thick, red] (-4.4, 0) .. controls (-4.4, -0.2) and (-4, -0.4) .. (-3.6, -0.4);
          \fill[white] (-4.13, -0.305) circle[radius = 0.125];
          \draw[thick, red] (4.4, 0) .. controls (4.4, -0.2) and (4, -0.4) .. (3.6, -0.4);
          \fill[white] (4.13, -0.305) circle[radius = 0.125];
          % complete set
          \draw[green!60!black] (-2.4, 0) ellipse (1.75 and 1.35);
          \fill[white] (-4.13, 0.305) circle[radius = 0.125];
          \draw[green!60!black] (2.4, 0) ellipse (1.75 and 1.35);
          \fill[white] (4.13, 0.305) circle[radius = 0.125];
          \draw[green!60!black] (-0.71, 0.4) to (0.71, 0.4);
          \draw (0, 0.4) to (0, 1.1);
          \fill[green!60!black] (-0.71, 0.4) circle[radius = 0.1];
          \fill[green!60!black] (0.71, 0.4) circle[radius = 0.1];
          \fill (0, 0.4) circle[radius = 0.1];
          % punctures
          \draw[thick] (-0.1, 1.2) to (0.1, 1);
          \draw[thick] (-0.1, 1) to (0.1, 1.2);
          % fine mesh
          \draw[thick, red] (-2.4, 0) ellipse (1.35 and 0.9);
          \draw[thick, red] (2.4, 0) ellipse (1.35 and 0.9);
          \draw[thick, red] (-1.05, 0) to (-0.775, 0);
          \draw[thick, red] (-0.525, 0) to (0.525, 0);
          \draw[thick, red] (0.775, 0) to (1.05, 0);
          \draw[thick, red] (-3.6, 0.4) .. controls (-4, 0.4) and (-4.4, 0.2) .. (-4.4, 0);
          \draw[thick, red] (3.6, 0.4) .. controls (4, 0.4) and (4.4, 0.2) .. (4.4, 0);
          % labels
          \node[scale = 0.8, above] at (-3.5, 1.6) {$L_{p_x}\boxtimes\overline{L}_{\overline{p}_x}$};
          \draw[->, > = stealth, bend left = 60] (-1.8, 2.3) to (-1.2, 1.1);
          \node[scale = 0.8, above] at (3.5, 1.6) {$L_{p_y}\boxtimes\overline{L}_{\overline{p}_y}$};
          \draw[->, > = stealth, bend right = 60] (1.8, 2.3) to (1.2, 1.1);
          \node[scale = 0.8, below] at (-1.5, -1.2) {$L_{p_z}\boxtimes\overline{L}_{\overline{p}_z}$};
          \draw[->, > = stealth, bend right = 20] (-1, -1.4) to (-0.4, 0.3);
          \node[scale = 0.8, below] at (2, -1.2) {$L_{p_w}\boxtimes\overline{L}_{\overline{p}_w}$};
          \draw[->, > = stealth, bend left = 20] (1, -1.4) to (0.4, 0.3);
        \end{tikzpicture}
        \lower2.1ex\hbox{\text{\scalebox{2}[4.5]{$\rangle$}}}. \label{eq: VTQFT path-integral for the wormhole with the condensable anyon}
      \end{align}
      The six-fold integral by $p_x,\,\overline{p}_x,\,\cdots,\,p_z,\,\overline{p}_z$ is due to the presence of both the chiral and the anti-chiral part.
      The subtleties arise from the outer boundary state with the network of condensable anyons $\mathcal{A}$ linking with the complete set of basis (the green lines) in non-trivial ways.
      The superposition rule decompose the outer boundary state into the nine-fold integral over each simple component of the diagonal condensable anyon
      \begin{align}
        \prod_{i=1}^9\bigg(\int_{0}^{\infty}dp_i\,\rho_0(p_i)\bigg)C_{134}C_{234}C_{129}C_{569}C_{578}C_{678}\lower2.1ex\hbox{\text{\scalebox{1.6}[4.5]{$|$}}}\,
        \begin{tikzpicture}[scale = 0.4, baseline = {([yshift=-.5ex]current bounding box.center)}]
          \draw[thick] (0, 1.35) .. controls (-0.8, 1.35) and (-1.5, 2) .. (-2.5, 2) .. controls (-4, 2) and (-4.7, 0.8) .. (-4.7, 0) .. controls (-4.7, -0.8) and (-4, -2) .. (-2.5, -2) .. controls (-1.5, -2) and (-0.8, -1.35) .. (0, -1.35) .. controls (0.8, -1.35) and (1.5, -2) .. (2.5, -2) .. controls (4, -2) and (4.7, -0.8) .. (4.7, 0) .. controls (4.7, 0.8) and (4, 2) .. (2.5, 2) .. controls (1.5, 2) and (0.8, 1.35) .. (0, 1.35);
          % 2 genus
          \draw[thick, bend right = 30] (-3.1, 0) to (-1.7, 0);
          \draw[thick, bend left = 30] (-3, -0.05) to (-1.8, -0.05);
          \draw[thick, bend left = 30] (3.1, 0) to (1.7, 0);
          \draw[thick, bend right = 30] (3, -0.05) to (1.8, -0.05);
          % fine mesh
          \draw (-4.4, 0) .. controls (-4.4, -0.2) and (-4, -0.4) .. (-3.6, -0.4);
          \fill[white] (-4.13, -0.305) circle[radius = 0.125];
          \draw (4.4, 0) .. controls (4.4, -0.2) and (4, -0.4) .. (3.6, -0.4);
          \fill[white] (4.13, -0.305) circle[radius = 0.125];
          % complete set
          \draw[green!60!black] (-2.4, 0) ellipse (1.75 and 1.35);
          \fill[white] (-4.13, 0.305) circle[radius = 0.125];
          \draw[green!60!black] (2.4, 0) ellipse (1.75 and 1.35);
          \draw (0, 0.4) to (0, 1.1);
          \draw[green!60!black] (2.4, 0) ellipse (1.75 and 1.35);
          \fill[white] (4.13, 0.305) circle[radius = 0.125];
          \draw[green!60!black] (-0.71, 0.4) to (0.71, 0.4);
          \fill[green!60!black] (-0.71, 0.4) circle[radius = 0.1];
          \fill[green!60!black] (0.71, 0.4) circle[radius = 0.1];
          \fill (0, 0.4) circle[radius = 0.1];
          % punctures
          \draw[thick] (-0.1, 1.2) to (0.1, 1);
          \draw[thick] (-0.1, 1) to (0.1, 1.2);
          % fine mesh
          \draw (-2.4, 0) ellipse (1.35 and 0.9);
          \draw (2.4, 0) ellipse (1.35 and 0.9);
          \draw (-1.05, 0) to (-0.775, 0);
          \draw (-0.525, 0) to (0.525, 0);
          \draw (0.775, 0) to (1.05, 0);
          \draw (-3.6, 0.4) .. controls (-4, 0.4) and (-4.4, 0.2) .. (-4.4, 0);
          \draw (3.6, 0.4) .. controls (4, 0.4) and (4.4, 0.2) .. (4.4, 0);
          \fill (-3.6, 0.4) circle[radius = 0.1];
          \fill (-3.6, -0.4) circle[radius = 0.1];
          \fill (3.6, 0.4) circle[radius = 0.1];
          \fill (3.6, -0.4) circle[radius = 0.1];
          \fill (-1.05, 0) circle[radius = 0.1];
          \fill (1.05, 0) circle[radius = 0.1];
          % labels
          \node[scale = 0.8, above] at (-4.8, 1.2) {$(1,1)$};
          \draw[->, > = stealth, bend left = 30] (-3.9, 1.9) to (-2.8, 1);
          \node[scale = 0.8, below] at (-4.8, -1.2) {$(2,2)$};
          \draw[->, > = stealth, bend right = 30] (-3.9, -1.9) to (-2.8, -1);
          \node[scale = 0.8, above] at (-1.6, 1.8) {$(3,3)$};
          \draw[->, > = stealth, bend left = 30] (-1.6, 2) to (-3.6, 0);
          \node[scale = 0.8, left] at (-4.2, 0) {$(4,4)$};
          \node[scale = 0.8, above] at (4.8, 1.2) {$(5,5)$};
          \draw[->, > = stealth, bend right = 30] (3.9, 1.9) to (2.8, 1);
          \node[scale = 0.8, below] at (4.8, -1.2) {$(6,6)$};
          \draw[->, > = stealth, bend left = 30] (3.9, -1.9) to (2.8, -1);
          \node[scale = 0.8, above] at (1.6, 1.8) {$(7,7)$};
          \draw[->, > = stealth, bend right = 30] (1.6, 2) to (3.6, 0);
          \node[scale = 0.8, right] at (4.2, 0) {$(8,8)$};
          \node[scale = 0.8, below] at (0, -1.8) {$(9,9)$};
          \draw[->, > = stealth] (0, -2.1) to (0, -0.1);
        \end{tikzpicture}
        \lower2.1ex\hbox{\text{\scalebox{2}[4.5]{$\rangle$}}}. \label{eq: the outer boundary state with thecondensable anyon}
      \end{align}
      Here we write $(i,i)$ instead of $L_{p_i}\boxtimes\overline{L}_{p_i}$ for visibility.
      In order to reduce it to a more handleable form, we apply the projector formula \eqref{eq: projector formula} to both links near the two genera, alternating $p_i\:(i=3,4,7,8)$ integrals to $p,\,p^{\prime}$ integrals as
      \begin{align}
        & \prod_{i\in\qty{3,4,7,8}}\bigg(\int_{0}^{\infty}dp_i\,\rho_0(p_i)\bigg)C_{134}C_{234}C_{578}C_{678}\lower2.1ex\hbox{\text{\scalebox{1.6}[4.5]{$|$}}}\,
        \begin{tikzpicture}[scale = 0.4, baseline = {([yshift=-.5ex]current bounding box.center)}]
          \draw[thick] (0, 1.35) .. controls (-0.8, 1.35) and (-1.5, 2) .. (-2.5, 2) .. controls (-4, 2) and (-4.7, 0.8) .. (-4.7, 0) .. controls (-4.7, -0.8) and (-4, -2) .. (-2.5, -2) .. controls (-1.5, -2) and (-0.8, -1.35) .. (0, -1.35) .. controls (0.8, -1.35) and (1.5, -2) .. (2.5, -2) .. controls (4, -2) and (4.7, -0.8) .. (4.7, 0) .. controls (4.7, 0.8) and (4, 2) .. (2.5, 2) .. controls (1.5, 2) and (0.8, 1.35) .. (0, 1.35);
          % 2 genus
          \draw[thick, bend right = 30] (-3.1, 0) to (-1.7, 0);
          \draw[thick, bend left = 30] (-3, -0.05) to (-1.8, -0.05);
          \draw[thick, bend left = 30] (3.1, 0) to (1.7, 0);
          \draw[thick, bend right = 30] (3, -0.05) to (1.8, -0.05);
          % fine mesh
          \draw (-4.4, 0) .. controls (-4.4, -0.2) and (-4, -0.4) .. (-3.6, -0.4);
          \fill[white] (-4.13, -0.305) circle[radius = 0.125];
          \draw (4.4, 0) .. controls (4.4, -0.2) and (4, -0.4) .. (3.6, -0.4);
          \fill[white] (4.13, -0.305) circle[radius = 0.125];
          % complete set
          \draw[green!60!black] (-2.4, 0) ellipse (1.75 and 1.35);
          \fill[white] (-4.13, 0.305) circle[radius = 0.125];
          \draw[green!60!black] (2.4, 0) ellipse (1.75 and 1.35);
          \fill[white] (4.13, 0.305) circle[radius = 0.125];
          \draw (0, 0.4) to (0, 1.1);
          \draw[green!60!black] (-0.71, 0.4) to (0.71, 0.4);
          \fill[green!60!black] (-0.71, 0.4) circle[radius = 0.1];
          \fill[green!60!black] (0.71, 0.4) circle[radius = 0.1];
          \fill (0, 0.4) circle[radius = 0.1];
          % punctures
          \draw[thick] (-0.1, 1.2) to (0.1, 1);
          \draw[thick] (-0.1, 1) to (0.1, 1.2);
          % fine mesh
          \draw (-2.4, 0) ellipse (1.35 and 0.9);
          \draw (2.4, 0) ellipse (1.35 and 0.9);
          \draw (-1.05, 0) to (-0.775, 0);
          \draw (-0.525, 0) to (0.525, 0);
          \draw (0.775, 0) to (1.05, 0);
          \draw (-3.6, 0.4) .. controls (-4, 0.4) and (-4.4, 0.2) .. (-4.4, 0);
          \draw (3.6, 0.4) .. controls (4, 0.4) and (4.4, 0.2) .. (4.4, 0);
          \fill (-3.6, 0.4) circle[radius = 0.1];
          \fill (-3.6, -0.4) circle[radius = 0.1];
          \fill (3.6, 0.4) circle[radius = 0.1];
          \fill (3.6, -0.4) circle[radius = 0.1];
          \fill (-1.05, 0) circle[radius = 0.1];
          \fill (1.05, 0) circle[radius = 0.1];
        \end{tikzpicture}
        \lower2.1ex\hbox{\text{\scalebox{2}[4.5]{$\rangle$}}} \nonumber \\
        \, \nonumber \\
        & \qquad=\int_{0}^{\infty} dp_{\alpha}dp_{\beta}\,\rho_0(p_{\alpha})\rho_0(p_{\beta})C_{1x\alpha}C_{2x\alpha}C_{5y\beta}C_{6y\beta} \label{eq: after use of projectors} \\
        & \qqqquad\qqquad\times\frac{\delta(p_x-\overline{p}_x)\delta(p_y-\overline{p}_y)}{\rho_0(p_x)\rho_0(p_y)}\lower2.1ex\hbox{\text{\scalebox{1.6}[4.5]{$|$}}}\,
        \begin{tikzpicture}[scale = 0.4, baseline = {([yshift=-.5ex]current bounding box.center)}]
          \draw[thick] (0, 1.35) .. controls (-0.8, 1.35) and (-1.5, 2) .. (-2.5, 2) .. controls (-4, 2) and (-4.7, 0.8) .. (-4.7, 0) .. controls (-4.7, -0.8) and (-4, -2) .. (-2.5, -2) .. controls (-1.5, -2) and (-0.8, -1.35) .. (0, -1.35) .. controls (0.8, -1.35) and (1.5, -2) .. (2.5, -2) .. controls (4, -2) and (4.7, -0.8) .. (4.7, 0) .. controls (4.7, 0.8) and (4, 2) .. (2.5, 2) .. controls (1.5, 2) and (0.8, 1.35) .. (0, 1.35);
          % 2 genus
          \draw[thick, bend right = 30] (-3.1, 0) to (-1.7, 0);
          \draw[thick, bend left = 30] (-3, -0.05) to (-1.8, -0.05);
          \draw[thick, bend left = 30] (3.1, 0) to (1.7, 0);
          \draw[thick, bend right = 30] (3, -0.05) to (1.8, -0.05);
          % finemesh
          \draw (-0.83, 0) to (0.83, 0);
          \fill[white] (-0.4, 0) circle[radius = 0.125];
          \fill[white] (0.4, 0) circle[radius = 0.125];
          \begin{scope}
            \clip (-2.4, 1.12) rectangle (-0.81, -1.12);
            \draw (-2.4, 0) ellipse (1.57 and 1.1);
          \end{scope}
          \begin{scope}
            \clip (0.81, 1.12) rectangle (2.4, -1.12);
            \draw (2.4, 0) ellipse (1.57 and 1.1);
          \end{scope}
          \begin{scope}
            \clip (-3, 1.4) rectangle (-2.4, 1.08);
            \draw (-2.4, 1.4) ellipse (0.6 and 0.3);
          \end{scope}
          \begin{scope}
            \clip (-3, -1.4) rectangle (-2.4, -1.08);
            \draw (-2.4, -1.4) ellipse (0.6 and 0.3);
          \end{scope}
          \begin{scope}
            \clip (3, 1.4) rectangle (2.4, 1.08);
            \draw (2.4, 1.4) ellipse (0.6 and 0.3);
          \end{scope}
          \begin{scope}
            \clip (3, -1.4) rectangle (2.4, -1.08);
            \draw (2.4, -1.4) ellipse (0.6 and 0.3);
          \end{scope}
          \fill (-3, 1.4) circle[radius = 0.1];
          \fill (-3, -1.4) circle[radius = 0.1];
          \fill (3, 1.4) circle[radius = 0.1];
          \fill (3, -1.4) circle[radius = 0.1];
          \fill (-0.83, 0) circle[radius = 0.1];
          \fill (0.83, 0) circle[radius = 0.1];
          % complete set
          \begin{scope}
            \clip (-2.4, 1.8) rectangle (-0.38, -1.8);
            \draw[green!60!black] (-2.4, 0) ellipse (2 and 1.7);
          \end{scope}
          \begin{scope}
            \clip (0.38, 1.8) rectangle (2.4, -1.8);
            \draw[green!60!black] (2.4, 0) ellipse (2 and 1.7);
          \end{scope}
          \begin{scope}
            \clip (-3, 1.72) rectangle (-2.4, 1.4);
            \draw[green!60!black] (-2.4, 1.4) ellipse (0.6 and 0.3);
          \end{scope}
          \begin{scope}
            \clip (3, 1.72) rectangle (2.4, 1.4);
            \draw[green!60!black] (2.4, 1.4) ellipse (0.6 and 0.3);
          \end{scope}
          \begin{scope}
            \clip (-3, -1.72) rectangle (-2.4, -1.4);
            \draw[green!60!black] (-2.4, -1.4) ellipse (0.6 and 0.3);
          \end{scope}
          \begin{scope}
            \clip (3, -1.72) rectangle (2.4, -1.4);
            \draw[green!60!black] (2.4, -1.4) ellipse (0.6 and 0.3);
          \end{scope}
          \draw[green!60!black] (-0.47, 0.4) to (0.47, 0.4);
          \draw (-3, 1.4) .. controls (-4.65, 1) and (-4.65, -1) .. (-3, -1.4);
          \draw (3, 1.4) .. controls (4.65, 1) and (4.65, -1) .. (3, -1.4);
          \draw (0, 0.4) to (0, 1.1);
          \fill[green!60!black] (-0.47, 0.4) circle[radius = 0.1];
          \fill[green!60!black] (0.47, 0.4) circle[radius = 0.1];
          \fill (0, 0.4) circle[radius = 0.1];
          % punctures
          \draw[thick] (-0.1, 1.2) to (0.1, 1);
          \draw[thick] (-0.1, 1) to (0.1, 1.2);
          % labels
          \node[scale = 0.8, left] at (-4, -0.4) {$(\alpha,\alpha)$};
          \node[scale = 0.8, right] at (4, -0.4) {$(\beta,\beta)$};
          \node[scale = 0.8, above] at (-4, 1.7) {$(x,x)$};
          \draw[->, > = stealth, bend left = 30] (-2.9, 2.4) to (-1.9, 1.7);
          \draw[->, > = stealth, bend right = 10] (-2.9, 2.4) to (-0.7, 0.65);
          \node[scale = 0.8, below] at (-3.8, -1.7) {$(2,2)$};
          \draw[->, > = stealth, bend left = 45] (-3.8, -1.9) to (-1.5, -0.8);
          \node[scale = 0.8, above] at (4, 1.7) {$(y,y)$};
          \draw[->, > = stealth, bend right = 30] (2.9, 2.4) to (1.9, 1.7);
          \draw[->, > = stealth, bend left = 10] (2.9, 2.4) to (0.7, 0.65);
          \node[scale = 0.8, below] at (3.8, -1.7) {$(6,6)$};
          \draw[->, > = stealth, bend right = 45] (3.8, -1.9) to (1.5, -0.8);
          \node[scale = 0.8, above] at (-1.2, 1.8) {$(z,\overline{z})$};
          \draw[->, > = stealth, bend left = 15] (-0.5, 2) to (-0.25, 0.5);
          \node[scale = 0.8, above] at (1.2, 1.8) {$(w,\overline{w})$};
          \draw[->, > = stealth, bend right = 15] (0.5, 2) to (0.25, 0.5);
          \node[scale = 0.8, left] at (-4.3, 1.2) {$(1,1)$};
          \draw[->, > = stealth, bend right = 20] (-4.6, 1) to (-1.1, 0.5);
          \node[scale = 0.8, right] at (4.3, 1.2) {$(5,5)$};
          \draw[->, > = stealth, bend left = 20] (4.6, 1) to (1.1, 0.5);
          \node[scale = 0.8, below] at (0, -1.8) {$(9,9)$};
          \draw[->, > = stealth] (0, -2.1) to (0, -0.2);
        \end{tikzpicture}
        \lower2.1ex\hbox{\text{\scalebox{2}[4.5]{$\rangle$}}}. \nonumber
      \end{align}
      The $p_i\:(i=1,2,5,6,9)$ integrals remain unchanged under this operation.
      $(z,\overline{z})$ in the graph is a shorthand for $L_{p_z}\boxtimes\overline{L}_{\overline{p}_z}$.
      The line $L_{p_x}\boxtimes\overline{L}_{\overline{p}_x}$ (resp.~$L_{p_y}\boxtimes\overline{L}_{\overline{p}_y}$) is restricted to the component $L_{p_x}\boxtimes\overline{L}_{p_x}$ (resp.~$L_{p_y}\boxtimes\overline{L}_{p_y}$) of the diagonal condensable anyon \eqref{eq: diagonal Lagrangian algebra object in VTQFT} by virtue of the projectors while $L_{p_z}\boxtimes\overline{L}_{\overline{p}_z}$, $L_{p_w}\boxtimes\overline{L}_{\overline{p}_w}$ are not.
      The result is still far from being refined, so let us delve further into a smaller part of the link as shown in Figure \ref{fig: left half}.
      There, we bisect the link in eq.\,\eqref{eq: after use of projectors} in the middle of the picture and only pick the left half.
      The strategy is to attempt by some means to create the Wilson bubble to apply the formula \eqref{eq: Wilson bubble}. The procedure breaks down into the three steps to be outlined next:
      \begin{figure}[t]
        \centering
        {

        }
        \caption{The left half of the link in eq.\,\eqref{eq: after use of projectors}. The container manifold $\Sigma_{2,0}$ is omitted for simplicity and the X mark represents the location of the left genus.}
        \label{fig: left half}
      \end{figure}
      \begin{gather}
        %
        \qqquad\qquad \label{eq: the third step} \\
        \qqqquad\qqquad\times e^{\pi i\qty{(h_a-\overline{h}_a)+(h_c-\overline{h}_c)-(h_b-\overline{h}_b)}}
        %
        . \nonumber
      \end{gather}
      In the final equation, we employ the $s$-$u$ crossing transformation formula \eqref{eq: s-u crossing transform}. Repeating the same procedure (except for the use of eq.\,\eqref{eq: fusion kernel to u-channel ver.2} in the final step) to the right half of the link in eq.\,\eqref{eq: after use of projectors} and applying the Wilson bubble formula \eqref{eq: Wilson bubble} twice and the Wilson triangle identity \eqref{eq: Wilson triangle identity} once result in
      \begin{align}
        & %
        . \label{eq: resolve three Wilson bubbles}
      \end{align}
      This equation, combined with the procedure \eqref{eq: the third step}, enables us to rephrase the state in the integrand in the third line of eq.\,\eqref{eq: after use of projectors} into
      \begin{align}
         & \lower2.1ex\hbox{\text{\scalebox{1.6}[4.5]{$|$}}}\,
        %
        \lower2.1ex\hbox{\text{\scalebox{2}[4.5]{$\rangle$}}}. \nonumber
      \end{align}
      Substituting this equation into eq.\,\eqref{eq: after use of projectors} and performing the integration w.\,r.\,t.~(i)\,$p_1$ and $p_5$ (ii)\,$p_2$ and $p_6$ in this order in eq.\,\eqref{eq: the outer boundary state with thecondensable anyon} with the aid of eq.\,\eqref{eq: version of invertibility} uncomplicates the outer boundary state in a way
      \begin{align}
         & \lower2.1ex\hbox{\text{\scalebox{1.6}[4.5]{$|$}}}\,
        %
        \lower2.1ex\hbox{\text{\scalebox{2}[4.5]{$\rangle$}}}. \nonumber
      \end{align}
      We further carry out the integration w.\,r.\,t.~$p_a$ and $p_d$, and then $p_9$ by eq.\,\eqref{eq: version of invertibility} and substitute the result to eq.\,\eqref{eq: VTQFT path-integral for the wormhole with the condensable anyon} to arrive at the final form
      \begin{align}
         & \lower2.1ex\hbox{\text{\scalebox{1.6}[4.5]{$|$}}}\,
        %
        \lower2.1ex\hbox{\text{\scalebox{2}[4.5]{$\rangle$}}}. \label{eq: genus two with Wilson line wormhole factorization}
      \end{align}
      Thus, we conclude that the partition function of the genus two wormhole $\Sigma_{2,0}\times\qty[0,1]$ with the Wilson line $L_p\boxtimes\overline{L}_{\overline{p}}$ connecting the two boundaries actually factorizes after condensing the diagonal condensable anyon \eqref{eq: diagonal Lagrangian algebra object in VTQFT}.
      The prefactor $\delta(p-\overline{p})$ derives from the existence of $\mathcal{A}$ seen as the topological boundary that forces $L_p\boxtimes\overline{L}_{\overline{p}}$ to be diagonal, as in the torus wormhole case.
      However, a significant difference arises when taking the limit $p,\overline{p}\to\mathbbm{1}$ and applying eq.\,\eqref{eq: structure constant limit}
      \begin{align}
         & \lim_{p,\overline{p}\to\mathbbm{1}}\lower2.1ex\hbox{\text{\scalebox{1.6}[4.5]{$|$}}}\,
        \begin{tikzpicture}[scale = 0.4, baseline = {([yshift=-.5ex]current bounding box.center)}]
          \draw[thick] (0, 1.35) .. controls (-0.8, 1.35) and (-1.5, 2) .. (-2.5, 2) .. controls (-4, 2) and (-4.7, 0.8) .. (-4.7, 0) .. controls (-4.7, -0.8) and (-4, -2) .. (-2.5, -2) .. controls (-1.5, -2) and (-0.8, -1.35) .. (0, -1.35) .. controls (0.8, -1.35) and (1.5, -2) .. (2.5, -2) .. controls (4, -2) and (4.7, -0.8) .. (4.7, 0) .. controls (4.7, 0.8) and (4, 2) .. (2.5, 2) .. controls (1.5, 2) and (0.8, 1.35) .. (0, 1.35);
          % 2 genus
          \draw[thick, bend right = 30] (-3.1, 0) to (-1.7, 0);
          \draw[thick, bend left = 30] (-3, -0.05) to (-1.8, -0.05);
          \draw[thick, bend left = 30] (3.1, 0) to (1.7, 0);
          \draw[thick, bend right = 30] (3, -0.05) to (1.8, -0.05);
          %inner boundary
          \draw[thick, dashed] (0, 0.75) .. controls (-0.8, 0.75) and (-1.5, 1.45) .. (-2.5, 1.45) .. controls (-3.7, 1.45) and (-4.25, 0.5) .. (-4.25, 0) .. controls (-4.25, -0.5) and (-3.7, -1.45) .. (-2.5, -1.45) .. controls (-1.5, -1.45) and (-0.8, -0.75) .. (0, -0.75) .. controls (0.8, -0.75) and (1.5, -1.45) .. (2.5, -1.45) .. controls (3.7, -1.45) and (4.25, -0.5) .. (4.25, 0) .. controls (4.25, 0.5) and (3.7, 1.45) .. (2.5, 1.45) .. controls (1.5, 1.45) and (0.8, 0.75) .. (0, 0.75);
          \draw[thick, dashed] (-2.4, 0) ellipse (1.3 and 0.9);
          \draw[thick, dashed] (2.4, 0) ellipse (1.3 and 0.9);
          % finemesh
          \draw[thick, red] (-2.4, 0) ellipse (1.1 and 0.65);
          \draw[thick, red] (2.4, 0) ellipse (1.1 and 0.65);
          \draw[thick, red] (-1.3, 0) to (-1.1, 0);
          \draw[dashed, thick, red] (-1.1, 0) to (1.1, 0);
          \draw[thick, red] (1.1, 0) to (1.3, 0);
          \draw[thick, red] (-3.25, 0.41) .. controls (-4, 0.38) and (-4.4, 0.1) .. (-4.4, 0);
          \draw[thick, red] (-4.4, 0) .. controls (-4.4, -0.07) and (-4.25, -0.14) .. (-4.23, -0.155);
          \draw[dashed, thick, red] (-4.23, -0.155) .. controls (-3.95, -0.315) and (-3.7, -0.345) .. (-3.58, -0.367);
          \draw[thick, red, bend left = 8] (-3.58, -0.367) .. controls (-3.51, -0.39) and (-3.28, -0.4) .. (-3.25, -0.41);
          \draw[thick, red] (3.25, 0.41) .. controls (4, 0.38) and (4.4, 0.1) .. (4.4, 0);
          \draw[thick, red] (4.4, 0) .. controls (4.4, -0.07) and (4.25, -0.14) .. (4.23, -0.155);
          \draw[dashed, thick, red] (4.23, -0.155) .. controls (3.95, -0.315) and (3.7, -0.345) .. (3.58, -0.367);
          \draw[thick, red, bend left = 8] (3.58, -0.367) .. controls (3.51, -0.39) and (3.28, -0.4) .. (3.25, -0.41);
          % Wilson line
          \draw (0, 0.3) to (0, 1.1);
          % punctures
          \draw[thick] (-0.1, 0.4) to (0.1, 0.2);
          \draw[thick] (-0.1, 0.2) to (0.1, 0.4);
          \draw[thick] (-0.1, 1.2) to (0.1, 1);
          \draw[thick] (-0.1, 1) to (0.1, 1.2);
        \end{tikzpicture}
        \lower2.1ex\hbox{\text{\scalebox{2}[4.5]{$\rangle$}}}\,\propto\int_{0}^{\infty}dp_xdp_ydp_z\,\rho_0(p_x)\rho_0(p_y)\rho_0(p_z)C_{xxz}C_{yyz} \nonumber \\
        & \qquad\times\int_{0}^{\infty}dp_{\alpha}dp_{\beta}dp_b\,\rho_0(p_{\alpha})\rho_0(p_{\beta})\rho_0(p_b) C_{\alpha\alpha b}C_{\beta\beta b}\lower2.1ex\hbox{\text{\scalebox{1.6}[4.5]{$|$}}}\,
        \begin{tikzpicture}[scale = 0.4, baseline = {([yshift=-.5ex]current bounding box.center)}]
          \draw[thick] (0, 1.35) .. controls (-0.8, 1.35) and (-1.5, 2) .. (-2.5, 2) .. controls (-4, 2) and (-4.7, 0.8) .. (-4.7, 0) .. controls (-4.7, -0.8) and (-4, -2) .. (-2.5, -2) .. controls (-1.5, -2) and (-0.8, -1.35) .. (0, -1.35) .. controls (0.8, -1.35) and (1.5, -2) .. (2.5, -2) .. controls (4, -2) and (4.7, -0.8) .. (4.7, 0) .. controls (4.7, 0.8) and (4, 2) .. (2.5, 2) .. controls (1.5, 2) and (0.8, 1.35) .. (0, 1.35);
          % 2 genus
          \draw[thick, bend right = 30] (-3.1, 0) to (-1.7, 0);
          \draw[thick, bend left = 30] (-3, -0.05) to (-1.8, -0.05);
          \draw[thick, bend left = 30] (3.1, 0) to (1.7, 0);
          \draw[thick, bend right = 30] (3, -0.05) to (1.8, -0.05);
          % Wilson lines
          \draw[green!60!black] (-2.4, 0) ellipse (1.4 and 1);
          \draw[green!60!black] (2.4, 0) ellipse (1.4 and 1);
          \draw[green!60!black] (-1, 0) to (1, 0);
          \fill (-1, 0) circle[radius = 0.1];
          \fill (1, 0) circle[radius = 0.1];
          % labels
          \node[scale = 0.8, above] at (-2.4, 0.8) {$(x, x)$};
          \node[scale = 0.8, above] at (2.4, 0.8) {$(y, y)$};
          \node[scale = 0.8, below] at (0, -0.25) {$(z,z)$};
        \end{tikzpicture}
        \lower2.1ex\hbox{\text{\scalebox{2}[4.5]{$\rangle$}}}\otimes\,\lower2.1ex\hbox{\text{\scalebox{1.6}[4.5]{$|$}}}\,
        \begin{tikzpicture}[scale = 0.4, baseline = {([yshift=-.5ex]current bounding box.center)}]
          \draw[thick] (0, 1.35) .. controls (-0.8, 1.35) and (-1.5, 2) .. (-2.5, 2) .. controls (-4, 2) and (-4.7, 0.8) .. (-4.7, 0) .. controls (-4.7, -0.8) and (-4, -2) .. (-2.5, -2) .. controls (-1.5, -2) and (-0.8, -1.35) .. (0, -1.35) .. controls (0.8, -1.35) and (1.5, -2) .. (2.5, -2) .. controls (4, -2) and (4.7, -0.8) .. (4.7, 0) .. controls (4.7, 0.8) and (4, 2) .. (2.5, 2) .. controls (1.5, 2) and (0.8, 1.35) .. (0, 1.35);
          % 2 genus
          \draw[thick, bend right = 30] (-3.1, 0) to (-1.7, 0);
          \draw[thick, bend left = 30] (-3, -0.05) to (-1.8, -0.05);
          \draw[thick, bend left = 30] (3.1, 0) to (1.7, 0);
          \draw[thick, bend right = 30] (3, -0.05) to (1.8, -0.05);
          % Wilson lines
          \draw (-2.4, 0) ellipse (1.4 and 1);
          \draw (2.4, 0) ellipse (1.4 and 1);
          \draw[orange] (-1, 0) to (1, 0);
          \fill (-1, 0) circle[radius = 0.1];
          \fill (1, 0) circle[radius = 0.1];
          % labels
          \node[scale = 0.8, above] at (-2.4, 0.8) {$(\alpha, \alpha)$};
          \node[scale = 0.8, above] at (2.4, 0.8) {$(\beta, \beta)$};
          \node[scale = 0.8, below] at (0, -0.25) {$(b,b)$};
        \end{tikzpicture}
        \lower2.1ex\hbox{\text{\scalebox{2}[4.5]{$\rangle$}}}
        \nonumber \\
         & \quad=\,
        \lower2.1ex\hbox{\text{\scalebox{1.6}[4.5]{$|$}}}\,
        \begin{tikzpicture}[scale = 0.4, baseline = {([yshift=-.5ex]current bounding box.center)}]
          \draw[thick] (0, 1.35) .. controls (-0.8, 1.35) and (-1.5, 2) .. (-2.5, 2) .. controls (-4, 2) and (-4.7, 0.8) .. (-4.7, 0) .. controls (-4.7, -0.8) and (-4, -2) .. (-2.5, -2) .. controls (-1.5, -2) and (-0.8, -1.35) .. (0, -1.35) .. controls (0.8, -1.35) and (1.5, -2) .. (2.5, -2) .. controls (4, -2) and (4.7, -0.8) .. (4.7, 0) .. controls (4.7, 0.8) and (4, 2) .. (2.5, 2) .. controls (1.5, 2) and (0.8, 1.35) .. (0, 1.35);
          % 2 genus
          \draw[thick, bend right = 30] (-3.1, 0) to (-1.7, 0);
          \draw[thick, bend left = 30] (-3, -0.05) to (-1.8, -0.05);
          \draw[thick, bend left = 30] (3.1, 0) to (1.7, 0);
          \draw[thick, bend right = 30] (3, -0.05) to (1.8, -0.05);
          % complete set
          \draw[thick, red] (-2.4, 0) ellipse (1.57 and 1.1);
          \draw[thick, red] (2.4, 0) ellipse (1.57 and 1.1);
          \draw[thick, red] (-0.83, 0) to (0.83, 0);
          % labels
          \node[scale = 0.8, above] at (-2.4, 0.9) {$\mathcal{A}$};
          \node[scale = 0.8, above] at (2.4, 0.9) {$\mathcal{A}$};
          \node[scale = 0.8, below] at (0, 0) {$\mathcal{A}$};
        \end{tikzpicture}
        \lower2.1ex\hbox{\text{\scalebox{2}[4.5]{$\rangle$}}}\,\otimes\lower2.1ex\hbox{\text{\scalebox{1.6}[4.5]{$|$}}}\,
        \begin{tikzpicture}[scale = 0.4, baseline = {([yshift=-.5ex]current bounding box.center)}]
          \draw[thick] (0, 1.35) .. controls (-0.8, 1.35) and (-1.5, 2) .. (-2.5, 2) .. controls (-4, 2) and (-4.7, 0.8) .. (-4.7, 0) .. controls (-4.7, -0.8) and (-4, -2) .. (-2.5, -2) .. controls (-1.5, -2) and (-0.8, -1.35) .. (0, -1.35) .. controls (0.8, -1.35) and (1.5, -2) .. (2.5, -2) .. controls (4, -2) and (4.7, -0.8) .. (4.7, 0) .. controls (4.7, 0.8) and (4, 2) .. (2.5, 2) .. controls (1.5, 2) and (0.8, 1.35) .. (0, 1.35);
          % 2 genus
          \draw[thick, bend right = 30] (-3.1, 0) to (-1.7, 0);
          \draw[thick, bend left = 30] (-3, -0.05) to (-1.8, -0.05);
          \draw[thick, bend left = 30] (3.1, 0) to (1.7, 0);
          \draw[thick, bend right = 30] (3, -0.05) to (1.8, -0.05);
          % complete set
          \draw[thick, red] (-2.4, 0) ellipse (1.57 and 1.1);
          \draw[thick, red] (2.4, 0) ellipse (1.57 and 1.1);
          \draw[thick, red] (-0.83, 0) to (0.83, 0);
          % labels
          \node[scale = 0.8, above] at (-2.4, 0.9) {$\mathcal{A}$};
          \node[scale = 0.8, above] at (2.4, 0.9) {$\mathcal{A}$};
          \node[scale = 0.8, below] at (0, 0) {$\mathcal{A}$};
        \end{tikzpicture}
        \lower2.1ex\hbox{\text{\scalebox{2}[4.5]{$\rangle$}}}. \label{eq: genus 2 wormhole factorization}
      \end{align}
      We observe no divergent ``constants" $\delta(0)$, $\int_{0}^{\infty}dp=\infty$ up to the $\displaystyle\frac{\delta(p-\overline{p})}{\rho_0(p)}$ factor, and the integrals in the first two lines include exactly the required number of factors $\rho_0$, $C_{ijk}$, culminating in the final line.
      This is consistent with the fact that $\Sigma_{2,0}\times[0,1]$ is hyperbolic for which VTQFT is sensible on it.
      The state factorization \eqref{eq: genus 2 wormhole factorization} implies the partition function factorization in the same way as the torus wormhole by taking the overlap with the moduli basis of the Hilbert space on each boundary $|\bm{m}_1\rangle$, $|\bm{m}_2\rangle$
      \begin{align}
        \big\langle\bm{m}_1,\bm{m}_2\big|\Sigma_{2,0}\times[0,1],\mathcal{A}\big\rangle=\frac{\delta(p-\overline{p})}{\rho_0(p)}\big\langle\bm{m}_1\big|\Sigma_{2,0},\mathcal{A}\big\rangle\big\langle\bm{m}_2\big|\Sigma_{2,0},\mathcal{A}\big\rangle,
      \end{align}
      Again, the symmetry TFT sandwich construction makes it possible to identify the partition function $\langle\bm{m}_i\big|\Sigma_{2,0},\mathcal{A}\big\rangle$ ($i=1,2$) with the Liouville CFT partition function on the genus 2 Riemann surface, where $\mathcal{A}$ serves as a topological boundary and the genus 2 boundary as a physical boundary.
      \par So far, we have seen that anyon condensation leads to the partition function factorization for $\Sigma_{g,n}\times[0,1]$ with $(g,n)=(1,1),(2,0),(2,1)$, but it should generalize to arbitrary pair of $(g,n)\in(\mathbb{N}\cup\{0\})\times(\mathbb{N}\cup\{0\})$ and ultimately extend to multi-boundary wormholes with arbitrary number of Wilson lines as long as the target geometry is hyperbolic.
      In Chern-Simons theory, the factorization for generic topologies is guaranteed by the fact that anyon condensation trivializes the partition function for arbitrary closed $3$-manifolds \cite{Benini:2022hzx}.
      We have not yet proven that this statement also holds in VTQFT, but we insist that the topological-boundary picture instead provides compelling evidence that factorization indeed occurs, at least for two-boundary wormholes.
      However, the explicit check for any $\Sigma_{g,n}\times\qty[0,1]$ with greater $(g,n)$ will be computationally quite strenuous.

    \section{Conclusions and discussions} \label{sec: conclusions and discussions}
      This paper begins with a review of Virasoro TQFT in Section \ref{sec: Virasoro TQFT}, summarizing the structure of the Hilbert space and methods for calculating partition functions using Heegaard splittings and compression bodies.
      We then examine that, when a solid torus is treated within the framework of VTQFT, conventional results such as the Hawking-Page phase transition and black hole entropy can be derived.
      Section \ref{sec: Anyon condensation in modular tensor category} provides an overview of anyon condensation in non-Abelian TQFT and introduces the concept of the diagonal condensable anyons when the symmetry category is the Drinfeld center $\mathcal{Z}(\mathcal{C})\cong\mathcal{C}\boxtimes\overline{\mathcal{C}}$.
      In Section \ref{sec: Anyon condensation in Virasoro TQFT}, we make an attempt to extend the results from Section \ref{sec: Anyon condensation in modular tensor category} to VTQFT, showing that the 2-boundary wormholes actually factorize.
      Let us conclude this paper with comments on the possible future direction.

      \paragraph{Mathematical formulation} ~ \\
        It is non-trivial to define mathematically a direct sum for continuous labels that appears in the diagonal condensable anyon \eqref{eq: diagonal Lagrangian algebra object in VTQFT}.
        The continuous spectrum also renders the symmetry category less well-behaved, as depicted in Figure \ref{fig: inclusion relation} in Appendix \ref{append: From monoidal category to fusion category and modular tensor category}.
        There is indeed a growing attention to \textit{non-semisimpel finite} tesnor categories in the recent literature, where each object in the category is possibly non-semisimple but the number of isomorphism class of simple objects must be finite $|\mathcal{I}(\mathcal{C})|<\infty$.
        However, the key challenge in VTQFT is that the symmetry category $\mathcal{C}=\operatorname{Rep}\big(\,\mathcal{U}_q(\mathfrak{sl}(2,\mathbb{C}))\big)$ is \textit{infinite}, much less non-semisimple.
        An intriguing future direction is whether we can properly define the infinite ``direct sum" for tensor categories that are not finite (Ref.\,\cite{chang2024modular} might be useful in this regard).
        A rigorous categorical formulation of VTQFT could deepen our understanding of the 3d AdS quantum gravity as a TQFT.

      \paragraph{Other types of condensable anyons and sum over topologies} ~ \\
        The diagonal condensable anyon \eqref{eq: diagonal Lagrangian algebra object in VTQFT} might not be the unique condensable anyon in VTQFT.
        In particular, if we restrict our attention to the commutative condition \eqref{eq: commutative condition} for condensable anyons, a similar argument to that in Subsection 3.3 of Ref.\,\cite{Benini:2022hzx} shows that for any $s$ and $t$ satisfying $s^2-t^2\in\mathbb{Z}$, the direct sum $\bigoplus_{s,t}L_s\boxtimes\overline{L}_t$ is commutative.
        In Chern-Simons theory, there are multiple ways to choose Lagrangian condensable anyons, and the partition functions for solid tori that result from condensing each of them corresponds to the ADE classification of the Wess-Zumino-Witten model \cite{Cappelli:1986hf, Cappelli:1987xt}.
        If such a classification of condensable anyons is completed in VTQFT, it would allow us to discuss the sum over all possible anyon condensations as mentioned in Ref.\,\cite{Dymarsky:2024frx}, which may correspond to bulk side description of the ensemble average of boundary theories.
        There exists a candidate non-diagonal CFT with continuous spectrum that could potentially give rise to a \textit{non-diagonal} condensable anyons via the topological boundary picture mentioned in Subection \ref{subsec: diagonal Lagrangian condensable anyon}.
        The CFT is recently constructed by taking the non-rational limit of the central charge of $D$-series minimal models \cite{Ribault:2019qrz}, and although the theory has a central charge $c<1$, it can be analytically continued to the half-open region $\Re c<13$.

      \paragraph{Relation to topological quantum computing} ~ \\
        Chern-Simons theory serves as the low-energy effective theory for the quantum Hall effect and has been shown to be universal as a qubit model \cite{Freedman:2000gwh}.
        The concept of topological quantum computing \cite{Kitaev:1997wr}  was initiated by A.\,Y.\,Kitaev, and there has been intensive research on anyon systems such as the Fibonacci anyon, $SU(N)$ Chern-Simons, and topological error-correcting codes like the toric code.
        In the case of VTQFT, the spectrum is continuously infinite posing the significant challenge of whether such a system can be realized in physical systems.
        Nevertheless, it remains an interesting question if this system can perform universal quantum computation. Concerning universality, one possible approach, similar to the proof for Fibonacci anyons, involves assigning the $\ket{0}$ and $\ket{1}$ states to a four-point block with four interacting anyons.
        However, it will cause an excessive number of unnecessary states $\ket{N}$ due to the infinite number of Wilson lines.
        While the braiding of the first and second anyons only introduces a phase $B_{p_3}^{p_1p_2}$ in eq.\,\eqref{eq: braiding phase}, the braiding of the second and third anyons leads to a complicated transformation as shown in eq.\,\eqref{eq: fusion kernel to u-channel}.
        It remains unclear whether this braiding can be expressed in a block-diagonal form. Furthermore, it is highly challenging whether the group generated by these two operations is dense within the set of unitary operation.
        One possibility is that this system is better suited for use as a general qudit model rather than as a qubit model.

    \acknowledgments
      The author is grateful to Toshiya Kawai and Yuma Furuta for comments during the weekly seminar held at the RIMS, and Masamichi Miyaji for comments on an early draft of this paper.
      He would also like to thank Kantaro Ohmori for drawing his attention to Ref.\,\cite{etingof2015tensor}, and Yusuke Taki, Takashi Tsuda from whom he learned so much about 3d quantum gravity and Liouville theory.
      He acknowledges the anonymous referees for identifying, among other points, a critical issue in the genus-2 wormhole analysis, whose insights were invaluable in refining our understanding.
      This work is done in partial fulfillment of the requirements for the master's degree at RIMS, Kyoto University.
      He special thanks to physics students Nagare Katayama, Jun Maeda, Eiji Muto, Tsubasa Ohishi, Keito Shimizu, Kotaro Shinmyo, Toi Tachibana, Ryo Takami, Kenya Tasuki, Tatsuya Wada, Shogo Yamada, Shiki Yoshikawa, Naoki Ogawa, Takahiro Waki, Yu-ki Suzuki, Taishi Kawamoto, Masaya Amo and Masashi Kawahira.
      The author also acknowledges he would not complete the master's course without his friends outside physics, Takaki, Chihiro, Hiroki, Kohei, Kyoka, Rei, Shinnosuke, Shun, Suzuka, Yuqi, Ai, Chihiro, Harumichi, Kensuke, Taisei, Go, Ichi, Koki, Taichi, Yuki, Yusei, Yushi, Naoki, just to name a few.

  \appendix

  \section{Crossing kernels and Moore-Seiberg consistency conditions} \label{append: Crossing kernels and Moore-Seiberg consistency conditions}
    In this appendix, we collect some crossing equation for Virasoro conformal blocks.
    Most of the results are excerpted from the review \cite{Eberhardt:2023mrq}.
    Although not all of them are used in the main body, we derive a number of practical formulae in detail for the reader's sake of grasping original constructions \cite{Collier:2023fwi,Collier:2024mgv} and related literature.
    For rational conformal field theories (RCFT), the number of primary operators is finite and the crossing transformations like fusion transformation, braiding and modular $S$-transformation are all expressed as finite size matrices satisfying hexagon identity, pentagon identity among others \cite{Moore:1988qv}.
    However, due to the continuously infinite number of primaries labeled by $p\in\mathbb{R}_{\geq 0}$ in Liouville theory, the crossing transformations are rather represented as integration kernels whose explicit form was first developed in Refs.\,\cite{Ponsot:1999uf,Ponsot:2000mt, Teschner:2013tqy}.
    \par Before advancing to tedious consistency equations, here is the easiest case that follows from a four point crossing relation without reference to them:
    \begin{align}
      \begin{tikzpicture}[scale = 0.5, baseline = {([yshift=-.5ex]current bounding box.center)}]
        \draw[red] (-2.5, 1.5) -- (-1, 0);
        \draw[dashed, red] (-2.5, -1.5) -- (-1, 0);
        \draw[red] (-1, 0) -- (1, 0);
        \draw[red] (1, 0) -- (2.5, 1.5);
        \draw[red] (1, 0) -- (2.5, -1.5);
        \node[scale = 0.8, above left] at (-2.5, 1.5) {$p_3$};
        \node[scale = 0.8, below left] at (-2.5, -1.5) {$\mathbbm{1}$};
        \node[scale = 0.8, above right] at (2.5, 1.5) {$p_2$};
        \node[scale = 0.8, below right] at (2.5, -1.5) {$p_1$};
        \node[scale = 0.8, above] at (0, 0) {$p_3$};
      \end{tikzpicture}
      =\int_{0}^{\infty}dp\,F_{p_3p}
      \begin{bmatrix}
        p_3         & p_2 \\
        \mathbbm{1} & p_1
      \end{bmatrix}
      \begin{tikzpicture}[scale = 0.5, baseline = {([yshift=-.5ex]current bounding box.center)}]
        \draw[red] (-1.5, 2.5) -- (0, 1);
        \draw[dashed, red] (-1.5, -2.5) -- (0, -1);
        \draw[red] (1.5, 2.5) -- (0, 1);
        \draw[red] (1.5, -2.5) -- (0, -1);
        \draw[red] (0, 1) -- (0, -1);
        \node[scale = 0.8, above left] at (-1.5, 2.5) {$p_3$};
        \node[scale = 0.8, below left] at (-1.5, -2.5) {$\mathbbm{1}$};
        \node[scale = 0.8, above right] at (1.5, 2.5) {$p_2$};
        \node[scale = 0.8, below right] at (1.5, -2.5) {$p_1$};
        \node[scale = 0.8, left] at (0, 0) {$p$};
      \end{tikzpicture}
      .
    \end{align}
    Recall that $p_4=\mathbbm{1}$ is used to denote $p_4=\pm i\frac{Q}{2}$.
    The diagrams on both sides are actually three point blocks, so we find that the fusion kernel $F$ reduces to
    \begin{align}
      F_{p_3p}
      \begin{bmatrix}
        p_3         & p_2 \\
        \mathbbm{1} & p_1
      \end{bmatrix}
      =\delta(p_1-p) \label{eq: F-kernel limit 2}.
    \end{align}
    Other crossing transformation with simple analytic expression is the torus modular $S$-kernel $S_{p_1p_2}\qty[\mathbbm{1}]$ deduced from the precise form of the Virasoro characters \eqref{eq: Virasoro character} and the relation $\chi_{p_1}(-\frac{1}{\tau})=\int_{0}^{\infty}dp_2\,S_{p_1p_2}\qty[\mathbbm{1}]\chi_{p_2}(\tau)$:
    \begin{gather}
      S_{p_1p_2}\qty[\mathbbm{1}]=2\sqrt{2}\cos(4\pi p_1p_2), \\
      S_{\mathbbm{1}p}\qty[\mathbbm{1}]=4\sqrt{2}\sinh\qty(2\pi bp)\sinh\qty(2\pi\frac{p}{b}).
    \end{gather}
    The second one is the Cardy density since it asymptotes to leading order to the Cardy formula
    \begin{align}
      \log S_{\mathbbm{1}p}[\mathbbm{1}]\approx 2\pi p\Big(b+\frac{1}{b}\Big)\approx 2\pi\sqrt{\frac{c}{6}\Big(h_p-\frac{c}{24}\Big)},
    \end{align}
    in the $p\to\infty$ limit (see eqs.\,\eqref{eq: central charge}, \eqref{eq: conformal weight}).
    \par We now explore the constraints on the general fusion kernel and the general modular $S$-kernel.
    The first three indicate that applying $F$ and $S$ twice will return the block to its original state.

    \begin{tcb}[\textcolor{black}{\textbf{Invertiblity of modular $S$-kernel}}]
      \begin{align}
        \int_{0}^{\infty}dp_2\,S_{p_1p_2}\qty[p_0]S_{p_2p_3}\qty[p_0]=e^{\pi ih_0}\delta(p_1-p_3) \label{eq: projective invertibility of S}
      \end{align}
      \begin{align}
        \int_{0}^{\infty}dp_2\,S_{p_1p_2}\qty[p_0]S_{p_2p_3}^*\qty[p_0]=\delta(p_1-p_3) \label{eq: unitarity of S-kernel}
      \end{align}
    \end{tcb}
    \begin{tcb}[\textcolor{black}{\textbf{invertibility of fusion kernel}}]
      \begin{align}
        \int_{0}^{\infty}dp_t\,F_{p_sp_t}
        \begin{bmatrix}
          p_3 & p_2 \\
          p_4 & p_1
        \end{bmatrix}
        F_{p_tp_u}
        \begin{bmatrix}
          p_4 & p_3 \\
          p_1 & p_2
        \end{bmatrix}
        =\delta(p_s-p_u) \label{eq: invertibility of F}
      \end{align}
    \end{tcb}
    $S$ satisfies two different equations depending on whether it is paired with itself or its complex conjugated version, while $F$ has no such mutations.
    The next two are not only highly powerful constraints but also come with a variety of insightful implications.
    \begin{tcb}[\textcolor{black}{\textbf{Hexagon identity}}]
      \begin{gather}
        \int_{0}^{\infty}dp_t\,e^{\pi i\qty(\sum_{i=1}^{4}h_i-h_s-h_t-h_u)}F_{p_sp_t}
        \begin{bmatrix}
          p_3 & p_2 \\
          p_4 & p_1
        \end{bmatrix}
        F_{p_tp_u}
        \begin{bmatrix}
          p_1 & p_3 \\
          p_4 & p_2
        \end{bmatrix}
        =F_{p_sp_u}
        \begin{bmatrix}
          p_3 & p_1 \\
          p_4 & p_2
        \end{bmatrix}
        \label{eq: hexagon 1} \\
        \iff \int_{0}^{\infty}dp_t\,e^{\pi i\qty(h_s+h_t+h_u-\sum_{i=1}^{4}h_i)}F_{p_sp_t}
        \begin{bmatrix}
          p_3 & p_2 \\
          p_4 & p_1
        \end{bmatrix}
        F_{p_tp_u}
        \begin{bmatrix}
          p_1 & p_3 \\
          p_4 & p_2
        \end{bmatrix}
        =F_{p_sp_u}
        \begin{bmatrix}
          p_3 & p_1 \\
          p_4 & p_2
        \end{bmatrix}
        \label{eq: hexagon 2}
      \end{gather}
    \end{tcb}
    From the first to the second line, we multiply both sides by $e^{\pi ih_u}F_{p_up_a}
      \begin{bmatrix}
        p_4 & p_1 \\
        p_2 & p_3
      \end{bmatrix}
    $, integrate over $p_u$ and use the identity \eqref{eq: invertibility of F}, then rewrite the subscript as $u\to t$, $a\to u$ and $1\leftrightarrow 2$. \\
    \begin{tcb}[\textcolor{black}{\textbf{Pentagon identity}}]
      \begin{align}
        \int_{0}^{\infty}dp_t\,F_{p_sp_t}
        \begin{bmatrix}
          p_3 & p_2 \\
          p_u & p_1
        \end{bmatrix}
        F_{p_up_v}
        \begin{bmatrix}
          p_4 & p_t \\
          p_5 & p_1
        \end{bmatrix}
        F_{p_tp_w}
        \begin{bmatrix}
          p_4 & p_3 \\
          p_v & p_2
        \end{bmatrix}
        =F_{p_sp_v}
        \begin{bmatrix}
          p_w & p_2 \\
          p_5 & p_1
        \end{bmatrix}
        F_{p_up_w}
        \begin{bmatrix}
          p_4 & p_3 \\
          p_5 & p_s
        \end{bmatrix}
        \label{eq: pentagon}
      \end{align}
    \end{tcb}
    Readers may wonder if we could obtain a new formula when multiplying both sides by $F_{p_wp_x}
      \begin{bmatrix}
        p_v & p_4 \\
        p_2 & p_3
      \end{bmatrix}
    $, integrate over $p_w$ and use the identity \eqref{eq: invertibility of F} in the same way as the second eq.\,\eqref{eq: hexagon 2}, but it turns out that the result is nothing but eq.\,\eqref{eq: pentagon} itself.
    If we set $p_s=p_3$, $p_u=\mathbbm{1}$, $p_4=p_5$ to apply the formula \eqref{eq: F-kernel limit 2}, and change subscripts as $1\to s$, $2\to 1$, $3\to 2$, $4\to 3$, $v\to 4$, $w\to t$, we obtain
    \begin{tcb}[Corollary 1: Tetrahedral symmetry of fusion kernel]
      \begin{align}
        F_{\mathbbm{1}p_4}
        \begin{bmatrix}
          p_3 & p_s \\
          p_3 & p_s
        \end{bmatrix}
        F_{p_sp_t}
        \begin{bmatrix}
          p_3 & p_2 \\
          p_4 & p_1
        \end{bmatrix}
        =F_{p_2p_4}
        \begin{bmatrix}
          p_t & p_1 \\
          p_3 & p_s
        \end{bmatrix}
        F_{\mathbbm{1}p_t}
        \begin{bmatrix}
          p_3 & p_2 \\
          p_3 & p_2
        \end{bmatrix}
        \label{eq: tetrahedral symmetry}
      \end{align}
    \end{tcb}
    A brief reflection with reference to eq.\,\eqref{eq: tetrahedral symmetry} uncovers a property of fusion kernel $F$ as follows.
    \begin{tcb}[Corollary 2]
      \begin{align}
        F_{\mathbbm{1}p}
        \begin{bmatrix}
          p_2 & p_1 \\
          p_2 & p_1
        \end{bmatrix}
        =\rho_0\qty(p)C_0\qty(p_1,p_2,p) \label{eq: F-kernel limit 1}
      \end{align}
    \end{tcb}
    In this identity, $C_0(p_1,p_2,p_3)$ is a certain totally symmetric function and $\rho_0(p)$ is some function of $p$ whose explicit forms are to be determined below.
    Taking into account the degenerate fusion rule and the shift relation (see Subsection 2.7 in Ref.\,\cite{Eberhardt:2023mrq}), it turns out that
    \begin{align}
      C_0\qty(p_1,p_2,p_3)=\frac{1}{\sqrt{2}}\frac{\Gamma_b\qty(2Q)}{\Gamma_b\qty(Q)^3}\frac{\Gamma_b\big(\tfrac{Q}{2}\pm ip_1\pm ip_2\pm ip_3\big)}{\prod_{a=1,2,3}\Gamma_b\qty(Q+2ip_a)\Gamma_b\qty(Q-2ip_a)}. \label{eq: universal three point coefficient}
    \end{align}
    We also see by taking the limit $p_1\to\mathbbm{1}$ in eq.\,\eqref{eq: F-kernel limit 1}, combined with eq.\,\eqref{eq: F-kernel limit 2} that
    \begin{tcb}[Corollary 3]
      \begin{align}
        \lim_{p_1\to\mathbbm{1}}C_0(p_1,p_2,p)=\frac{1}{\rho_0(p)}\delta(p-p_2) \label{eq: structure constant limit}
      \end{align}
    \end{tcb}
    Starting with the invertibility of the fusion kernel $F$ \eqref{eq: invertibility of F} and using the tetrahedral symmetry \eqref{eq: tetrahedral symmetry},
    \begin{align}
      & \delta(p_s-p_u)=\int_{0}^{\infty}dp_t\,F_{p_sp_t}
      \begin{bmatrix}
        p_3 & p_2 \\
        p_4 & p_1
      \end{bmatrix}
      F_{p_tp_u}
      \begin{bmatrix}
        p_4 & p_3 \\
        p_1 & p_2
      \end{bmatrix}
      \nonumber \\
      & \qquad\quad=\int_{0}^{\infty}dp_t\,\frac{\rho_0(p_t)C_0(p_t,p_2,p_3)}{\rho_0(p_4)C_0(p_4,p_s,p_3)}F_{p_2p_4}
      \begin{bmatrix}
        p_t & p_1 \\
        p_3 & p_s
      \end{bmatrix}
      F_{p_tp_u}
      \begin{bmatrix}
        p_4 & p_3 \\
        p_1 & p_2
      \end{bmatrix}
      \nonumber \\
      & \qquad=\int_{0}^{\infty}dp_t\,\frac{\rho_0(p_t)C_0(p_t,p_2,p_3)}{\rho_0(p_4)C_0(p_4,p_s,p_3)}F_{p_2p_4}
      \begin{bmatrix}
        p_1 & p_t \\
        p_s & p_3
      \end{bmatrix}
      F_{p_tp_u}
      \begin{bmatrix}
        p_4 & p_3 \\
        p_1 & p_2
      \end{bmatrix}
      \nonumber \\
      & \quad=\int_{0}^{\infty}dp_t\,\frac{\rho_0(p_t)C_0(p_t,p_2,p_3)\cancel{\rho_0(p_4)}C_0(p_t,p_1,p_4)}{\cancel{\rho_0(p_4)}C_0(p_4,p_s,p_3)\rho_0(p_s)C_0(p_2,p_1,p_s)}F_{p_tp_s}
      \begin{bmatrix}
        p_4 & p_3 \\
        p_1 & p_2
      \end{bmatrix}
      F_{p_tp_u}
      \begin{bmatrix}
        p_4 & p_3 \\
        p_1 & p_2
      \end{bmatrix}
      .
    \end{align}
    We can rearrange this into a version of fusion kernel invertibility formula \eqref{eq: invertibility of F}:
    \begin{tcb}[Corollary 4]
      \begin{align}
        & \int_{0}^{\infty}dp_t\,\rho_0(p_t)C_0(p_1,p_4,p_t)C_0(p_2,p_3,p_t)F_{p_tp_s}
        \begin{bmatrix}
          p_1 & p_2 \\
          p_4 & p_3
        \end{bmatrix}
        F_{p_tp_u}
        \begin{bmatrix}
          p_1 & p_2 \\
          p_4 & p_3
        \end{bmatrix}
        \nonumber \\
        & \qquad\qquad\qquad\qquad\qquad\qquad=\rho_0(p_s)C_0(p_1,p_2,p_s)C_0(p_3,p_4,p_s)\delta(p_s-p_u) \label{eq: version of invertibility}
      \end{align}
    \end{tcb}
    Given all these ingredients, the shift equation (3.22) in \cite{Eberhardt:2023mrq} demonstrates that in terms of the Barnes double gamma function $\Gamma_b(z)$ and the double sine function $S_b(z):=\frac{\Gamma_b(z)}{\Gamma_b(Q-z)}$,
    \begin{align}
      F_{p_sp_t}
      \begin{bmatrix}
        p_1 & p_2 \\
        p_3 & p_4
      \end{bmatrix}
      & =\frac{\Gamma_b(Q\pm 2ip_s)}{\Gamma_b(\pm 2ip_t)}\frac{\Gamma_b(\tfrac{Q}{2}\pm ip_t\pm ip_3-ip_4)\Gamma_b(\tfrac{Q}{2}\pm ip_t+ip_1\pm ip_2)}{\Gamma_b(\tfrac{Q}{2}\pm ip_s\pm ip_2-ip_4)\Gamma_b(\tfrac{Q}{2}\pm ip_s+ip_1\pm ip_3)} \nonumber \\
      & \quad\times(-i)\int_{\tfrac{Q}{4}+i\mathbb{R}}dz\,\frac{S_b(z+(\pm ip_2-ip_4))S_b(z+(ip_1\pm ip_3))}{S_b(z+(\tfrac{Q}{2}\pm ip_t+ip_1-ip_4))S_b(z+(\tfrac{Q}{2}\pm ip_s))}. \label{eq: fusion kernel}
    \end{align}
    It is worth noting that $F_{p_sp_t}$ is real for $p_1,\cdots,p_4,p_s,p_t\in\mathbb{R}_{\geq 0}$.
    \par Lastly, we now move on to introduce the other consistency equation to determine $S_{p_1p_2}\qty[p_0]$.
    The equation involves all three basic crossing moves on a 2-punctured torus.
    \begin{tcb}[\textcolor{black}{\textbf{Consistency on the two-punctured torus}}]
      \begin{align}
        & S_{p_1p_2}\qty[p_3]\int_{0}^{\infty}dp_4\,F_{p_3p_4}
        \begin{bmatrix}
          p_2 & p_0 \\
          p_2 & p_0
        \end{bmatrix}
        e^{2\pi i(h_4-h_2)}F_{p_4p_5}
        \begin{bmatrix}
          p_0 & p_0 \\
          p_2 & p_2
        \end{bmatrix}
        \nonumber \\
        & \qqquad\qquad=\int_{0}^{\infty}dp_6\,F_{p_3p_6}
        \begin{bmatrix}
          p_1 & p_0 \\
          p_1 & p_0
        \end{bmatrix}
        F_{p_1p_5}
        \begin{bmatrix}
          p_0 & p_0 \\
          p_6 & p_6
        \end{bmatrix}
        e^{\pi i(2h_0-h_5)}S_{p_6p_2}\qty[p_5] \label{eq: consistency on the two-punctured torus}
      \end{align}
    \end{tcb}
    If we take the limit $p_3\to i\frac{Q}{2}$, $p_1\to i\frac{Q}{2}$ in this order, the first fusion kernel $F$ in the integrand of the second line converges to the limit
    \begin{align}
      \lim_{p_1\to i\frac{Q}{2}}\lim_{p_3\to i\frac{Q}{2}}F_{p_3p_6}
      \begin{bmatrix}
        p_1 & p_0 \\
        p_1 & p_0
      \end{bmatrix}
      =\lim_{p_1\to i\frac{Q}{2}}\rho_0(p_6)C_0(p_0,p_1,p_6)=\delta(p_0-p_6),
    \end{align}
    which reduces eq.\,\eqref{eq: consistency on the two-punctured torus} to
    \begin{align}
      S_{\mathbbm{1}p_2}\qty[\mathbbm{1}]\int_{0}^{\infty}dp_4\,F_{\mathbbm{1}p_4}
      \begin{bmatrix}
        p_2 & p_0 \\
        p_2 & p_0
      \end{bmatrix}
      e^{2\pi i(h_4-h_2)}F_{p_4p_5}
      \begin{bmatrix}
        p_0 & p_0 \\
        p_2 & p_2
      \end{bmatrix}
      =F_{\mathbbm{1}p_5}
      \begin{bmatrix}
        p_0 & p_0 \\
        p_0 & p_0
      \end{bmatrix}
      e^{\pi i(2h_0-h_5)}S_{p_0p_2}\qty[p_5]. \label{eq: reduced two-punctured torus consistency}
    \end{align}
    After renaming $1\to 2$, $5\to 0$ and $4\to a$, we land on the relation
    \begin{tcb}[Corollary 5: Relation between $S$-kernel and $F$-symbol]
      \begin{align}
        S_{p_1p_2}\qty[p_0] & =S_{\mathbbm{1}p_2}\qty[\mathbbm{1}]\int_{0}^{\infty}dp_a\,\frac{\rho_0\qty(p_a)C_0(p_1,p_2,p_a)}{\rho_0(p_0)C_0(p_1,p_1,p_0)}e^{\pi i(2h_a-2h_2-2h_1+h_0)}F_{p_ap_0}
        \begin{bmatrix}
          p_1 & p_1 \\
          p_2 & p_2
        \end{bmatrix}
        \label{eq: relation between S and F}\\
        & =\int_{0}^{\infty}dp_a\,\frac{\rho_0\qty(p_2)C_0(p_0,p_2,p_2)}{C_0(p_a,p_1,p_2)}e^{\pi i(2h_a-2h_2-2h_1+h_0)}F_{p_0p_a}
        \begin{bmatrix}
          p_2 & p_1 \\
          p_2 & p_1
        \end{bmatrix}
      \end{align}
    \end{tcb}
    This corollary states that the modular $S$-kernel is fully determined by the fusion kernel $F$, so in order to tell the exact form of them it suffices to compute $F$, which is already obtained in  eq.\,\eqref{eq: fusion kernel}.
    We also see from the first identity that the r.\,h.\,s.~of
    \begin{align}
      \frac{S_{p_1p_2}\qty[p_0]}{S_{\mathbbm{1}p_2}\qty[\mathbbm{1}]}C_0(p_1,p_1,p_0)=\int_{0}^{\infty}dp_a\,\frac{\rho_0\qty(p_a)C_0(p_1,p_2,p_a)}{\rho_0(p_0)}e^{\pi i(2h_a-2h_2-2h_1+h_0)}F_{p_ap_0}
      \begin{bmatrix}
        p_1 & p_1 \\
        p_2 & p_2
      \end{bmatrix}
      ,
    \end{align}
    is symmetric under $p_1\leftrightarrow p_2$, leading to the following result:
    \begin{tcb}[Corollary 6: Symmetric property of $S$-kernel]
      \begin{align}
        \frac{S_{p_1p_2}\qty[p_0]}{S_{\mathbbm{1}p_2}\qty[\mathbbm{1}]}C_0(p_1,p_1,p_0)=\frac{S_{p_2p_1}\qty[p_0]}{S_{\mathbbm{1}p_1}\qty[\mathbbm{1}]}C_0(p_2,p_2,p_0) \label{eq: S-kernel lael exchange}
      \end{align}
    \end{tcb}
    Taking the limit $p_0\to\mathbbm{1}$ and using the property \,\eqref{eq: structure constant limit} lead to the equality
    \begin{tcb}[Corollary 7]
      \begin{align}
        \rho_0(p)=S_{\mathbbm{1}p}\qty[\mathbbm{1}]=4\sqrt{2}\sinh\qty(2\pi bp)\sinh\qty(2\pi\frac{p}{b}) \label{eq: Cardy density}
      \end{align}
    \end{tcb}
    The relation between modular $S$-kernel and fusion kernel \eqref{eq: relation between S and F}, along with eq.\,\eqref{eq: fusion kernel} leads to the precise form
    \begin{align}
      & S_{p_1p_2}\qty[p_0]=S_b\qty(\tfrac{Q}{2}-p_0)\rho_0\qty(p_2)\frac{\Gamma_b\qty(Q\pm 2p_1)\Gamma_b\qty(\tfrac{Q}{2}-p_0\pm 2p_2)}{\Gamma_b\qty(Q\pm 2p_2)\Gamma_b\qty(\tfrac{Q}{2}+p_0\pm 2p_1)} \nonumber \\
      & \qquad\times e^{\frac{\pi i}{2}\qty(-p_0Q-2p_0^2)+2\pi i\qty(p_1^2+p_2^2)}\cdot\frac{1}{2i}\int_{0}^{\infty}dp\,e^{-2\pi ip^2}\frac{S_b\qty(\tfrac{Q}{4}+\frac{p_0}{2}\pm p_1\pm p_2\pm p)}{S_b\qty(\pm 2p)}. \label{eq: modular S-kernel}
    \end{align}
    By this identity it also holds that $S_{p_1p_2}^*\qty[p_0]=e^{-\pi ih_0}S_{p_1p_2}\qty[p_0]$ in accordance with eqs.\,\eqref{eq: projective invertibility of S}, \eqref{eq: unitarity of S-kernel}.

  \section{From monoidal category to fusion category and modular tensor category} \label{append: From monoidal category to fusion category and modular tensor category}
    In lower-dimensional TQFT, the mathematics of primary importance includes category theory, particularly fusion categories and modular tensor categories, both of which are grounded in the theory of tensor categories.
    A tensor category is a specific version of a more fundamental structure, a monoidal category that is a category with a product structure (\textit{tensor product}) and a special object (\textit{unit object}).
    Although the terms ``monoidal category" and ``tensor category" are sometimes used interchangeably, we treat them as distinct concepts in line with the polished textbook \cite{etingof2015tensor}.
    The rigorous construction from monoidal categories to fusion categories and modular tensor categories through tensor categories is exquisitely explored in Refs.\,\cite{etingof2015tensor, kong2014anyon}.
    However, the construction is fairly lengthy due to the need for a number of interweaving definitions such as ``braided", ``locally finite", ``sovereign", ``spherical" and others, which may introduce unnecessary complexities when interpreting physical implications.
    Here we elucidate their relations rather than to just itemize the cluster of precise definitions.
    \par Let us examine the categorical concepts that constitute fusion categories and modular tensor categories (MTC).
    2d TQFTs are characterized by fusion categories, while 3d TQFTs are by MTC.
    There are three significant classes of fundamental categories needed to formulate tensor categories (see Definition 4.1.1 of \cite{etingof2015tensor}), namely (i) monoidal category (ii) Abelian category (iii) $\mathbb{C}$-linear category depicted by bold rectangles in Figure \ref{fig: inclusion relation}.
    \begin{figure}[t]
      {
        \begin{tikzpicture}[scale = 0.9]
          % monoidal category
          \draw[ultra thick] (-5, 4) rectangle (6, -4);
          \node[fill = white] at (0.5, 4) {\textbf{monoidal category}};
          % rigid category
          \draw (-4.5, 3.5) rectangle (4.5, -3.5);
          \node[fill = white] at (0, 3.5) {rigid};
          % braided category
          \draw (-3.5, 3) rectangle (5.5, -3);
          \node[fill = white] at (1, 3) {braided};
          % ribon category
          \draw (-3.25, 2.5) rectangle (3.75, -2.5);
          \node[fill = white] at (0.25, 2.55) {ribbon};
          % $\mathbb{C}$-linear category
          \draw[ultra thick] (-4, 2) rectangle (7, -5.75);
          % $\dim\Hom_{\mathcal{C}}(\mathbbm{1},\mathbbm{1})=1$
          \draw (-2.5, 1.5) rectangle (6.5, -4.75);
          \node[fill = white] at (2, 1.4) {$\dim_{\mathbb{C}}\Hom_{\mathcal{C}}(\mathbbm{1},\mathbbm{1})=1$};

          % Abelian category
          \draw[ultra thick] (-7, 1) rectangle (4.25, -6.75);
          \node[fill = white] at (-1.375, -6.75) {\textbf{Abelian category}};
          % semisimple
          \draw (-6.5, -0.5) rectangle (3.25, -6.25);
          \node[fill = white] at (-1.625, -6.25) {semisimple};

          % locally finite
          \draw (-3, 0.5) rectangle (4, -5.25); % label is attached lastly below
          % finite
          \draw (-2.75, 0) rectangle (3.5, -5);
          \node[fill = white] at (0.375, 0) {finite};

          % tensor category
          \draw[very thick, red] (-2.47, 0.47) rectangle (3.97, -3.47);
          \draw[->, > = stealth, red, bend right = 30] (6.5, -6.5) to (4.1, -2);
          \node[red, below] at (6.5, -6.5) {\textbf{tensor category}};
          % fusion category
          \draw[very thick, blue] (-2.42, -0.53) rectangle (3.22, -3.42);
          \draw[->, > = stealth, blue, bend left = 25] (-5.5, 2) to (-1.5, -0.4);
          \node[blue, left] at (-5.5, 2) {\textbf{fusion category}};
          % modular tensor category
          \draw[very thick, olive] (-2.2, -0.75) rectangle (3, -2.3);
          \node[olive] at (0.4, -1.225) {\textbf{modular tensor}};
          \node[olive] at (0.4, -1.825) {\textbf{category}};
          % VTQFT
          \draw[very thick, green!70!black] (-2.47, 0.94) rectangle (3.72, 0.53);

          % "$\mathbb{C}$-linear category" label
          \node[fill = white] at (1.5, 2) {\textbf{$\mathbb{C}$-linear category}};
          % "locally finite" label
          \draw[green!70!black] (-2.47, 0.94) rectangle (3.72, 0.53);
          \node[fill = white] at (0.5, 0.5) {locally finite};
          \draw[olive] (-2.2, -0.75) rectangle (3, -2.3);
        \end{tikzpicture}
      }
      \caption{Inclusion relation for various categorical definitions. Surrounded by the red rectangle is the family of tensor categories.
      The fusion categories are marked inside the blue rectangle and categories in the olive region satisfying the modularity condition are MTC.
      The green region is where the symmetry category of VTQFT is located. All the colored regions are inside the intersection of (i)\,monoidal categories (ii)\,Abelian categories (iii)\,$\mathbb{C}$-linear categories.}
      \label{fig: inclusion relation}
    \end{figure}

    \subsection*{Monoidal category}
      \begin{definition}[Monoidal category]
        A \bit{monoidal category} $\mathcal{C}$ is a category $\mathcal{C}$ endowed with the following data:
        \begin{itemize}
          \item A bifunctor $\otimes:\mathcal{C}\times\mathcal{C}\to\mathcal{C}$ called the \bit{tensor product}
          \item A natural isomorphism $\alpha$ called the \bit{associator} (\textit{associativity isomorphism}) $\alpha_{X,Y,Z}:(X\otimes Y)\otimes Z\overset{\sim}{\longrightarrow}X\otimes(Y\otimes Z)$ for all $X,Y,Z\in\Obj\mathcal{C}$
          \item An object $\mathbbm{1}\in\Obj\mathcal{C}$ called the \bit{unit}
        \end{itemize}
        satisfying the \bit{pentagon identity} and the \bit{unit axiom} (see Definition 2.1.1 of \cite{etingof2015tensor} for details).
      \end{definition}
      The central idea is the existence of a tensor product and a unit comensurate with particle fusion and the identity line.
      Two specific classes of monoidal categories capture a physically relevant structure in TQFT.
      \begin{definition}[Rigid category]
        A \bit{rigid category} is a monoidal category where every object has left and right duals (see Definition 2.10.11 of \cite{etingof2015tensor} for details).
      \end{definition}
      In most physical examples, the ``dual" of an object is the object itself, so the ``left" and ``right" are unimportant.
      \begin{definition}[Braided monoidal category]
        A \bit{braided monoidal category} is a monoidal category equipped with a natural transformation $R_{X,Y}:X\otimes Y\overset{\sim}{\to}Y\otimes X$ called \bit{braiding} satisfying the \bit{hexagon identity} (see Definition 8.1.1 of \cite{etingof2015tensor} for details).
      \end{definition}

    \subsection*{Abelian category}
      \begin{definition}[Additive category]
        An additive category $\mathcal{C}$ is a category abiding by the conditions
        \begin{enumerate}
          \item $\Hom_{\mathcal{C}}(X,Y)$ is an abelian group for all $X,Y\in\Obj\mathcal{C}$.
          \item There exists an object $0$ called the \bit{zero} such that $\Hom_{\mathcal{C}}(0,0)=0$
          \item There exists a bifunctor $\oplus:\mathcal{C}\times\mathcal{C}\to\mathcal{C}$ satisfying additional conditions (see Definition 1.2.1 of \cite{etingof2015tensor} for details).
        \end{enumerate}
      \end{definition}
      The existence of direct sum is of primary importance since it corresponds to particle superposition.
      An \bit{Abelian category} is an additive category with a well-defined notion of kernel $\ker$ and image $\im$. We omit the details since they are immaterial in physics (see Definition 1.3.1 in \cite{etingof2015tensor}).
      \begin{definition}[Semisimple category]
        An additive category $\mathcal{C}$ is semisimple if every $X\in\Obj\mathcal{C}$ is a direct sum of simple objects (see Definition 1.5.1 of \cite{etingof2015tensor}).
      \end{definition}
      VTQFT does not satisfy the assumption in this definition when it comes to a continuous superposition of Wilson lines.
      Semisimplicity is essential for formulating the direct sum in a mathematically rigorous manner.

    \subsection*{$\mathbb{C}$-linear category}
      \begin{definition}[$\mathbb{C}$-linear category]
        A \bit{$\mathbb{C}$-linear category} is a $\cat{Vect}_{\mathbb{C}}$-enriched category, that is, a category where $\Hom_{\mathcal{C}}(X,Y)$ is a $\mathbb{C}$-vector space for all $X,Y\in\Obj\mathcal{C}$.
      \end{definition}
      If $\mathcal{C}$ is additionally an Abelian category, the condition $\dim_{\mathbb{C}}\Hom_{\mathcal{C}}(\mathbbm{1},\mathbbm{1})=1$ is equivalent to saying that $\mathbbm{1}$ is a simple object in $\mathcal{C}$.
    ~ \\
    \par There are other definitions shown in Figure \ref{fig: inclusion relation}, say \bit{local finiteness} (Definition 1.8.1) and \bit{finiteness} (Definition 1.8.5).
    Readers may refer to the indicated definition number in \cite{etingof2015tensor}.
    A \bit{tensor category} is a locally finite $\mathbb{C}$-linear Abelian rigid monoidal category with $\dim\Hom_{\mathcal{C}}(\mathbbm{1},\mathbbm{1})=1$, displayed by the red rectangle in Figure \ref{fig: inclusion relation}.
    A \bit{fusion category} is a finite semisimple tensor category presented by the blue rectangle. A {modular tensor category} is a ribbon fusion category with an additional condition (modularity condition, see Definition A.12 in Ref.\,\cite{kong2014anyon}) shown by the olive rectangle.

  \bibliographystyle{JHEP}
  \bibliography{main}

\end{document}